\documentclass[referee]{aa} 
\usepackage[varg]{txfonts}
\usepackage{graphicx}
\usepackage{natbib}
\usepackage{amssymb}
\usepackage[usenames]{color}

\usepackage{etex}
\reserveinserts{18}
\usepackage{morefloats}

\begin{document}
   \title{Accelerated post-AGB evolution, initial-final mass relations, and
     the star-formation history of the Galactic bulge\thanks{Based on
       observations collected at the European Organisation for Astronomical
       Research in the Southern Hemisphere, Chile (proposal 075.D-0104) and
       HST (program 9356)} }

   \subtitle{}

   \author{K. Gesicki\inst{1} \and A. A. Zijlstra\inst{2}
           \and M. Hajduk\inst{1,3} \and C. Szyszka\inst{1,2}
	  }

   \institute{Centre for Astronomy, 
              Faculty of Physics, Astronomy and Informatics,
              Nicolaus Copernicus University,
              Grudziadzka 5, 
	      PL-87-100 Torun, 
	      Poland
              \\
              \email{Krzysztof.Gesicki@astri.uni.torun.pl}
              \and
              Jodrell Bank Centre for Astrophysics,
              School of Physics \&\ Astronomy,
	      University of Manchester,
              Oxford Road,
              Manchester M13\,9PL, UK \\
              \email{a.zijlstra@manchester.ac.uk}
               \and
              Nicolaus Copernicus Astronomical Center, 
              ul. Rabia\'{n}ska 8, 87-100 Torun, Poland  
             }

\titlerunning{Post-AGB evolution and the star-formation history of the Galactic bulge}
\authorrunning{K.Gesicki et al.}


  \abstract
   {}
     {We study the star-formation history of the Galactic bulge, as
   derived from the age distribution of the central stars of planetary
   nebulae that belong to this stellar population. }
     { The high resolution imaging and spectroscopic observations of
      31  compact planetary nebulae are used to derive their central
      star masses. We use  the Bl\"ocker  post asymptotic giant branch
      (post-AGB) evolutionary models, which are accelerated by a
      factor of three in this case to better fit the white dwarf mass
      distribution and asteroseismological masses. Initial-final mass
      relations (IFMR) are derived using white dwarfs in clusters.
      These are applied to determine original stellar masses and
      ages.  The age distribution is corrected for observational bias
      as a function of stellar mass.  We predict that there are about
      2000 planetary nebulae in the bulge.  }
{The planetary nebula population points at a young bulge population
  with an extended star-formation history. The Bl\"ocker tracks with
  the cluster IFMR result in ages, which are unexpectedly young. We
  find that the Bl\"ocker post-AGB tracks need to be accelerated by a
  factor of three  to fit the local white dwarf masses. This
  acceleration extends the age distribution.  We adjust the IFMR as a
  free parameter to map the central star ages on the full age range of
  bulge stellar populations.  This fit requires a steeper IFMR than
  the cluster relation. We find a  star-formation rate in the Galactic
  bulge, which is approximately constant between 3 and 10\,Gyr ago. 
  The result indicates that planetary nebulae are mainly associated
  with the younger and more metal-rich bulge populations.  }
   {The constant rate of star-formation between 3 and 10\,Gyr
    agrees with suggestions that the metal-rich component of the
    bulge is formed during an extended process, such as a bar
    interaction.}

   \keywords{ISM: planetary nebulae: general -- 
             Stars: AGB and post-AGB -- Galaxy: bulge
            }
   \maketitle
%

\section{Introduction}

The origin and evolution of the Galactic bulge (GB) of the Milky Way
is an area of active research. Bulges were believed to be similar to
elliptical galaxies with a proposed origin in minor mergers. This was
supported by the relation between bulge/spheroid luminosities and
central black-hole masses \citep{G2007}. However, some bulges are more
disk-like in their properties and these have been termed
'pseudo-bulges': their origin is likely unrelated to those of the
spheroidal bulges \citep{KK2004}. Pseudo-bulges can form by secular
evolution over a longer period, while classical bulges are
predominantly an early product of galaxy formation.

Whether the Milky Way bulge is a classical or pseudo-bulge is
disputed. There is evidence for old stellar populations. \citet{V2009}
compared OGLE and 2MASS data to the galactic model TRILEGAL to find
good fits for models with a starburst from 8 to 10\,Gyr ago with the
best fit for the younger age. The metallicity is close to solar with a
distribution extending not very far towards lower values. In recent
years, however, the evidence for a younger stellar component has been
growing.  In a recent review, \citet{Babu2012} concluded that the
other characteristics point toward a mix of stellar populations
although the main shape of the GB seems to be driven by secular
evolution.  This multi-component structure is also seen in the shape
of the GB \citep{Robin2012}, which may be a combination of a classical
and a pseudo (box-shape) bar.

Regarding the age and duration of star-formation in GB,
\citet{Tsuji2012} concluded that the metal-poor and metal-rich
populations are formed at different times with the former extending
over 2\,Gyr and the latter over 4\,Gyr.  This separation was also
found by \citet{Bensby2010, Bensby2011, Bensby2013}. They carried out
spectroscopy of GB dwarf stars during gravitational lensing
amplification events.  Stellar parameters were derived based on LTE
model atmospheres, and stellar ages were interpolated from a grid of
isochrones. Bensby et al. found that stars with sub-solar
metallicities are predominantly old (around 10\,Gyr) as expected in
the GB. At super-solar metallicities, however, they find a wide range
of ages with a peak around 5\,Gyr and a tail towards higher ages. They
conclude that the origin of the GB is still poorly constrained and
speculate that the young stars might be the manifestation of the inner
thin disk.  \citet{Gonzalez2011} also found evidence for a separate,
high metallicity component of the bulge, which they attribute to stars
originating from the thin disk.

Planetary nebulae (PN), which are bright and easy to identify, can
help us to address the question of the age of the Galactic bulge. The
PNe trace populations which have ages ranging from $<1$ to 10\,Gyr. In
contrast, red clump data analysis, as used by \citet{Robin2012},
cannot discriminate any ages over 5\,Gyr.  However, deriving ages for
the central stars of individual PNe has been a challenge.

In this paper, we investigate the relation between the stellar
populations and the planetary nebula population of the Galactic bulge
by deriving the initial masses and, therefore, the ages of the PN
central stars. This makes use of new Hubble Space telescope (HST)
images and complementary Very Large Telescope (VLT) high-resolution
spectra to find the mass distribution of the central stars of bulge
planetary nebulae. These are converted to initial masses using
initial-final mass relations (IFMR); the stellar ages are found using
the stellar mass-age relations. The central star mass distribution can
be compared with a well-known distribution of white dwarf masses and
the age distribution can be compared with ages of bulge stars.

The steps are critically dependent on post asymptotic giant branch
(post-AGB) stellar evolution models and the IFMR. We show that a
recalibration of the speed in the post-AGB evolution is required to
reproduce the well-known distribution of white dwarf masses.  The IFMR
is used as a free parameter to fit known bulge ages.  The final
conversion to a star-formation history also requires corrections for
observational bias.  After these corrections, we find that the results
are consistent with an extended epoch of bulge formation.

The paper is organized as follows. Section\,\ref{obse_anal} presents
the collected observational material and its analysis with
photoionization models. In Sect.\,\ref{final_m}, masses of central
stars of PNe are derived and compared to the known white dwarf masses.
The discrepancies can be resolved with a proposed acceleration of
post-AGB evolution. In Sect.\,\ref{ini_fin_m}, the stellar initial
masses and ages from the zero-age main sequence (ZAMS) are derived
using a newly-derived IFMR and some evolutionary model tracks. 
Section\,\ref{pn_popul} considers the effect of selection and PN
visibility bias. Finally, the adjusted age distribution of the GB PNe
is derived in Sect.\,\ref{sfh_b}, where we use the IFMR as a free
parameter. When compared to other published bulge age determinations,
it points towards a rather steep IFMR.

\section{Observations and analysis}
\label{obse_anal}

\subsection{The HST images}

We obtained SNAPshot HST/WFPC2 images of  compact bulge PNe (program
9356). The data were obtained in 2002 and 2003 and comprised of images
in three different filters: F656N, F547M, and F502N for a total of one
HST orbit (96 min including overheads) per object. The pixel scale of
the camera is 0.0455\,arcsec.

To get as close as possible to a randomly selected sample for this
SNAPshot survey, we first identified all compact bulge PNe in the
ESO/Strasbourg catalogue \citep{Acker1992} with a reported diameter of
less than 5\,arcsec or of unknown diameter. Sixty objects were
randomly selected from this list, which is  about half the total
number. These were entered in the SNAPshot survey without further
prioritization. Based only on schedulability, 37 of these were
actually observed.  One target turned out to be too faint for further
analysis. This leaves a sample of 36 compact objects, representing the
GB PNe population. It forms an unbiased selection from the compact PNe
in the Acker catalogue and is not based on any brightness or
morphological information. Inevitably, it is still affected by
discovery biases inherent in the original catalogue. 

\subsection{VLT/UVES long-slit spectra}

A series of 600-second VLT/UVES \citep{Dekker2000} echelle spectra
were obtained in 2005 with a slit 0.5\,arcsec wide and 11\,arcsec
long. The slit was put through the centres of the nebulae, which are
aligned with the brightest structures. In most cases, the slit was
aligned with the minor axis, and the spectra therefore represent the
densest part of the nebula.  The observations cover a spectral region
approximately from 3300 to 6600\,\AA.  The spectral resolution of
60\,000 resolves the line profiles, allowing us to determine the
velocity structure of the expanding nebula.

The ESO Common Library (CPL) pipeline was applied (version 4.1.0),
which included correction for bias, flat field, wavelength
calibration, and order merging. Special attention was required for the
order merging process, as there is usually no continuum in PN spectra.
The long-slit spectra are spatially resolved in most cases, however,
only the central arcsecond was extracted for further analysis as the
most representative for the whole PN.

\subsection{Photoionization modelling of the nebulae}
\label{phi_mod}

To model the spectra and line profiles, we applied the Torun
photo-ionization codes \citep{GZ2003, GZAGGW2006}. The star is assumed
to be a black body with luminosity and an effective temperature. The
nebula is approximated as a spherical shell defined by a radial
density distribution and a radial velocity field. The chemical
composition and the observed line intensities were adopted from the
literature. For most of the PNe, the chemical abundances were recently
published by \citet{CGSB2009} and \citet{GCSC2009}, while some data
were used from \citet{SRC1998}. These abundances were used in the
photo-ionization model. The oxygen data are given in
Table\,\ref{pne_data}.

The photo-ionization code calculates a 1-dimensional (radial)
emissivity profile. Assuming spherical symmetry, this is converted to
the observed projected intensity profile and fitted to the HST images.
For this purpose, slices through the images were extracted by running
across the brightest areas, which usually correspond to the minor axis
(Details are presented in the online Appendix.). This gave not only
the outer radius but also an approximation to the density profile.
Once a good fit is obtained from the images and the line flux ratios,
we add a radial velocity profile to the model nebula. The emissivities
are integrated along the line of sight to produce a model line profile
at each position. Profiles of lines with different ionization states,
which trace different radii within the nebula, are fitted to obtain
the radial velocity profile of the nebula. Finally, an average
expansion velocity $V_{\rm aver}$ is calculated weighted over the
ionized mass at each radius.

Four objects indicated turbulent motions within the nebula, which
complicates the velocity interpretation. The lines of different
species show similar line widths with a more Gaussian line shape than
normally seen. The widening can be modelled with a turbulent velocity
component, but the physical meaning of this is not clear. The presence
of extra velocity components makes the derivation of the mass-averaged
expansion velocity more uncertain.  The four objects were therefore
excluded from the sample. One object, Th4-1, was unresolved in HST,
and we were unable to converge on an acceptable fit. This object was
also removed.

The remaining sample of 31 PNe is uniform and covered by high quality
data, which benefits the modelling. The objects are at nearly the same
distance, which allows for robust luminosity and dimensions
estimations. All objects have spatially resolved HST images. Because
of the very high angular resolution of these images, the radial
intensity profiles are well determined, and these, in turn, are used
to derive the radial dependency of the density.  The high resolution
spectra provide a number of emission lines, which cover a reasonably
wide range of ionization and excitation.

For all selected PNe, the most representative observations with the
fitted models are presented in the online Appendix. The HST images,
VLT/UVES emission lines, and model radial runs are shown in
Figs.\,\ref{mod_fit_1}--\ref{mod_fit_last}. The online
Table\,\ref{pne_ratios} presents more details of the fitted
photoionization models.

\subsection{Nebular age determination}

The kinematical age of the nebula can be obtained from the outer
radius divided by its expansion velocity. The outer radius is well
determined through the HST images (in the H$\alpha$ line observed with
filter F656N, see the Figs.\,\ref{mod_fit_1}--\ref{mod_fit_last}). 
However, the velocity at this precise point cannot be measured
spectroscopically. Instead, we use the mass-averaged expansion
velocity, which is well-determined from the observations and models.
It is derived from the hydrogen density distribution and the velocity
field of the ionized region.

The expansion rate of the PN outer radius accelerates over time, at
least as long as the central star heats up. This has been shown in
hydrodynamical modelling by \citet{SJS2005} and confirmed
observationally for bulge PNe by \citet{Richer2008, Richer2010}.  The
velocity field inside the nebular shell typically shows increasing
velocities with radius with the highest gas velocity found at the
outer edge.  The outer radius is the location of a shock and expands
at the shock propagation speed. The mass-weighted expansion velocity
is thus lower than the expansion rate of outer radius. We use the
hydrodynamical models of \citet{Perinotto2004} to calibrate the
relation between the average expansion velocity, the radius, and the
age.  These models present self-consistent calculations of the
evolution of a PN on a post-AGB track with full coverage of radiation
transport, ionization, and hydrodynamics.  We calculate the {\em mean
propagation speed} by dividing the outer radius by the age of the
model nebula. The ratio between this and the mass-averaged velocities
gives the correction factor, which should be applied.

We applied this procedure to a number of models (kindly provided by
the authors) with a nebula evolving around stars of mass
$0.565\,M_{\sun}$ and $0.605\,M_{\sun}$ which are the most comparable
to our sample. A consistent and well-defined ratio of $1.4\pm0.1$ is
found. This agrees well with the correction factor to the expansion
parallaxes of \citet{SJS2005}, which are derived from the shell where
most of the gas resides. The mass-averaged velocity follows the
general acceleration trend and is well-approximated by a constant
correction factor within this accuracy. Thus,

\begin{equation}
  t_{\rm kin} = \frac{R_{\rm out}}{1.4 \times V_{\rm aver}}.
 \label{eq_age}
\end{equation}

The results of the analysis for the 31 objects are listed in
Table\,\ref{pne_data}. The temperatures refer to the black-body
photo-ionization temperatures for the stars. The nebular radius is
measured from the HST images by assuming a distance to the bulge of 8
kpc.

\begin{table*}
\caption{Galactic bulge PNe model results and literature data for
  analysed nebulae. The black-body temperature comes from
  photoionization modelling. The outer radius corresponds to the
  equatorial direction seen in the images observed with HST. The
  kinematical reconstruction provided the mass-averaged velocity and
  kinematical age. The methods for obtaining cspn masses in standard
  and accelerated cases are discussed in the text. }
\begin{flushleft}
\begin{tabular}{llllrllllllll}
\hline\hline
\noalign{\smallskip}
PN\,G & name & $\log T_{\rm bb}$ & $R_{\rm out}$ &  $V_{\rm
  aver}$  & $t_{\rm kin}$ & $M_{\rm cspn}$ &{ $M_{\rm cspn}^{\rm accele}$ } & O/H & ref.  \\
  &   & [K] & [pc] & [km\,s$^{-1}$] & [kyr] & [$M_{\sun}$] &
[$M_{\sun}$]  & 12+log$x$ &    \\
\noalign{\smallskip}
\hline
\noalign{\smallskip}
001.7-04.4   &  H 1-55       &  4.53    &   0.07   &    26    &  1.88  & 0.603  &  0.565  &  8.45    & 1  \\
002.3-03.4   &  H 2-37       &  5.07    &   0.12   &    29    &  2.89  & 0.628  &  0.590  &  8.06    & 1  \\
002.8+01.7   &  H 2-20       &  4.50    &   0.065  &    29    &  1.57  & 0.607  &  0.569  &  9.26    & 2  \\
002.9-03.9   &  H 2-39       &  5.04    &   0.11   &    39    &  1.97  & 0.639  &  0.601  &  8.41    & 1  \\
003.1+03.4   &  H 2-17       &  4.49    &   0.08   &    24    &  2.33  & 0.592  &  0.555  &  7.44    & 2  \\
003.6+03.1   &  M 2-14       &  4.63    &   0.045  &    16    &  1.96  & 0.609  &  0.571  &  8.56    & 3  \\
003.8+05.3   &  H 2-15       &  4.89    &   0.12   &    17    &  4.93  & 0.596  &  0.558  &  8.66    & 2  \\
004.1-03.8   &  KFL 11       &  4.76    &   0.08   &    22    &  2.54  & 0.609  &  0.572  &  8.39    & 1  \\ 
004.8+02.0   &  H 2-25       &  4.49    &   0.10   &    18    &  3.88  & 0.575  &  0.537  &  8.33    & 3  \\
006.1+08.3   &  M 1-20       &  4.81    &   0.044  &    16    &  1.92  & 0.623  &  0.585  &  8.53    & 3  \\
006.3+04.4   &  H 2-18       &  4.71    &   0.11   &    46    &  1.67  & 0.620  &  0.583  &  8.56    & 1  \\
006.4+02.0   &  M 1-31       &  4.68    &   0.04   &    19    &  1.47  & 0.623  &  0.585  &  8.91    & 1  \\
008.2+06.8   &  He 2-260     &  4.44    &   0.02   &    12    &  1.16  & 0.613  &  0.575  &  8.00    & 3  \\
008.6-02.6   &  MaC 1-11     &  4.80    &   0.08   &    28    &  2.00  & 0.621  &  0.583  &  -       &    \\
351.1+04.8   &  M 1-19       &  4.71    &   0.06   &    23    &  1.82  & 0.618  &  0.580  &  8.62    & 3  \\
351.9-01.9   &  Wray 16-286  &  4.92    &   0.06   &    13    &  3.22  & 0.613  &  0.575  &  8.74    & 3  \\
352.6+03.0   &  H 1-8        &  4.83    &   0.07   &    23    &  2.13  & 0.621  &  0.583  &  8.76    & 3  \\
354.5+03.3   &  Th 3-4       &  4.68    &   0.016  &    19    &  0.59  & 0.655  &  0.617  &  8.21    & 3  \\
354.9+03.5   &  Th 3-6       &  4.63    &   0.08   &    16    &  3.49  & 0.589  &  0.551  &  -       &    \\
355.9+03.6   &  H 1-9        &  4.55    &   0.04   &    21    &  1.33  & 0.617  &  0.579  &  8.30    & 1  \\
356.1-03.3   &  H 2-26       &  5.18    &   0.11   &    20    &  3.84  & 0.627  &  0.588  &  8.34    & 2  \\
356.5-03.6   &  H 2-27       &  4.81    &   0.12   &    25    &  3.35  & 0.604  &  0.566  &  -       &    \\
356.8+03.3   &  Th 3-12      &  4.51    &   0.03   &     9    &  2.33  & 0.594  &  0.556  &  -       &    \\
356.9+04.4   &  M 3-38       &  5.09    &   0.02   &     9    &  1.55  & 0.651  &  0.613  &  8.39    & 1  \\
357.1-04.7   &  H 1-43       &  4.26    &   0.04   &    27    &  1.03  & 0.604  &  0.566  &  8.12    & 1  \\
357.2+02.0   &  H 2-13       &  4.95    &   0.09   &     8    &  7.86  & 0.585  &  0.547  &  -       &    \\
358.5-04.2   &  H 1-46       &  4.75    &   0.04   &    14    &  2.00  & 0.618  &  0.579  &  8.27    & 1  \\
358.5+02.9   &  Al 2-F       &  4.95    &   0.08   &    28    &  2.00  & 0.632  &  0.594  &  -       &    \\
358.7+05.2   &  M 3-40       &  4.73    &   0.06   &    22    &  1.90  & 0.618  &  0.580  &  -       &    \\
358.9+03.4   &  H 1-19       &  4.67    &   0.05   &    17    &  2.05  & 0.610  &  0.572  &  8.26    & 3  \\
359.2+04.7   &  Th 3-14      &  4.32    &   0.04   &    15    &  1.86  & 0.588  &  0.550  &  -       &    \\
\noalign{\smallskip}
\hline
\noalign{\smallskip}
\multicolumn{11}{l}{Abundance references:~~ (1) \citet{GCSC2009}; (2) \citet{SRC1998}; (3) \citet{CGSB2009}. }\\
\end{tabular}
\end{flushleft}
\label{pne_data}
\end{table*}

\section{Stellar final masses }
\label{final_m}

In this section, we discuss the method of deriving masses of central
stars of PNe (cspn). We pay special attention to the derivation of the
lowest masses, where some known cspn masses are lower than predicted
by models.

\subsection{Post-AGB model tracks}

The post-AGB evolution shows rapid heating of the star.  The heating
rate can be measured from the stellar temperature and the nebula age.
The post-AGB models show that this heating rate is a sensitive measure
of the stellar mass.

We use the standard set of evolutionary tracks for post-AGB stars
obtained by \citet{B1995}, which are supplemented with lower-mass
models from \citet{Schoenberner1983}. Alternative post-AGB model
tracks are presented by \citet{VW94} and \citet{WF2009}.  We assume
that the post-AGB stars evolve as hydrogen burners. Stars in the
helium burning phase have lower luminosities and evolve slower. Only a
minor fraction of post-AGB stars are expected to be helium burners.

The circumstellar matter is ejected at the tip of the asymptotic giant
branch. The cessation of this ejection is defined as the end of the
AGB.  The time between the end of the AGB and the time when the star
reaches $T_{\rm eff}=10^4\,$K is commonly called the {\em transition
time}. Values for the transition times differ enormously between
different models.  The Bl\"ocker tracks predict transition times that
agree with observational constraints (referring mainly to IR
observations, see \citet{B1995}  and the review of
\citet{SchoBlo1993}). The models of \citet{WF2009} are particularly
fast and the \citet{VW94} (VW94) tracks are very slow.\footnote{The
VW94 start the post-AGB clock at the stellar temperature of $10^4$\,K;
the transition time should be added for comparison with other
models.}  The transition time is not always predicted by models:
sometimes, it is necessary to remove the envelope 'manually' and
resume the models from $T_{\rm eff}=10^4\,$K \citep{WF2009}.  

\begin{figure}
\resizebox{\hsize}{!}{\includegraphics{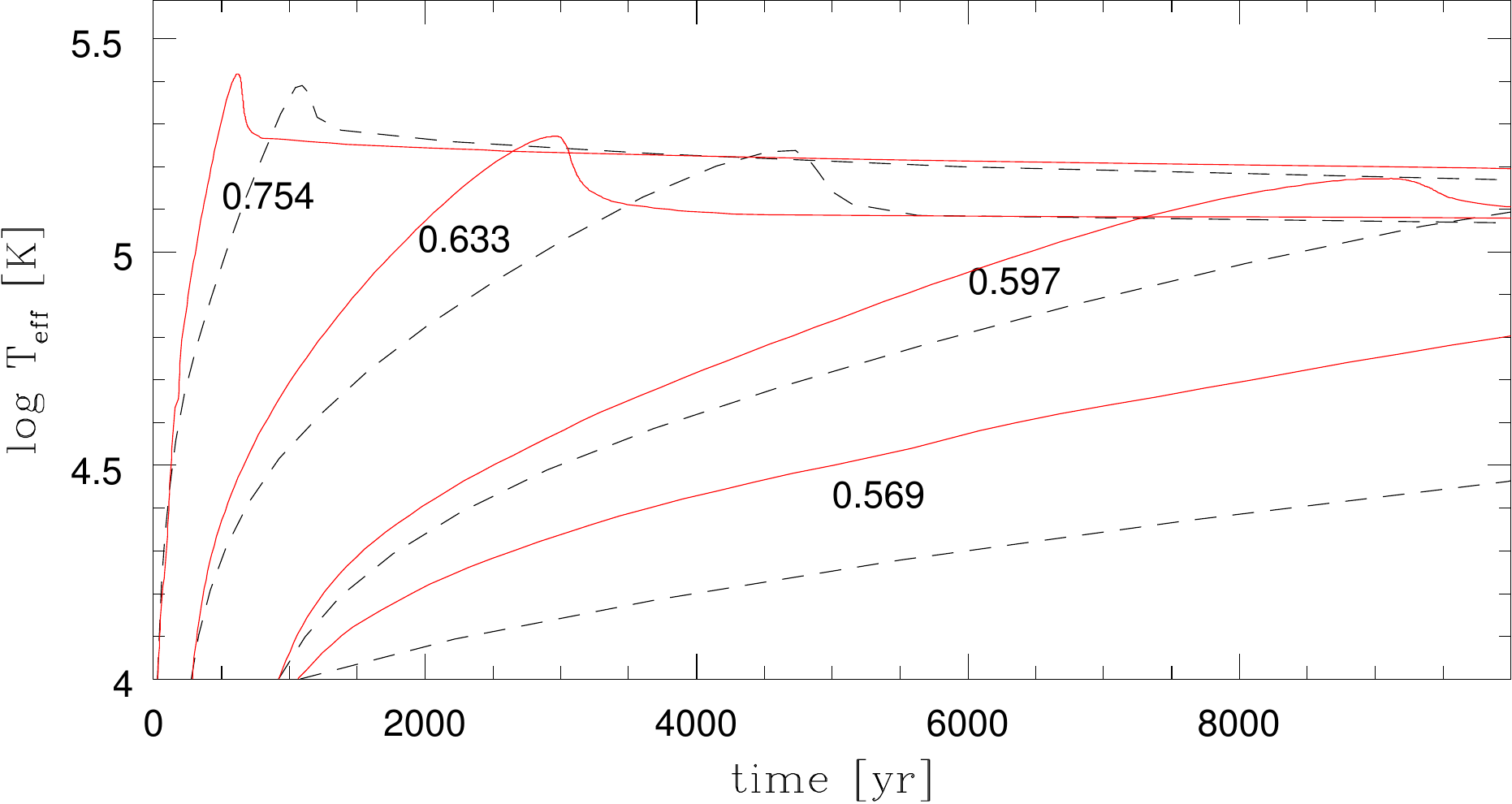} }
      \caption{Comparison of four interpolated Bl\"ocker 1995 post-AGB
	tracks (solid, red lines) with \citet{VW94} post-AGB tracks
	(dashed, in black) of the same mass. Each set of models are
	labelled with the core mass.  The VW94 tracks are shifted to
	the same age as the corresponding Bl\"ocker track at $T_{\rm
	eff}=10^4$\,K to correct for the different transition times. }
    \label{vw94}
\end{figure}

The heating rate after the transition time depends on the assumed
post-AGB mass loss, where the Bl\"ocker tracks and VW94 used a
different approach.  Figure\,\ref{vw94} compares the post-AGB
Bl\"ocker tracks with the tracks VW94 for four different stellar
masses.  The VW94 are shifted in time to start at the same age as the
Bl\"ocker track at $T=10^4$\,K.  In all four cases, the VW94 tracks
show slower heating rates for the same mass. In other words, the VW94
tracks require higher core masses for the same heating rate
\citep{Zijlstra2008}.

\subsection{Empirical evolution}

\subsubsection{Asteroseismological masses}

To constrain the models, it is important to directly measure masses of
central stars of planetary nebulae to a high accuracy. This has only
recently become possible, using asteroseismology on pulsating central
stars. \citet{Althaus2010} lists eight such central stars of PNe with
measured masses. They are reproduced in Table\,\ref{seismo}. Five
stars have masses around 0.53--0.55\,$M_{\sun}$. The remainder are
more massive. 

Because PNe are seen around these stars the age since the end of the
AGB must be $\lesssim 20\,$kyr. For two of the stars in
Table\,\ref{seismo}, we can calculate the kinematical age. For A\,43,
we find 8.5\,kyr and for VV\,47 13\,kyr, using the data of
\citet{FRR2011} and \citet{Weinb1989}, respectively. For the last
object, we adopt the higher expansion velocity derived from
[\ion{N}{ii}] rather than lower value from [\ion{O}{iii}], as assumed
earlier by \citet{SchNap1990} because [\ion{N}{ii}] emission better
probes the outer shell regions \citep[e.g.,][]{SJS2005}. We do not
have a mass-averaged velocity for either object, and the kinematical
ages are less certain than for the modelled nebulae. However, the
numbers confirm that the ages are less than 20\,kyr.

\begin{table}
\caption{Asteroseismology masses of PNe central stars of pulsation type
  PG\,1159 from \citet{Althaus2010}.  }
\begin{flushleft}
\begin{tabular}{llll}
\hline\hline
\noalign{\smallskip}
object  & name & $T_{\rm eff}$ [K] & $M_{\rm cspn}$ [$M_{\sun}$] \\
\noalign{\smallskip}
\hline
\noalign{\smallskip}
PN\,G\,274.3+09.1 & Longmore\,4     &  120 &  0.63 \\
PN\,G\,036.0+17.6  & Abel\,43       &  110 &  0.53 \\
PN\,G\,066.7$-$28.2  & NGC\,7094       &  110 &  0.53 \\
PN\,G\,118.8$-$74.7 & NGC\,246        &  150 &  0.75 \\
PN\,G\,080.3$-$10.4 & RX\,J2117.1+3412 & 170 &  0.72 \\
PN\,G\,094.0+27.4   & K\,1-16          &  140 &  0.54 \\
PN\,G\,104.2$-$29.6   & Jn\,1           &  150 &  0.55 \\
PN\,G\,164.8+31.1   & VV\,47          &  130 &  0.53 \\
\noalign{\smallskip}
\hline
\noalign{\smallskip}
\end{tabular}
\end{flushleft}
\label{seismo}
\end{table}

The Bl\"ocker models do not predict PNe around such low mass stars.
Their post-AGB evolution is extremely slow with the 0.546\,$M_{\sun}$
track \citep{Schoenberner1983} taking $3 \times 10^5$\,yr to reach a
peak temperature. The stars of Table \ref{seismo} are all PG1159
stars, which are hydrogen-poor and believed to have undergone a late
thermal pulse during the post-AGB evolution \citep{Miller2006}. Their
current evolution therefore differs from that of the hydrogen-burning
tracks considered here. However, although the late pulse rejuvenates
the star, it does not rejuvenate the nebula. The presence of the old
PN still sets a minimum speed of evolution during the preceding
hydrogen-burning post-AGB phase, and the issue remains that the
nebulae are much younger than predicted by the model tracks.

\subsubsection{White dwarf masses}

A separate constraint on the evolutionary tracks can be obtained by
comparing the mass distribution of central stars derived from the
Bl\"ocker tracks with the known mass distribution of white dwarfs.
White dwarf masses are not known for the Galactic bulge because their
low luminosity allows for spectroscopic analysis of objects that are
not very distant. The local DA white dwarf mass distribution peaks at
$0.576\,M_{\sun}$ \citep{Kepler2007}. Recently, this value has been
confirmed by \citet{Giamm2012}: their histogram was plotted with large
bins of $0.05\,M_{\sun}$; however, the published catalogue reveals a
pronounced maximum at $0.58\,M_{\sun}$. In contrast, PN central star
masses, as derived above and in \citet{GZ2007}, peak at values above
0.6\,$M_{\sun}$. It has proven difficult to reconcile the
distributions. The systematic error in the core masses
\citep[discussed in][]{GZ2007} is around 0.02\,$M_{\sun}$; however the
masses in that paper were already shifted down by assuming slow
expansion of the nebulae. 

To reconcile the models and observations, one can (i) assume a
systematic error in the white dwarf and pulsational masses, (ii) limit
PN formation to the more massive progenitor stars, or (iii) accelerate
the post-AGB tracks to allow lower masses stars to form PNe. Option
(i) is less likely, while option (ii) contradicts the presence of PNe
around low-mass PG1159 stars, the presence of PNe in the old stellar
populations of elliptical galaxies \citep{Arnaboldi2011}, and the
Galactic bulge.

In the next subsections, we therefore investigate a way to (iii)
accelerate the post-AGB model tracks.

\subsection{ Bl\"ocker track interpolation }

The Bl\"ocker and Sch\"onberner tracks are calculated for a set of
distinct masses.  \citet{GZ2007} interpolated between the individual
tracks to derive a denser model grid. In this interpolation, the
strongest constraint on the low-mass central stars comes from the
Sch\"onberner model of 0.546\,$M_{\sun}$. This model evolves so slow
that it is off the scale when compared to our kinematical ages, and in
the interpolation, this sets a lower limit for the central stars
masses at 0.55\,$M_{\sun}$. \citet{Schoenberner1983} comments that the
0.546\,$M_{\sun}$ model shows evidence for helium burning. The
0.546\,$M_{\sun}$ model is an early-AGB model before the onset of
thermal pulses, when helium burning contributes significantly to the
luminosity \citep{Schoenberner1983}. The very slow evolution is a
consequence of this.

We therefore leave out the 0.546\,$M_{\sun}$ model from the
interpolation. Instead, we use the higher mass tracks to extrapolate
towards the low-mass regime.  The interpolated formula for cspn masses
uses a linear fit in the log(time) -- log(temperature) plane:
\begin{equation} 
M_{\rm cspn} =  0.074 \times \log(T_{\rm eff}) -0.08 \times \log(t_{\rm kin}) + 0.29. 
\label{m_cspn}
\end{equation} 
This is fitted to the horizontal part of four Bl\"ocker tracks (0.565,
0.605, 0.625, and 0.696\,$M_{\sun}$).  The limited number of data
points warrants only a low-order extrapolation. The uniformity of the
published tables is very helpful, as each track has the same number of
points up to the maximum stellar temperature of each track. The Pikaia
genetic algorithm was used to locate the coefficients of the best fit.
The maximum difference between the track data points at their
horizontal parts and the masses from the simple formula is less than
2\%. This formula neglects the cooling part of the evolutionary 
tracks. However, its advantage is that it allows for easy re-scaling,
for example to check how any acceleration of the evolution might
affect the derived masses.

\subsection{Envelope mass}
\label{env_mass}

The second adjustment we can make is to systematically accelerate the
evolutionary tracks. This reduces the derived core masses and can thus
improve the match to the white dwarf mass distribution.

A physical mechanism for an acceleration of the tracks comes from
adjusting  the envelope mass at the end of the AGB. This is one of the
parameters, which comes from the chosen mass-loss prescription, and is
not known {\em a priori}.  The heating rate is set by the rate at
which this envelope mass is removed through hydrogen burning ($\dot
M_{\rm H}$) and a post-AGB ($\dot M_{\rm w}$) wind:

\begin{equation}
 \Delta t_{\rm H} \approx \frac{M_{\rm env}^{\rm pAGB} - M_{\rm env}^{T\rm
     max}}{\dot M_{\rm H} + \dot M_{\rm w}},
\end{equation}

\noindent where the envelope mass is measured at the end of the
transition time, $M_{\rm env}^{\rm pAGB}$, and the time of maximum
temperature (the 'knee' in the HR diagram), $M_{\rm env}^{T\rm max}$.

The relation of the maximum temperature, $T_{\rm max}$, (in Kelvin) to
the envelope mass $M_{\rm env}^{T\rm max}$ (in solar masses) is
obtained from a fit to Fig. 4 of \citet{B1995}:

\begin{equation}
 \log {T_{\rm max}} \approx - 0.37 \times
  \log { M_{\rm env}^{T \rm max} } + 3.83.
\end{equation}

The Bl\"ocker tracks also indicate an approximate relation between the
core mass and $T_{\rm max}$.  We extend the mass range covered by the
Bl\"ocker models by using the PN central stars with
asteroseismological masses, which reach temperatures of 130\,kK for
$M_{\rm c} = 0.53\,M_{\sun}$ \citep{Althaus2010}. (The lower mass
models in \citet{Schoenberner1983} do not reach such high
temperatures.)  From Fig.\,4 in \citet{B1995} and the data in
\citet{Althaus2010}, we find

\begin{equation}
   \log T_{\rm max} \approx 1.2 \times M_{\rm c} + 4.5
\end{equation}

\noindent for $T$ in Kelvin and $M_{\rm c}$ in solar masses. Combining
this with the previous equation give:

\begin{equation}
 \log M_{\rm env}^{T \rm max} \approx -3.24\times M_{\rm c} - 1.81.
\end{equation}

\noindent All models in \citet{B1995} show a decrease by about $\Delta
\log M_{\rm env} = -0.6$ (about a factor of 1/4) from the end of the
transition time to the time of maximum temperature:

\begin{equation}
 M_{\rm env}^{\rm pAGB} = 4\times M_{\rm env}^{\rm Tmax}.
\end{equation}

\begin{figure}
\resizebox{\hsize}{!}{\includegraphics{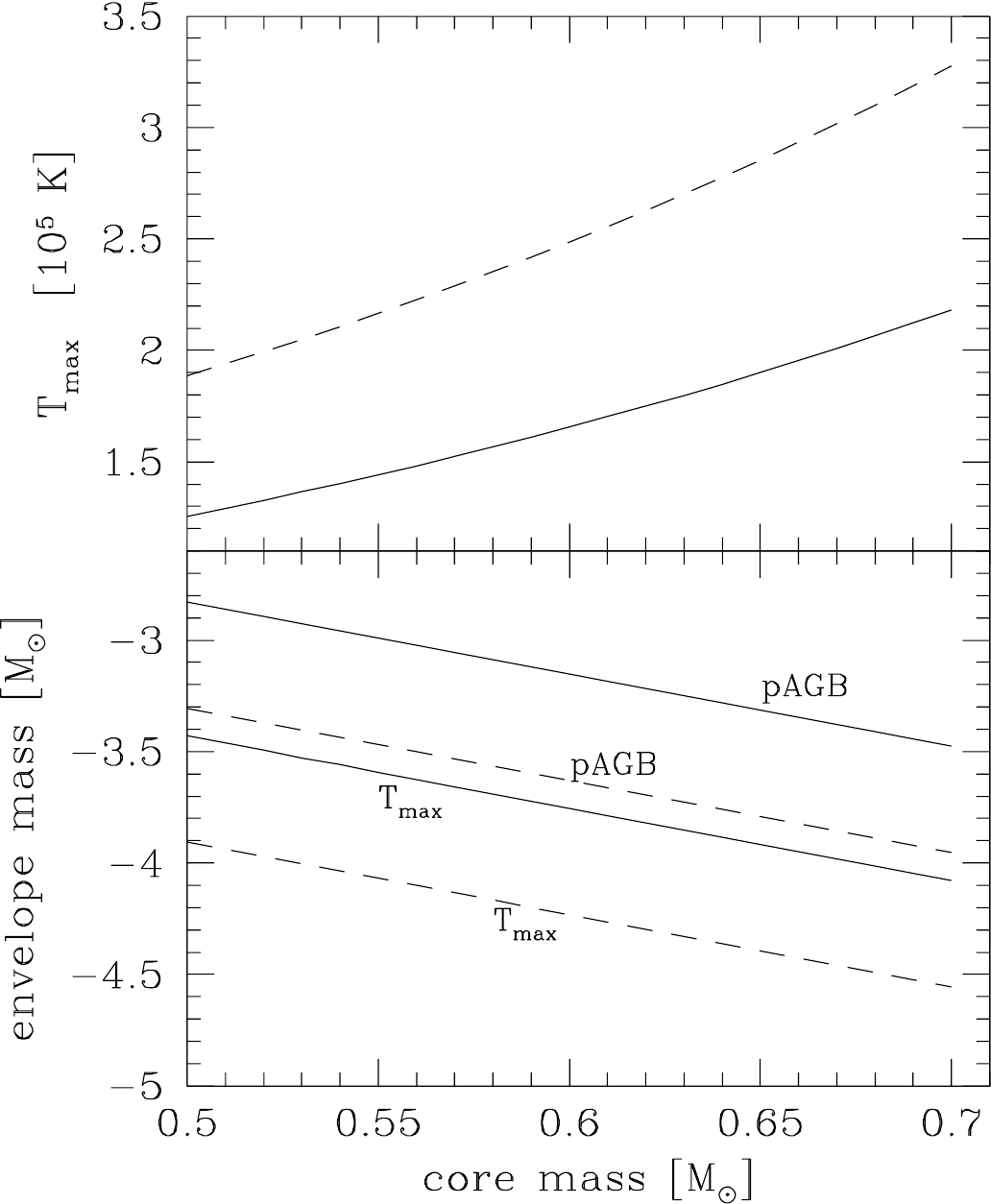} }
      \caption{Dependence of the post-AGB maximum temperatures $T_{\rm
	  max}$ and envelope masses on the core mass, based on the
	relations in Section \ref{env_mass}. The solid lines show the
	parametrization, which are fitted to the Bl\"ocker models. 
	The short-dashed lines show the result of reducing the
	envelope masses at the end of the transition phase by a factor
	of three. In the bottom panel, we show the envelope mass at
	the start of the post-AGB evolution and at the time of maximum
	stellar temperature. }
    \label{p_parms}
\end{figure}

These dependencies of $T_{\rm max}$, $M_{\rm env}^{\rm pAGB}$ and
    $M_{\rm env}^{T \rm max}$  are illustrated in Fig.\,\ref{p_parms}
    (solid lines).

If we assume that post-AGB mass loss rates are much less than the
nuclear burning rates, as found for the lower mass models by
\citet{B1995},  then the envelope mass $M_{\rm env}$ decreases with
the rate,
\begin{equation}
\dot M_{\rm env} \propto L.
\end{equation}

\noindent The core-mass luminosity relation (in solar
units) is given by \citep{Herwig1998}:

\begin{equation}
 {L} = 62200 \left( {M_{\rm c}} -  0.487 \right).
\end{equation}

Finally, from a fit of Fig. 4 of \citet{B1995}, we find that the
envelope mass at $T= 10^5$\,K  is approximately related to the
envelope mass at peak temperature: 

\begin{equation}
 \log \frac{ M_{\rm env}^{T5}}{ M_{\rm env}^{T \rm max} }=
 -0.65\times\left( \log T^{\rm  max}-5  \right),
\label{temp}
\end{equation}

\noindent where $M_{\rm env}^{T5}$ is the  envelope mass at $T=10^5$\,K.

\begin{figure}
\resizebox{\hsize}{!}{\includegraphics{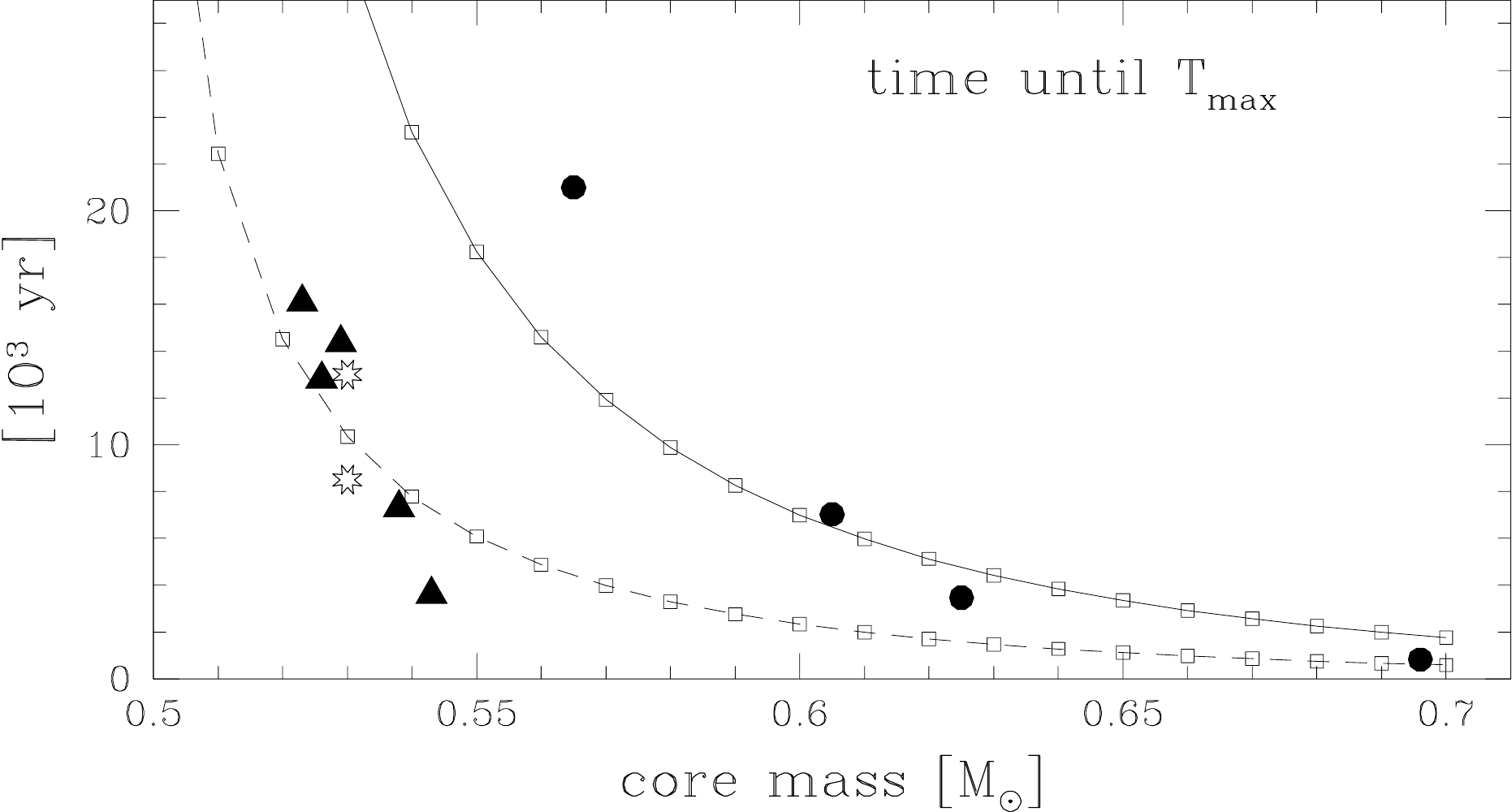} }
      \caption{Dependence of the post-AGB time scales on core mass,
	based on the relations in Section \ref{env_mass}. The vertical
	axis shows the time at which maximum temperature is reached. 
	The drawn line (with squares) shows the parametrization,
	fitted to the 0.605\,$M_{\sun}$ Bl\"ocker model. The
	short-dashed line (with squares) shows the result of reducing
	the envelope masses at the end of the transition phase by a
	factor of three. The filled circles show the Bl\"ocker model
	set, the filled triangles models from \citet{WF2009}, and the
	open octagons show the locations of A\,43 and VV\,47. }
    \label{p_time_1}
\end{figure}

\begin{figure}
\resizebox{\hsize}{!}{\includegraphics{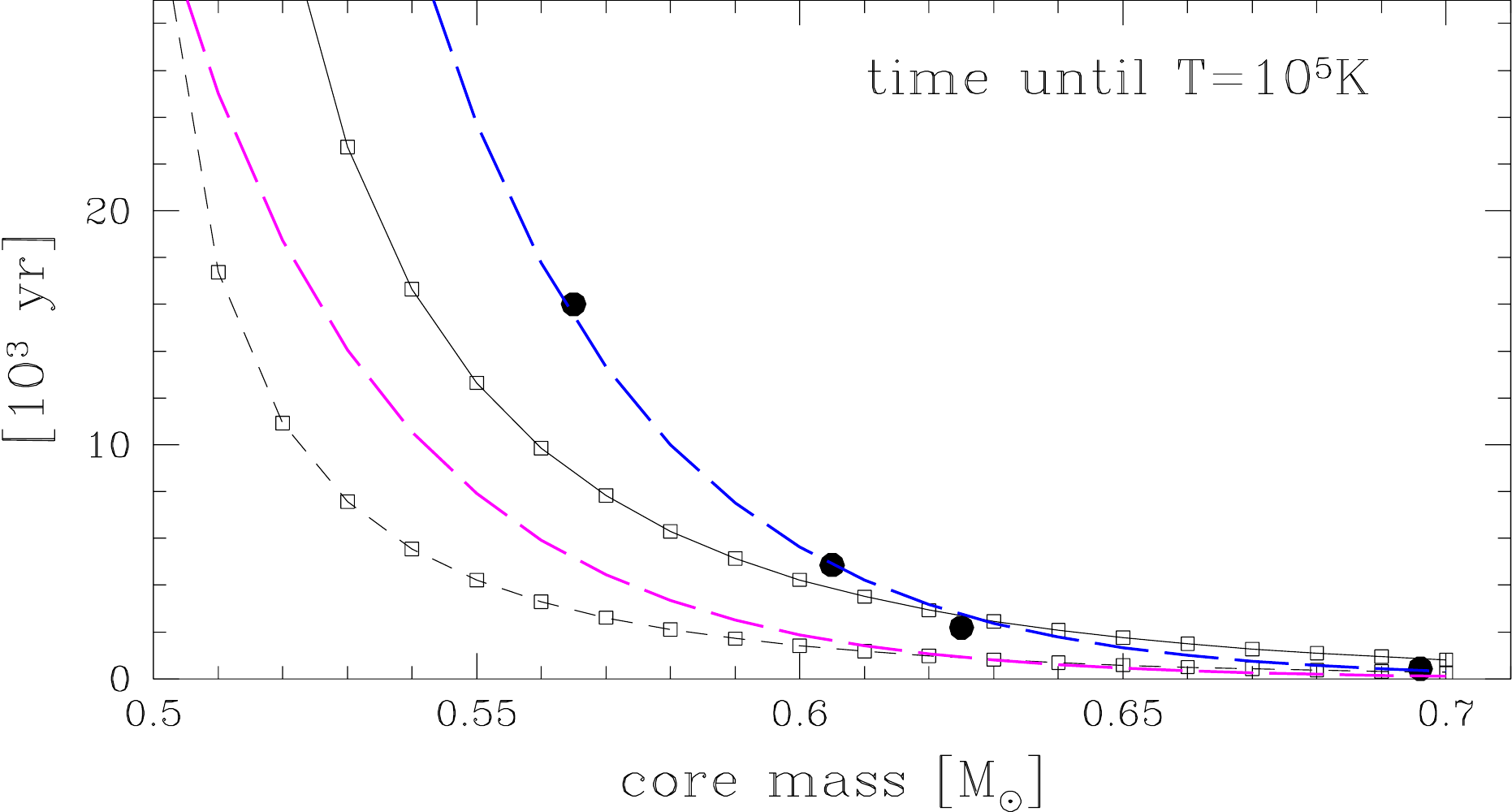} }
      \caption{The post-AGB time scale for the star to reach
	$T=10^5\,$K is plotted against core mass.  The filled circles
	show the Bl\"ocker model set.  The upper long-dash line (blue)
	correspond to the interpolated formula of Eq.\,(\ref{m_cspn}),
	and the lower long-dashed line (magenta) is for the
	accelerated evolution of Eq. (\ref{m_cspn_acc}). The solid
	line (with squares) shows the envelope-mass parametrization of
	Section \ref{accele}, and the short-dashed line (with squares)
	shows this parametrization with envelope masses at the end of
	the transition phase reduced by a factor of three.}
    \label{p_time_2}
\end{figure}

Combining the equations above, we can now calculate how the time until
peak temperature depends on core mass and envelope mass. The result is
shown in  Fig.\,\ref{p_time_1}, where the time until peak temperature
is plotted against core mass. Note that the time includes the
transition time and is measured from the zero point of the models. The
parametrization is scaled to the 0.605\,$M_{\sun}$ Bl\"ocker model.
The filled circles show the set of Bl\"ocker models: they show a
somewhat steeper dependence of the time versus core mass than shown by
our simple description. This is because of our neglect of post-AGB
mass loss: in the Bl\"ocker tracks, post-AGB mass loss increases the
evolutionary speed for masses above 0.6\,$M_{\sun}$ by as much as a
factor of two.

The dashed lines in Figs. \ref{p_parms} and \ref{p_time_1} show the
effect of reducing the initial envelope mass by a factor of three. The
peak temperature is increased by a factor of 1.67. The equations above
show that the time until peak temperature reduces linearly with
envelope mass, which is a factor of three. This would be sufficient
for the presence of an old PN around a PG1159 (post-VLTP) stars of
mass 0.53\,$M_{\sun}$.  The filled triangles in Fig. \ref{p_time_1}
show the recent models of \citet{WF2009}, and the open octagons are
the locations of Abell\,43 and VV\,47: both are well fitted with this
reduced envelope mass.  We note that the locations of Abell\,43 and
VV\,47 may shift down further (faster stellar evolution) if they have
experienced a late thermal pulse.

The peak temperature is not an observable (it is reached at a time
when evolution is extremely rapid), and it is not used in the
interpolation formula (Eq.\,(\ref{m_cspn})). We instead use the time
when a temperature of $10^5$\,K is reached.  We use Eq. (\ref{temp}),
assuming that the envelope mass is consumed at a constant rate. The
result is shown in Fig.\,\ref{p_time_2}.  The solid line is drawn
assuming that the rate of evolution increases with increasing
temperature as $d(\log T) / d(\log t) \approx 1$ (see
Eq.\,(\ref{m_cspn})) to scale from a temperature of $10^5$\,K. The
short-dashed line is this parametrization, accelerated by a factor of
three. This parametrization neglects post-AGB mass loss.

\subsection{Accelerated evolutionary tracks}
\label{accele}

The parametrization is not a perfect representation of Bl\"ocker
models, as shown by the slow evolution of the lower-mass Bl\"ocker
model as compared to that derived here. We therefore do not use this
parametrization directly. Instead, we use the acceleration factor,
which approximately fits the asteroseismological masses to adjust the
Bl\"ocker tracks.

The interpolated formula (Eq.\,(\ref{m_cspn})) can be easily
re-calibrated to  agree with the accelerated evolution and to provide
new values of cspn masses. Adding a number to the logarithm of time is
equivalent to accelerating the post-AGB evolution by a given factor.
The new equation shifts the masses up by 0.038\,$M_\sun$:

\begin{equation} 
M_{\rm cspn} =  0.074 \times \log(T_{\rm eff}) -0.08 \times \log(t_{\rm kin}) + 0.33. 
\label{m_cspn_acc}
\end{equation}

Figure\,\ref{p_time_2} shows the time when a stellar temperature
$T_{\rm eff}=10^5$\,K is reached.  The long-dashed (blue) line is
taken from Eq.\,(\ref{m_cspn}), and it goes as expected through the
filled circles representing the Bl\"ocker models. The lower
long-dashed (magenta) line was obtained with Eq.\,(\ref{m_cspn_acc}),
assuming post-AGB evolution accelerated by a factor of three, which
approximates the reduced-envelope-mass evolution discussed above.

\subsection{Final (core) masses}

\begin{figure}
\resizebox{\hsize}{!}{\includegraphics{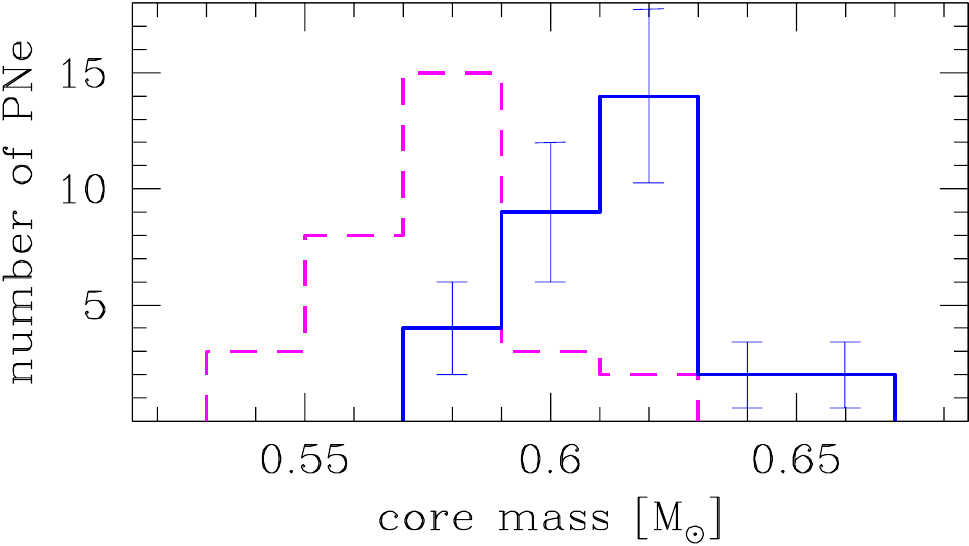} }
      \caption{The masses of the PN central stars for the 31 objects
	studied here with HST and VLT data. The solid (blue) line
	shows the data obtained from equation \ref{m_cspn}; the dashed
	(magenta) line presents the masses obtained from accelerated
	post-AGB evolution (see further in the text). One-sigma errors
	are indicated. }
    \label{cm}
\end{figure}

The stellar core masses are derived from Eq.\,(\ref{m_cspn}) by using
the nebular ages obtained from Eq.\,(\ref{eq_age}), and the stellar
temperature from the photo-ionization model. The values of the
kinematical age $t_{\rm kin}$ and the central star mass $M_{\rm cspn}$
are given in Table\,\ref{pne_data}. The histogram of the derived
masses is shown in Fig.\,\ref{cm} and is plotted with the solid (blue)
line. We adopted a bin width of 0.02\,$M_{\sun}$, which is comparable
to the estimated error in the core masses \citep{GZ2007}.  Wider bins
produce histograms that are too sparse, while narrower bins make the
histograms too noisy with only several objects in each bin. The
accelerated version of the formula, Eq. (\ref{m_cspn_acc}), was used
to obtain another set of cspn masses, which are given in column 9
($M_{\rm cspn}^{\rm accele}$) of Table\,\ref{pne_data} and are shown
as the dashed (magenta) histogram in Fig.\,\ref{cm}.

This histogram shows that the proposed acceleration of post-AGB
evolution can produce cspn masses in a good agreement with white dwarf
masses: the peak of the mass distribution has shifted from 0.62 to
0.58\,$M_\sun$. In addition, the accelerated tracks yield some masses
as low as those obtained from pulsating PG1159.

Applying the same acceleration factor for all masses is the simplest
adjustment possible. There are too few data points to search for more
complex dependencies. The acceleration dependence on stellar mass and
envelope mass should be verified with evolutionary codes, but this is
beyond the scope of this paper.

\section{Stellar initial masses and ages}
\label{ini_fin_m}

The masses derived above refer to the final (white dwarf) mass of the
evolving star.  It is a challenge to relate this final mass to the
initial mass of the progenitor star. The IFMR depends on the mass loss
on the giant branches \citep{MJZ2011}.  The model relations are very
sensitive to the way mass loss is treated on the AGB and RGB.
Theoretical isochrones incorporate different mass-loss assumptions and
predict very different final masses. The IFMR is best calibrated for
the lowest-mass stars using globular clusters and for intermediate
mass stars using white dwarfs in open clusters. Both populations have
different metallicities so their IFMR might differ (see
Fig.\,\ref{m_if} below). In between the two populations, the relation
is also quite uncertain. 

Here we use theoretical isochrones and empirical cluster white-dwarf
masses to relate our final masses to progenitor masses. Given the
uncertainties involved, the resulting mass should be seen as
indicative only.

Having estimated the initial masses, we can later relate them to the
total stellar ages, since the ZAMS. This relation is much better
known, as discussed further in the text, and these ages are compared
later with the Galactic bulge age.

\subsection{ Empirical initial--final mass relations}

\begin{figure}
\resizebox{\hsize}{!}{\includegraphics{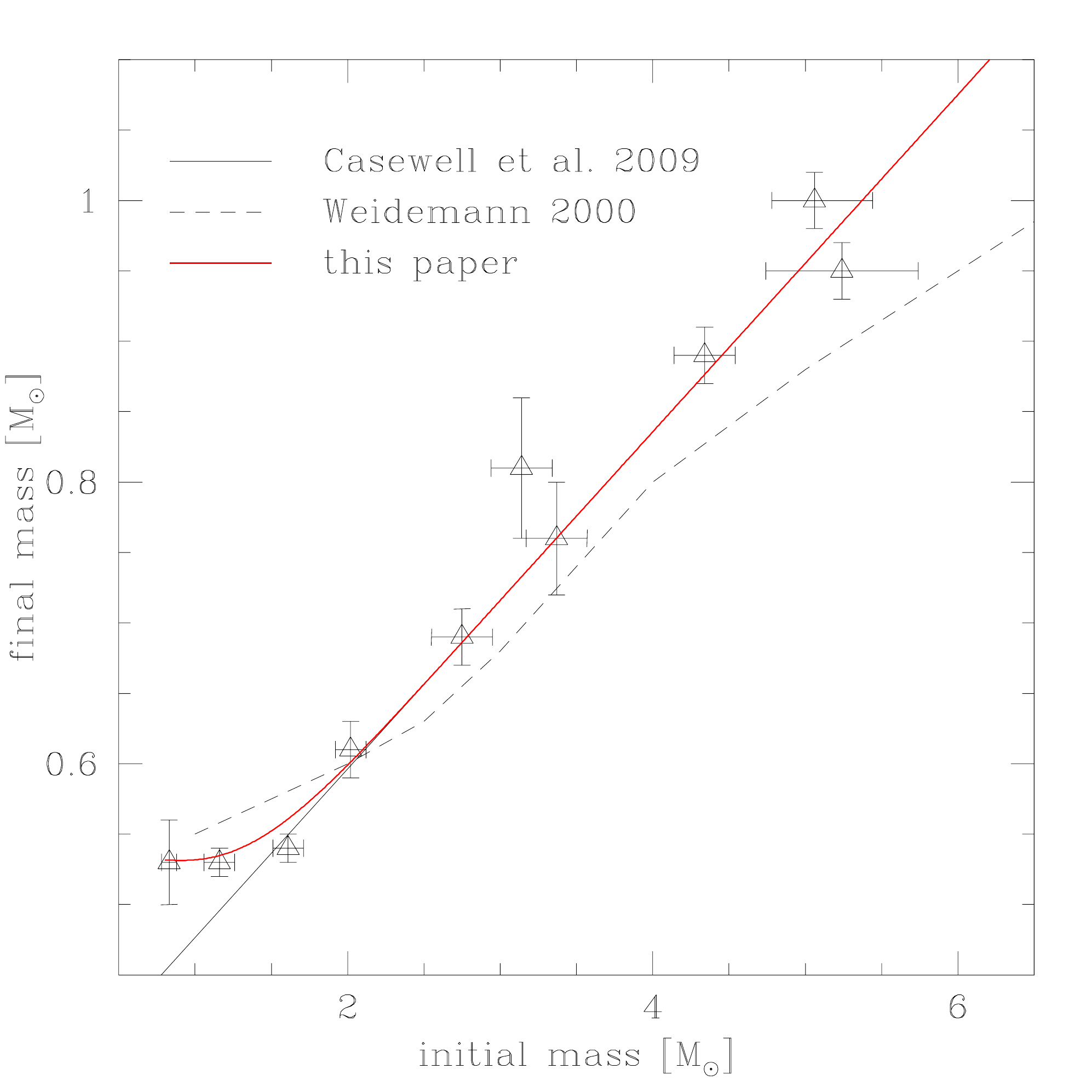} }
      \caption{Initial-final masses for stellar clusters. For each
      cluster, an average initial and final mass are shown, based on
      data from \citet{Casewell2009} and references therein. For
      NGC2516 we used \citet{Dobbie2012}, and we also included the
      globular cluster omega Cen \citep{MJZ2011}. The drawn line gives
      the linear fit of Casewell et al, whilst the dashed line is the
      older relation of \citet{Weidemann2000}. The red line shows the
      relation derived here, Eq. (\ref{IFM}). }
    \label{cluster_IFMR}
\end{figure}

Empirical relations have been determined using white dwarfs in open
clusters, which combine spectroscopic masses, model cooling ages, and
cluster ages. We have used the data from \citet{Casewell2009}, and
references therein, as supplemented with recent data from
\citet{MJZ2011} and \citet{Dobbie2012}, to calculate an average white
dwarf mass and initial mass per cluster. For each cluster, we left out
white dwarfs with significantly higher masses than the minimum white
dwarf masses to avoid cooling age uncertainties.  The resulting data
are listed in Table\,\ref{ifmr_table} and plotted in
Fig.\,\ref{cluster_IFMR}. For the initial mass, we used a minimum
uncertainty of 0.2\,$M_{\sun}$, or 0.1\,$M_{\sun}$ for masses of
$\lesssim 2\,M_{\sun}$ (apart for omega Cen); otherwise the
uncertainties are estimated from the scatter between individual stars
in the cluster.

The initial-final relation is well represented by the linear relation
of \citet{Casewell2009} at higher initial masses, but this relation
undershoots for the lowest masses (where the older
\citet{Weidemann2000} relation is higher). The closest initial mass to
the Sun is provided by the cluster NGC6791, where white dwarf masses
are around 0.53\,$M_{\sun}$. For omega Cen, 0.8\,$M_{\sun}$
progenitors have evolved into 0.53\,$M_{\sun}$ white dwarfs
\citep{MJZ2011} (where the lower metallicity of omega Cen should be
considered; however, solar metallicity stars of this mass are longer
lived and have not yet formed white dwarfs). The Casewell relation
predicts white dwarf masses in globular clusters well below
0.5\,$M_{\sun}$, which would require helium white dwarfs.  Instead,
the presence of AGB stars in globular clusters shows that they do
yield some C/O white dwarfs. The Casewell relation therefore seems too
low for the lowest initial masses.

To better fit the empirical relation, we add a term to the Casewell
relation:

\begin{equation}
  M_{\rm f} = 0.3569 + 0.1197\times M_{\rm i} + 0.15\times \exp[-(M_{\rm
    i})^2]\ {M_{\sun}}
 \label{IFM}
\end{equation}

\noindent (valid for $(M_{\rm i}>0.8\,M_{\sun} $). This relation is
shown by the red line in Fig. \ref{cluster_IFMR}. It removes the
mismatch at the lowest masses. 

\begin{table}
\caption{Average initial and final masses from cluster white dwarfs,
  based on data in \citet{Casewell2009, MJZ2011, Dobbie2012}. Minimum
  uncertainties on the initial mass are taken as 0.2\,$M_{\sun}$ or
  0.1\,$M_{\sun}$ for masses of $\lesssim 2\,M_{\sun}$ (apart for
  omega Cen); otherwise, the uncertainties are estimated from the
  scatter between individual stars in the cluster.}
\begin{flushleft}
\begin{tabular}{llllc}
\hline\hline
\noalign{\smallskip}
cluster & $M_{\rm f}$  & $\delta M_{\rm f}$ & $M_{\rm i}$ & $\delta M_{\rm i} $ \\
\noalign{\smallskip}
\hline
\noalign{\smallskip}
N6819    & 0.54  & 0.01  &  1.61 & +0.1 $-$0.1 \\
N7789    & 0.61  & 0.02  &  2.02 & +0.1 $-$0.1 \\
N6791    & 0.53  &  0.01 &  1.16 & +0.1 $-$0.1 \\
N2168    & 0.89  & 0.02  &  4.34 & +0.2 $-$0.2  \\
N2516    & 0.95  & 0.02  &  5.24 & +0.5 $-$0.5   \\
N6633    & 0.81  & 0.05  &  3.14 & +0.2 $-$0.2   \\ 
Sirius B & 1.00  & 0.02  &  5.06 & +0.4 $-$0.3  \\
Praesepe & 0.76  & 0.04  &  3.37 & +0.2 $-$0.2   \\
Hyades   & 0.69  & 0.02  &  2.75 & +0.2 $-$0.2   \\
Omega Cen & 0.53 & 0.03  &  0.83 & +0.05 $-$0.05   \\
\noalign{\smallskip}
\hline
\noalign{\smallskip}
\end{tabular}
\end{flushleft}
\label{ifmr_table}
\end{table}

\subsection{Stellar life times}

\begin{figure}
\resizebox{\hsize}{!}{ \includegraphics{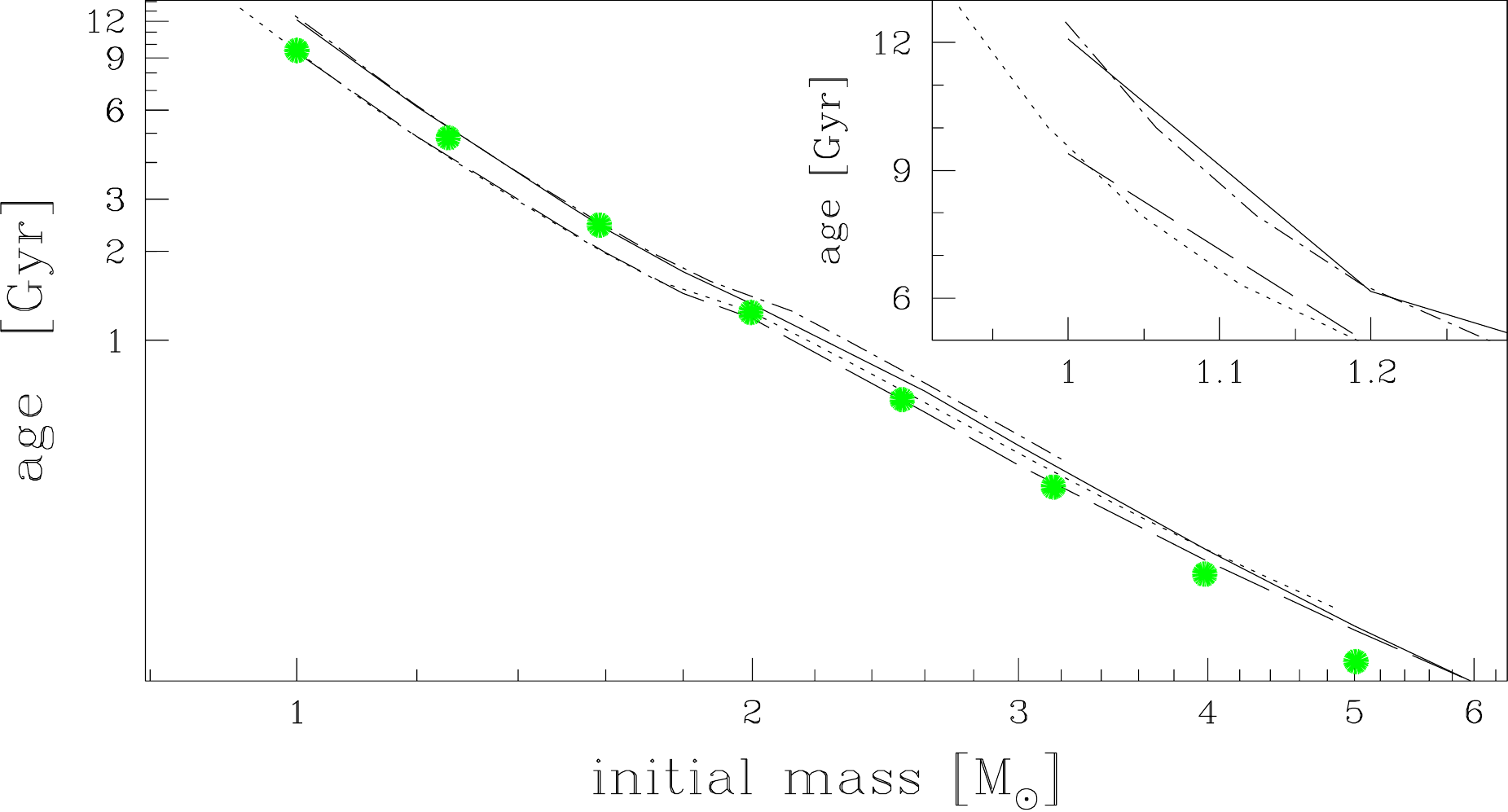} }
      \caption{The initial mass vs. the age from the ZAMS to the end
	of the AGB for the \citet{WF2009} model at $Z=0.02$ (drawn
	line) and $Z=0.008$ (dashed line) and for the Padova
	isochrones at $ Z=0.019 $ and $Z=0.008$ (dot-dashed and dotted
	lines). The large (green) dots indicate the approximated
	relation.}
    \label{m_a}
\end{figure}

In Fig.\,\ref{m_a}, we plot the stellar life times from the ZAMS until
the end of the TP-AGB versus the initial stellar mass. The curves are
taken from \citet{WF2009} for $Z=0.02$ and $Z=0.008$ (drawn and dashed
lines) and from \citet{GBBC2000} for $Z=0.019 $ and $Z=0.008$
(dot-dashed and dotted lines). There is better agreement between
different models than for the IFMR discussed above. The insert shows
the area of our interest on a linear scale.

In the following analysis, we use the linear (in the log--log plane)
relation shown by filled circles in Fig. \ref{m_a} which is defined as
\begin{equation}
      \log ( {\rm age} ) = 9.98 - 2.95 \times \log ( {\rm initial~mass} ).
\label{log_am}
\end{equation}
This is our approximation to the set of fairly similar theoretical
dependencies. These model relations show a small deviation near ZAMS
mass of 1.8\,M$_{\sun}$: our straight line was chosen to better
approximate the low-mass regime (between  1 and 2\,M$_{\sun}$) and
misses the better defined high mass end. Tests have shown that
different choices for the fitted line, if  restricted to the range
covered by different model relations, have only a small effect on the
derived ages.

\subsection{ZAMS masses and ages}
\label{mi}

\begin{figure*}
\sidecaption
\includegraphics[width=12cm]{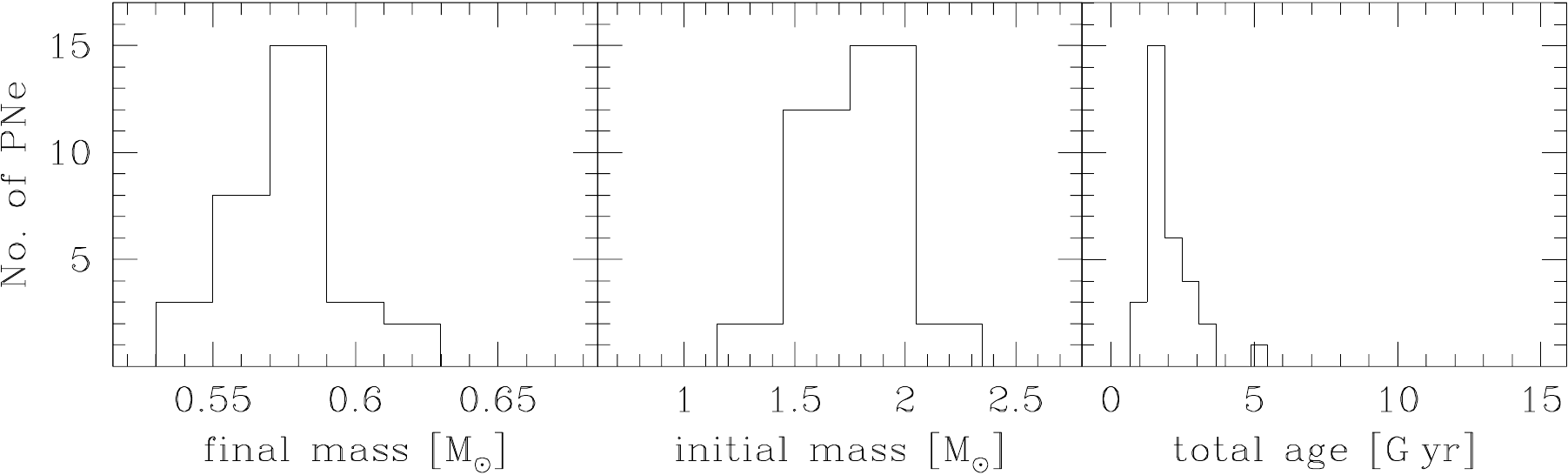}
\caption{The histograms of the interpolated data. The left panel shows
  the ``accelerated'' cspn masses; the central panel shows the initial
  ZAMS masses from empirical IFMR, and the right panel shows the ages
  since the ZAMS. }
\label{h_31}
\end{figure*}

To derive stellar ages, we follow the procedure derived above. From
the kinematical models we derive a nebular age and stellar
temperature. This parameter pair is fitted to a post-AGB stellar
evolution model, which yields (the ``accelerated'' case) a core mass
through Eq.\,(\ref{m_cspn_acc}). The initial mass is derived using an
IFMR defined by Eq.\,\ref{IFM}. Finally, the stellar age is derived
from the mass, and from the condition that the star is in the post-AGB
phase evolution, what means that it has finished its main sequence and
giant branches evolution  (Eq.\,(\ref{log_am}) and Fig.\,\ref{m_a}).
The stellar age is called the `total age' or `age since ZAMS' to
distinguish it from the earlier derived kinematical age of the nebula.

In Fig.\,\ref{h_31}, we show the results of our interpolation routines
in the form of histograms. The left-most panel shows the final masses
derived from our interpolation routines; the central panel shows
initial ZAMS masses, and the right panel shows the total ages of the
stars.

The derived stellar ages are surprisingly low.  The peak corresponds
to ages as young as 1--2\,Gyr. Without the derived acceleration of the
post-AGB tracks, the ages would be even lower. These ages should be
compared to what is known about the stellar populations for the
bulge.

\subsection{The age of the bulge}
\label{agbul}

Age determinations of the GB are mostly based on comparison between
observed and computed colour-magnitude diagrams. Different
observations and computed isochrones generally agree and indicate that
the stars in the GB are about 8--10\,Gyr \citep{V2009} or 10--12\,Gyr
\citep{Z2009} old. These are about 2\,Gyr younger than the oldest
stellar populations of the Milky Way \citep{MZ2010}. 

There is  evidence that there is also a younger population towards the
bulge. The VLT/MAD observations of the globular cluster Terzan\,5,
which are located near the Galactic centre, show the existence of two
horizontal branches. These are interpreted as two populations
(comparable in number) that differ in metallicities by a factor of 3
and with ages of 6 and 12\,Gyr \citep{FDM2009}. They suggest a
relation to a disruption of a dwarf galaxy that contributed to the
formation of the Galactic bulge. 

A younger bulge population has been found by \citet{Bensby2010,
Bensby2011, Bensby2013} as mentioned in the Introduction.
\citet{Uttenthaler2007} find evidence of the third dredge-up in four
out of 27 bulge AGB stars based on the presence of radioactive
technetium.  The third dredge-up is not expected to operate in
1\,$M_{\sun}$ stars and requires a younger population. Finally,
\citet{NUGP2010} derive an enhanced helium enrichment from the red
giant branch bump of the GB. They state that if confirmed this would
lead to a downward revision of the 10+\,Gyr age of the GB. These
results indicate that younger stars are a minor component of the
bulge. The dominant population is older with ages $ > 5$\,Gyr.

Another uncertainty is the duration of the star-formation episode
\citep{Nataf2012}. The assumption of rapid gravitational infall yields
a bulge formation time of about 0.5\,Gyr. The assumption that the
bulge is formed from accreted stellar clumps gives a bulge formation
time of 2.0\,Gyr.  However, star counts and radial velocities find
that bulge kinematics are consistent with a pure N-body bar that
evolved from secular instabilities, which would imply that bulge stars
are just disk stars on bar orbits and would suggest a duration of
star-formation with a time frame as extended as that of the inner
disk.

This information on the bulge age(s) imposes strong constraints on 
the  ages of central stars of GB PNe and suggests ages older than
those shown in Fig.\,\ref{h_31}. This can be accommodated by using the
IFMR as a free parameter.

\section{The PN population of the Galactic bulge}
\label{pn_popul}

Before concluding on the IFMR, we need to discuss whether our sample
can be compared with the dominant GB population and search for
possible factors disturbing the derived histograms.

\subsection{Birth rate of planetary nebulae in the bulge}
\label{sample}

To interpret the results above in terms of bulge evolution, we first
need to show that planetary nebulae trace the bulge population and
that our selected objects are representative.

The mass-loss event, which ejects the planetary nebulae occurs for all
low and intermediate-mass stars. The mass is ejected while the star is
evolving up the asymptotic giant branch, and is ionized while evolving
towards the white dwarf phase.  The lowest- and highest-mass stars may
avoid visibility as a PN (Low mass stars evolve so slowly that the
ejecta have dispersed before the star can ionize them, which is the
so-called ``lazy PN'', while high-mass stars evolve too fast and only
emit high levels of ionizing radiation for a brief period.), but apart
from this, all stars are expected to be surrounded by a detectable PN
for a significant length of time.

Evidence that the oldest stellar populations suffer from the ``lazy
PN'' process comes from globular clusters, where there are fewer PNe
detected than expected based on the number of stars
\citep{Jacoby1997}.  In contrast, PNe are not rare in the bulge: about
a quarter of known Galactic PNe are located towards the Galactic
bulge.  This already suggests that the bulge population, either in
part or in its entirety, is not as old as the globular clusters.

\begin{figure}
\resizebox{\hsize}{!}{\includegraphics{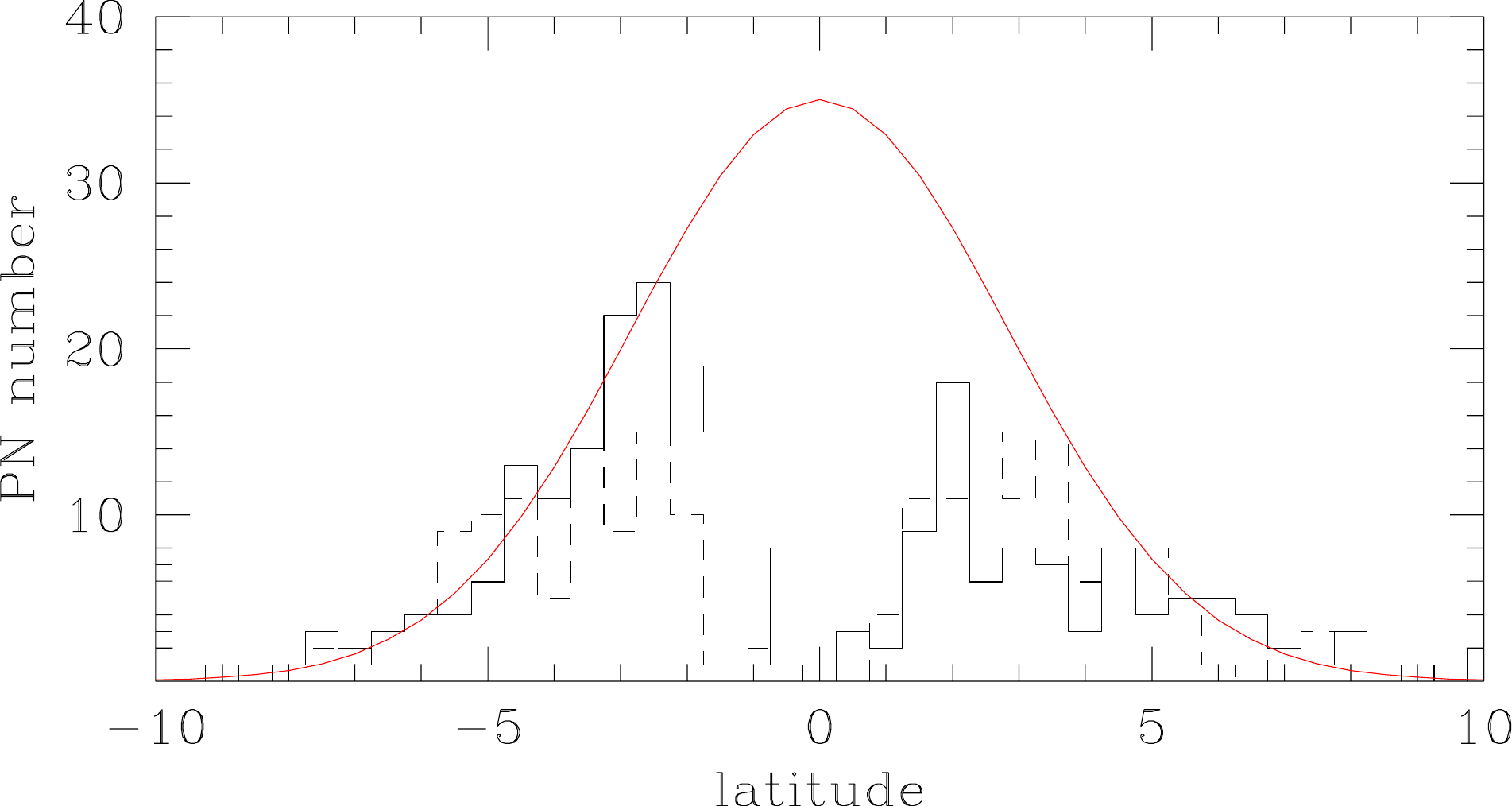} }
\caption{The galactic latitude ($b$) distribution of PNe between 0 and
  4 degree longitude (solid line) and between 356 and 0 degree
  longitude (dashed line), as fitted with a Gaussian of form
  $\exp(-b^2/4^2) $.  }
    \label{lat}
\end{figure}

To quantify the bulge PN population, there are currently 785 PNe
catalogued in the direction of the Galactic bulge, which have a
Galactic latitude and longitude within 10 degrees of the Galactic
Centre \citep{Acker1992, mash1, mash2}.  The distribution is patchy
and due to the high extinction, none are seen within a degree of the
Galactic plane.  Figure\,\ref{lat} shows the latitude distribution of
PNe within $0^{\circ}<l<4^{\circ}$. The distribution is fitted with a
Gaussian. Integrating under the Gaussian indicates that half the
objects are missing. The deficit is a bit larger at negative
longitudes.  Correcting the known number of PNe for this deficit
suggests that approximately 2000 PNe are present in the bulge.

We can convert this to a birth rate if we know the visibility time. 
Most known PNe in the bulge are intrinsically bright: faint nebulae
have been difficult to detect against the crowded stellar background.
The brightest phase of a PN ends when either the nebula becomes
optically thin to ionizing radiation or the star leaves the horizontal
evolutionary track and enters the white dwarf cooling track. This
happens after $\sim5000$\,yr (see \citet{B1995}).  \citet{GZAGGW2006}
find that bulge PNe tend to have kinematic ages less than 6000\,yr.
Combining this observational life time of 6000\,yr with the predicted
number of 2000, such PNe, gives an indicative PN birth rate in the
bulge of $\sim0.3\,\rm yr^{-1} $.

\begin{figure}
\resizebox{\hsize}{!}{\includegraphics{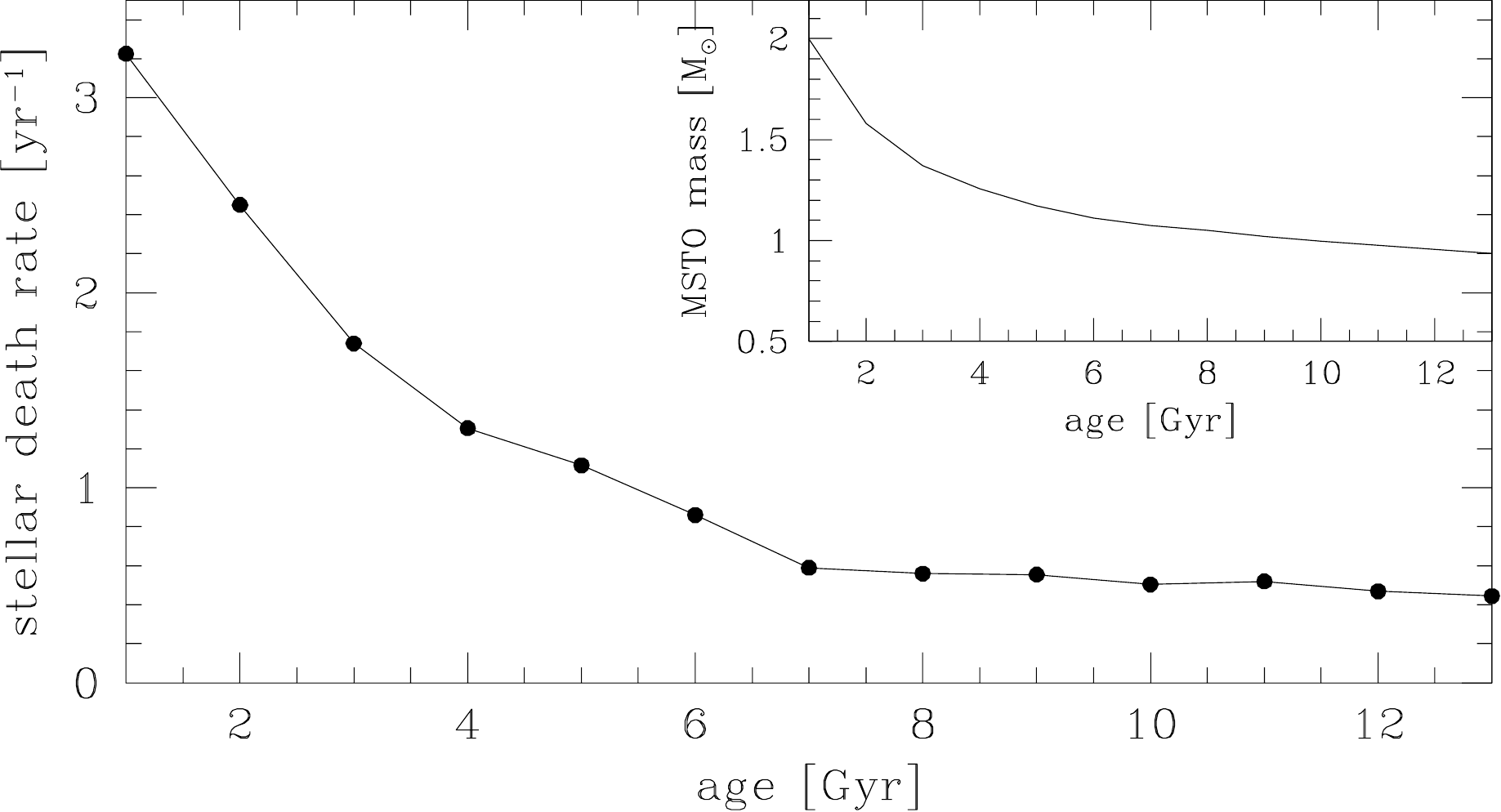} }
      \caption{For an assumed original stellar mass of $10^{10}$
	$M_{\sun}$ formed in a burst, the plot shows the stellar death
	rate (number of stars leaving the AGB per year) as a function
	of age since the star burst.  The top panel shows the
	corresponding main-sequence turn-off mass.}
    \label{birth}
\end{figure}

The rate at which stars leave the AGB (the stellar death rate) can be
calculated for a single age (star burst) stellar population. We use
the models of \citet{Maraston1998}, which are based on the FRANEC
isochrones, for solar metallicity to find the change in turn-off mass
per year and relate this to a number of stars by assuming a Salpeter
initial mass function starting from 0.25\,$M_{\sun}$.  For a total
original stellar mass in the bulge of $10^{10}\,M_{\sun}$
\citep{McGaugh2008, Flynn2006}, the stellar death rate as function of
age of the burst is shown in Fig.\,\ref{birth}. The death rate is
almost independent of age if the burst is older than 7\,Gyr at around
0.6--0.4 stars per year.

There is good consistency (within a factor 1.5--2) between the PN
birth rate and the stellar death rate for a $10^{10}\,M_{\sun}$ bulge
population for a stellar age $>7\,$Gyr. For younger populations, the
PN birth rate is a factor of 2 or more below.  Planetary nebulae
therefore trace a significant stellar population of the bulge with
some bias towards the younger stars. If the bulge is entirely old, the
birth and death rates imply that most bulge stars produce visible PNe:
this would put limits on any very old population ($>10^{10}\,$yr),
which would be deficient in PNe (see section \ref{sect_vis}). If there
is a younger population in the bulge, this would dominate the PN
population as long as this younger population is large enough (of
order 10--50\%\ depending on age).

\subsection{PN visibility}
\label{sect_vis}

The most uncertain part of this calculation is the visibility time. 
Predicted PNe numbers in the Solar neighbourhood suggest visibility
times of 25\,000 -- 50\,000 years \citep{ZP1991, MdM2006} (the latter
predict 3600 PNe within the bulge).  Our assumed visibility time for
bulge PNe is much less because the crowded stellar field combined with
the relatively low angular resolution of the original discovery plates
renders faint extended nebulae as poorly detectable.  Our visibility
time assumes that the {\it observed} bulge population is dominated by
bright, compact PNe with central stars on the horizontal part of the
track (prior to their initial rapid fading when entering the cooling
track) and dense, optically thick nebulae.

The models of \citet{SJSS2007} show that the majority of PNe on the
horizontal part are expected to be optically thin and therefore, are
already decreasing in nebular luminosity. However, the spectra for our
sample of the bulge PNe show evidence for low ionization lines, such
as [\ion{O}{i}], for 24 out of 31 objects, which indicates that these
must be optically thick in some directions \citep[see
also][]{Guzman2011}. This discrepancy between the models and data may
be explained by asphericities.  Our spectra were taken mostly along
the minor axis where the highest densities are found. An ionization
front may be trapped along the minor axis, whilst the polar directions
are already optically thin.  However, optically thin nebulae show only
moderate fading before rapid fading on the cooling track
\citep{SJSS2007}, so this  issue should not affect the visibility time
significantly.

\begin{figure}
\resizebox{\hsize}{!}{\includegraphics{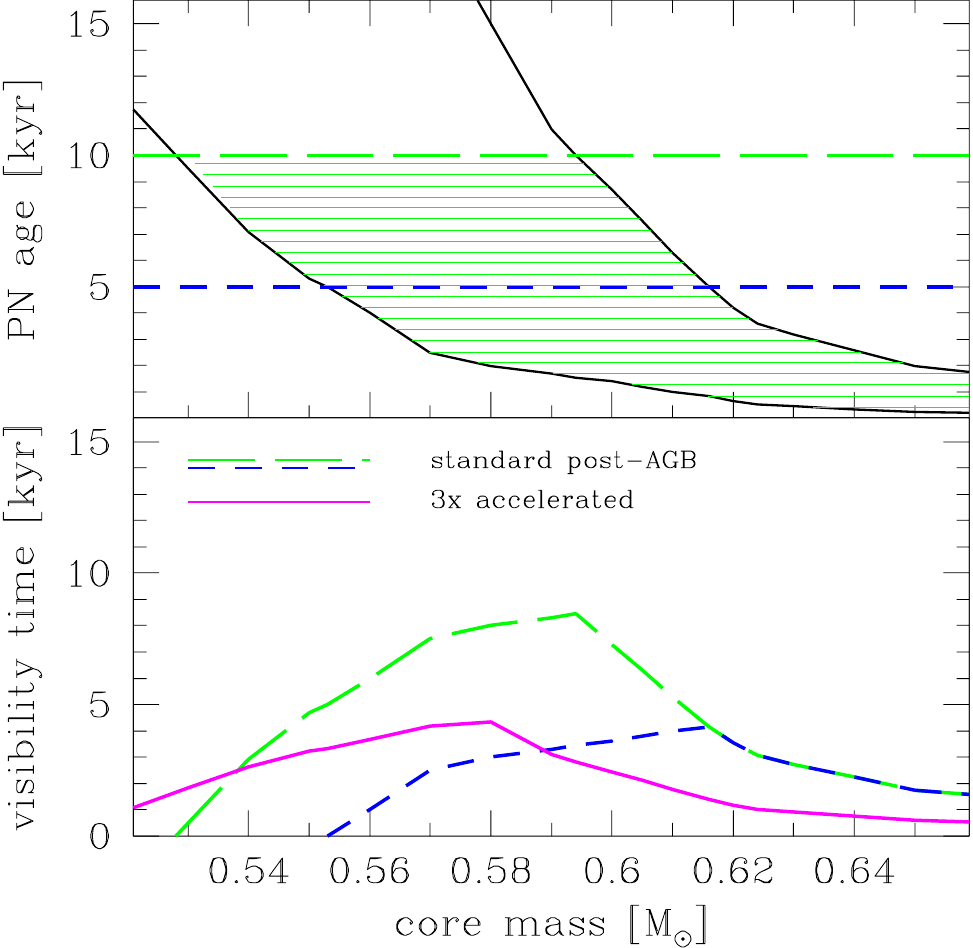} }
      \caption{The top panel shows the kinematic (nebular) ages
      derived from the interpolated Bl\"ocker tracks (without the
      0.546\,$M_\sun$ model) for the point (i) when the star reaches
      20\,kK (lower solid line) and (ii) when the star has faded by a
      factor of 10 in luminosity (upper solid line). The dashed lines
      show the 10\,kyr age when the nebula is assumed to become
      optically thin, and the 5\,kyr age corresponds to the size limit
      of the HST-selected sample of 5 arcsec.  The hashed area defines
      the visibility time of a bright PN.\newline  The lower panel
      shows the corresponding visibility time. The long-dashed (green)
      line shows the visibility time for the 10\,kyr limit. The dashed
      (blue) line shows the visibility time for the HST-selected
      sample (5\,kyr limit). The solid line shows the same but for the
      accelerated tracks. }
    \label{life1}
\end{figure}

Because PNe trace the dominant bulge populations, there is no
implication that they trace all populations equally. If we take a
limiting radius of 0.2\,pc for a bright PN (10 arcsec diameter at the
distance of the bulge,  which is twice the limit of our survey), this
corresponds to an expansion age of 10\,000\,yr for an expansion
velocity of 20\,km\,s$^{-1}$) or 5\,000\,yr for 40\,km\,s$^{-1}$. The
higher value is found in models \citep{SJS2005}; our data shows about
equal numbers of lower and higher velocities, correcting for the outer
edge, which expands 1.4 times faster than the mass-averaged velocity
derived from the spectra.  A PN becomes observable once the central
star reaches a temperature of 20\,kK and fades when either the nebula
becomes optically thin to ionizing radiation or the star begins to
fade. 

We define the end of the visibility time as the moment when the
stellar luminosity has declined by a factor of 10, or the nebula has
reached an age of 10\,kyr, whichever occurs sooner. For the newly
interpolated Bl\"ocker tracks, we plot these times in
Fig.\,\ref{life1}, top panel. The hashed region indicates when a
bright PN is visible, according to this criterion.  The lower panel
shows the corresponding visibility time for both the interpolated and
the accelerated Bl\"ocker tracks.

Above 0.61\,$M_{\sun}$ (0.58 for the accelerated tracks), the
visibility time is limited by the fading of the star and shortens
rapidly at higher masses. Below 0.56\,$M_{\sun}$, the nebula is
already older than 5000\,yr by the time ionization starts, and this
reduces the visibility time. For the accelerated tracks, these limits
are 0.58\,$M_\sun$ and 0.52\,$M_\sun$, respectively. The bright PN
population will be dominated by the stars with core masses in the
range 0.56--0.61\,$M_{\sun}$ (0.52--0.58\,$M_{\sun}$ for the
accelerated tracks), where the visibility time is maximum.  If much
fainter PNe are considered, the mass limits become relaxed.

More accurate visibility times can be found from the work of
\citet{SJSS2007}, where several evolutionary sequences of model PNe
were presented and computed with a 1-D radiation-hydrodynamic code.
They calculate total line fluxes as function of time.  For the
[\ion{O}{iii}] flux, the visibility times at $M(5007)<-1$ are shown in
Fig.\,15 of \citet{SJSS2007}. A similar result would be obtained for
H$\beta$. We note that these are optically thin models, which expand
faster than the value of 20\,km\,s$^{-1}$ assumed above; the
transition times are also included.  From their figures, it can be
clearly seen that if the faint limit is relaxed to $M=0$, the
visibility times of the faint models increase significantly. If
relaxed further to $M>1$, the more massive stars contribute for long
periods of time. This confirms how sensitive the life times (and
therefore number densities) are to the survey flux limits.

The observed  H$\beta$ line flux for our HST bulge sample varies
between $-11.9$ and $-14.2$ in units of ergs\,cm$^{-2}$\,s$^{-1}$ (or
mW\,m$^{-2}$). After correcting for extinction, the range is between
$-10.0$ and $-12.7$, which spans a range of 2.7\,dex. However, the
large majority fall within a range of a factor of 10 from $-10.5$ to
$-11.5$, which is consistent with the assumption that the known bulge
population is significantly flux limited.

\subsection{Selection bias}
\label{selbi}

The discussion in the previous subsection shows that the relation
between the PNe and the underlying stellar population depends on how
the sample is selected. A bright sample is biassed towards lower core
masses, while a fainter sample is sensitive to higher core masses - a
somewhat counter-intuitive effect. 

The selection effects are more complex. The presence in the PN
catalogues mainly depends on the [\ion{O}{iii}] fluxes, which depends
on stellar luminosity, size, and density of the nebula and foreground
extinction. In dense star fields, there may be a bias against highly
extended objects. High-resolution spectroscopy requires a high surface
brightness due to the narrow slits. The surface brightness declines as
$\sim r^{-3}$--$r^{-3.5}$ \citep{FP2006} for nebulae larger than 0.1pc
in radius. This declines introduces a bias against extended objects.

The HST sample was selected as an unbiased survey of compact PNe and
was taken from a random selection of nebulae of less than 5 arcsec
diameter (0.2 pc at the distance of the bulge).  There is no flux
limit imposed other than those in the original PNe surveys. Selecting
compact nebulae  negates the problem caused by decreasing surface
brightness of larger nebulae. Echelle spectra were obtained for all
objects observed by HST, which are independent of brightness, thus
removing a second potential bias.  The size limit corresponds to an
age of 2\,500--5\,000\,yr, depending on assumed expansion velocity
($V_{\rm aver} = 15$--30\,km\,s$^{-1}$: Table \ref{pne_data}).  The
corresponding visibility curves are shown in Fig.\,\ref{life1}.  The
visibility time for inclusion in the sample is shown by the dashed
line in the bottom panel. Compared to the original curve, the
distribution of visibility time versus core mass is much more uniform,
which varies by no more than a factor of two over the range of
masses   considered here (depicted is a wider range). The visibility
time for a core mass of 0.696\,$M_{\sun}$ (not shown) is down by a
factor of four compared to the peak at about 0.62\,$M_\sun$ and is
about 830\,yr.

For the accelerated tracks, the visibility time increases at lower
masses and decreases at higher masses; it varies only by about a
factor four over the considered range. Even for these models, the HST
sample is less biased than other samples currently available and is
sensitive to a wide range of core masses, but stars at the lowest and
highest core masses may still be under-represented. The derived ages
distribution of PN central stars can be corrected for this bias.

\subsection{Constraints by chemical composition}

\begin{figure}
\resizebox{\hsize}{!}{\includegraphics{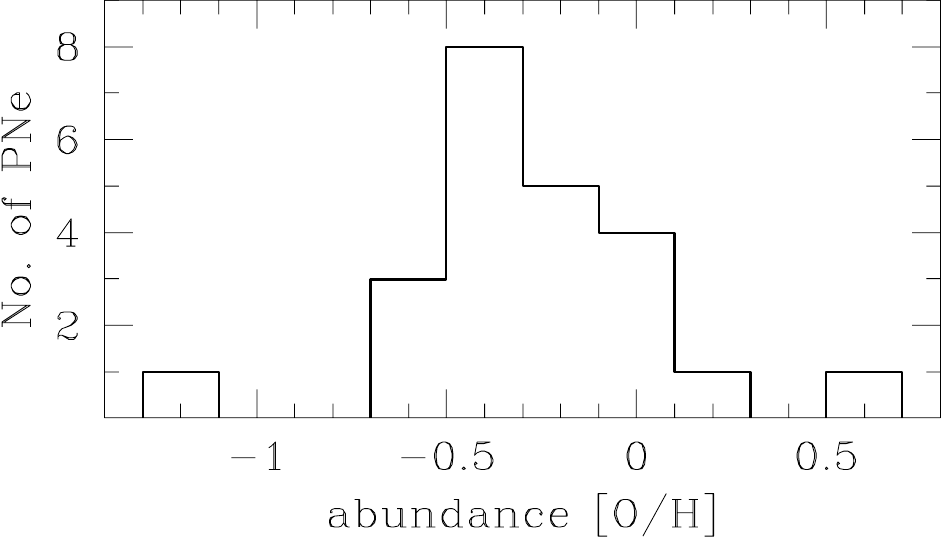} }
      \caption{The distribution of oxygen abundance within our sample. The data taken from Table\,\ref{pne_data} are presented relative to solar value (8.69 adopted from \citet{asplund2009}).}
    \label{abu_histo}
\end{figure}
 
\citet{ness2013} recently analysed the metallicity of the GB. The
sample of 28\,000 stars allowed them to define a number of subsets
with different [Fe/H] values. The three dominant components (named A,
B, and C in their  plots and tables) cover the range of [Fe/H] between
$-1.0$ and +0.5. We do not have iron abundance data, but we show the
oxygen data instead in Fig.\,\ref{abu_histo}, which is presented
relative to the solar abundance so it can be compared directly to the
plots of \citet{ness2013}. Our sample shows the same trend by
indicating a slightly sub-solar metallicity in general. Scarcity of
data does not allow for more detailed comparison; however, these data
do not contradict our previous statement (see Sect.\,\ref{sample})
that our sample corresponds to the dominant GB population.


\section{The star-formation history of the bulge}
\label{sfh_b}

Knowing that the best approximation to the empirical IFMR resulted in
total ages that are too young we use the IFMR as a free parameter and
adjust it trying to achieve better agreement. An alternative would be
to apply a further adjustment to the Bl\"ocker model: either of these
possibilities can increase the ages.

\subsection{Model initial-final mass relations}

\begin{figure}
\resizebox{\hsize}{!}{
\includegraphics{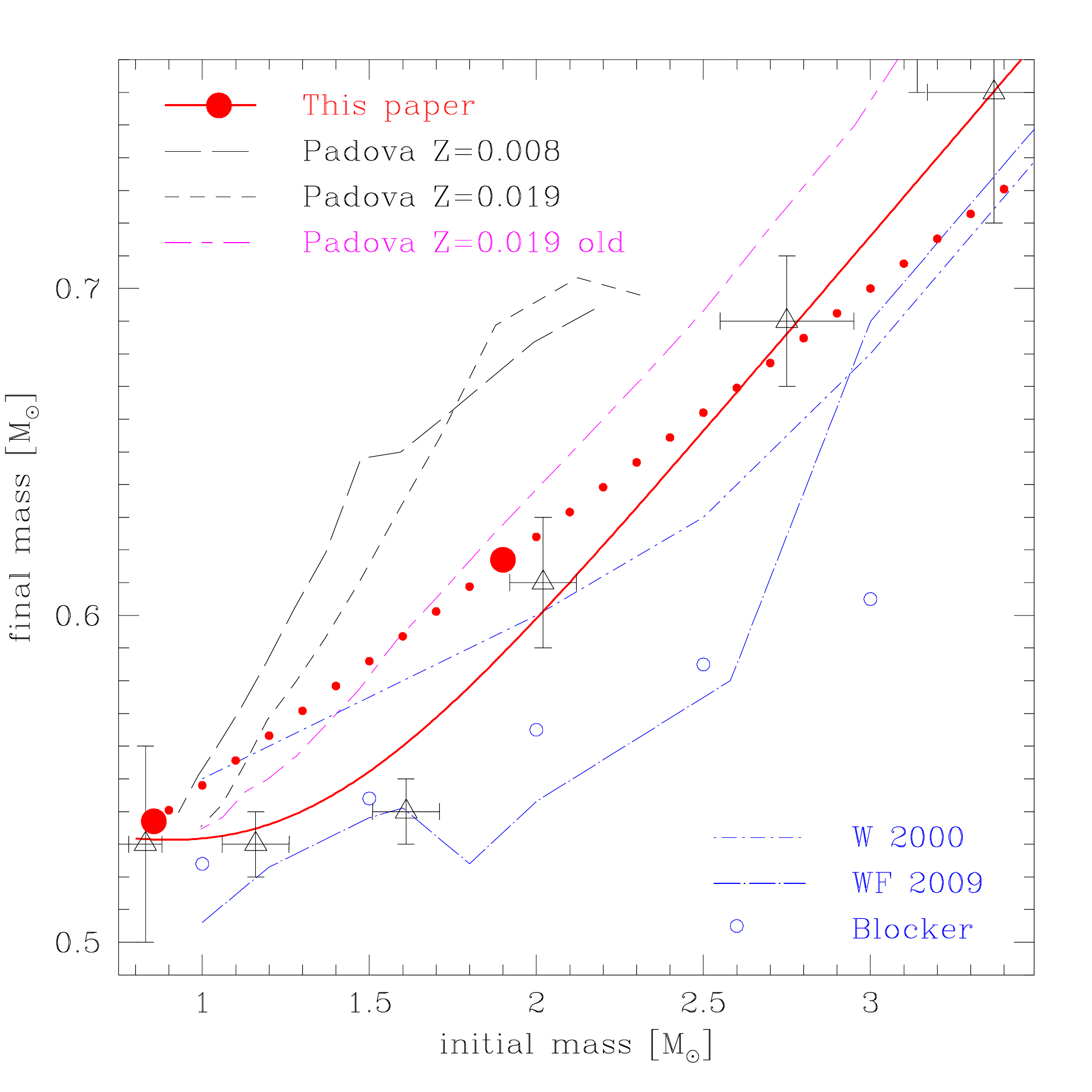} }
      \caption{The initial vs. final mass relation plotted for several
	models: Padova tracks (two sets at solar and one at subsolar
	metallicity); \citet{Weidemann2000}; \citet{WF2009} (solar
	metallicity).  The open circles follow the IFMR interpolated
	from \citet{B1995}.  The solid line follows IFMR as defined by
	Eq.\,(\ref{IFM}); the filled dots correspond to the best-fit
	IFMR described in the text.}
    \label{m_if}
\end{figure}

In Fig.\,\ref{m_if}, the empirical relations are compared to those
obtained from evolutionary models. From the ``Padova database of
stellar evolutionary tracks and isochrones'', we show (long-short-dash
magenta line) the set from \citet{GBBC2000} (with a simplified
thermal-pulsing AGB treatment). For solar metallicity we also show a
more recent set from \citet{MGB2008} (with a state-of-the-art
description of the thermal-pulsing AGB phase), which predicts a
steeper IFMR (the higher short-dash black line). At lower metallicity,
the relation is even steeper (long-dashed black line). Generally the
new data show larger scatter with no unambiguous IFMR. The Padova
isochrones are being improved and extended \citep{Bressan2012},
however for the moment; the isochrones do not include the TP-AGB.
\citet{WF2009} have also published a grid of stellar models for stars
on the AGB and post-AGB. Their data also show significant scatter
below initial masses of 2\,$M_{\sun}$.

Comparing these relations to the empirical IFMR above, the agreement
is poor. The old Padova set lies above the Casewell relation by some
0.05\,$M_{\sun}$. The new set lies even higher. The Bl\"ocker IFMR, in
contrast, lies well below the empirical relation, and the
\citet{WF2009} models are even lower.  The \citet{Weidemann2000}
relation still provides a better match than any of these models.

In the models, the final masses are largely determined by the
mass-loss prescription. The \citet{WF2009} models show low core masses
due to the use of the \citet{W2002} formalism leading to higher
mass-loss rates. We still lack a formalism, which reproduces the IFMR
measured in stellar clusters, and the scatter between the models (and
data) is not a surprise.

\subsection{Adjusting the IFMR to fit bulge ages}

\begin{figure}
\resizebox{\hsize}{!}{\includegraphics{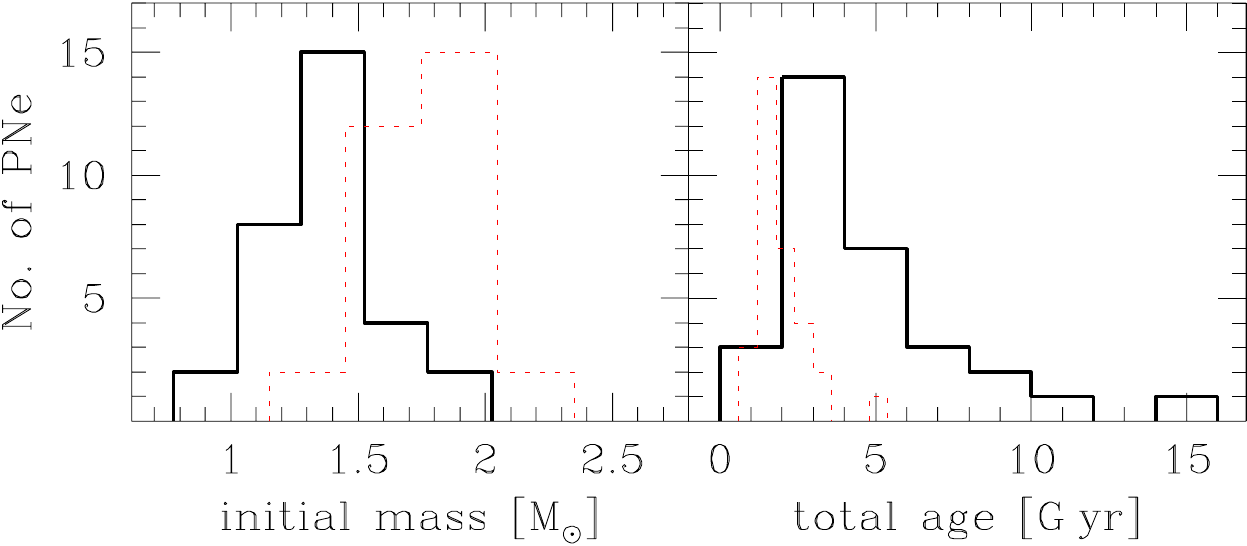} }
\caption{ The histograms of the interpolated data. The left panel
  shows the initial ZAMS masses, and the right panel shows the ages
  since the ZAMS was used later to derive a star-formation history
  (see Fig.\,\ref{sfr2}). The solid lines represent the data obtained
  with IFMR defined by Eq.\,(\ref{fin_IFMR}), while the thin dotted
  (red) lines repeat the histograms from Fig.\,\ref{h_31} for 
  comparison. }
\label{h_3final}
\end{figure}

From a selection of published ages of GB discussed in
Sect.\,\ref{agbul}, we focus on the data of \citet{Bensby2013}. They
have the advantage that they are based on a detailed and homogeneous
analysis of 58 individual stars; the data are presented in the tables,
can be grouped according to metallicity, and visualized as histograms.
Most other data comes from isochrones fitted to colour--magnitude
diagrams. 

The cspn ``accelerated'' masses obtained in Sect.\,\ref{final_m}
should be mapped to the desired age distribution with the use of an
appropriate IFMR. From Fig.\,\ref{m_if}, it can be seen that the
situation is complicated, and using a relation of higher order than
linear is not justified. 

We anchor the linear IFMR to the youngest and oldest objects from the
sample of \citet{Bensby2013}, which is up to about 1.6 and 14.7\,Gyr.
These should correspond to our most- and least-massive cspn. Applying
the relation for age--mass (Eq.\,(\ref{log_am})), we obtain an initial
mass of 1.90\,$M_\sun$ and age $1.4 \times 10^9$\,yrs for the youngest
object (that of final mass of 0.617\,$M_\sun$), while obtaining an
initial mass of 0.855\,$M_\sun$ and age $14.0 \times 10^9$\,yrs for
the oldest (having final mass of 0.537\,$M_\sun$). These two points
are shown as large dots in Fig.\,\ref{m_if}, while small dots follow
the proposed linear approximation to the IFMR:  \begin{equation}
M_{\rm i} = ( M_{\rm f} - 0.472) / 0.076. \label{fin_IFMR}
\end{equation} This relation was used to convert our final masses into
total ages. The results are shown in Fig.\,\ref{h_3final}. The solid
line represents the data obtained with IFMR defined by
Eq.\,(\ref{fin_IFMR}). For a comparison, these are overplotted (with a
thin dotted red line) with the histograms from Fig.\,\ref{h_31}
obtained previously with the IFMR defined by Eq.\,(\ref{IFM}). We see
now that the ages correspond far better to what is expected for the
bulge objects.

The large discrepancies between the various model IFMRs and the
cluster IFMR remain a concern. It would help this research if the IFMR
were better established.  We also note that our choice to adjust the
IFMR is non-unique. A higher acceleration of the Bl\"ocker tracks
could also increase the ages, and with a known IFMR, it would be
possible to constrain the post-AGB evolution.

\subsection{The star-formation history of the bulge from PNe}
 
To obtain the star-formation history from the age distribution, we
need to correct for observational biases.  The number of objects is
corrected for the difference in visibility time as a function of core
mass (see the discussion in Sect.\,\ref{selbi}) using the relation for
the accelerated tracks. The visibility correction reduces the peak and
raises the wings of the distribution.  Finally, the curves are divided
by the PN birth rate (stellar death rate) as function of stellar age
(Fig.\,\ref{birth}). This correction reduces the star-formation rates
for younger populations (which produce more PNe per unit mass).  The
correction becomes negligible for ages more than 7\,Gyr, but is large
for ages less than 4\,Gyr.

\begin{figure}
\resizebox{\hsize}{!}{\includegraphics{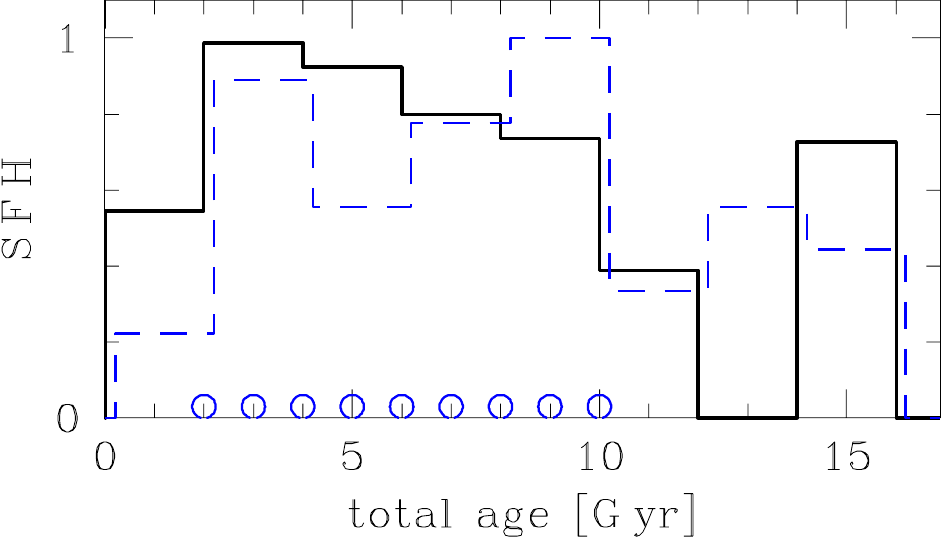} }
      \caption{The star-formation history for the Galactic bulge
	derived from the right panel of Fig. \ref{h_3final}, which is
	corrected for visibility time and PNe birth rate. A dashed
	(blue) line, which is slightly shifted for a better
	presentation, shows the overplotted histogram from
	\citet{Bensby2013} for the bulge stars, excluding the most
	metal-poor objects. The open circles represent the points at
	which a constant SFH was backward-corrected for visibility
	time and PNe birth rate, as shown in Fig.\,\ref{sfr2d1}. }
    \label{sfr2}
\end{figure}

The resulting histogram of the star-formation rate, normalized to
unity, is shown in Fig.\,\ref{sfr2} as a solid line. It is derived
from the right panel of Fig.\,\ref{h_3final}. The applied corrections
flatten the peak near 3\,Gyr and elevate the single oldest object. 

\begin{figure}
\resizebox{\hsize}{!}{\includegraphics{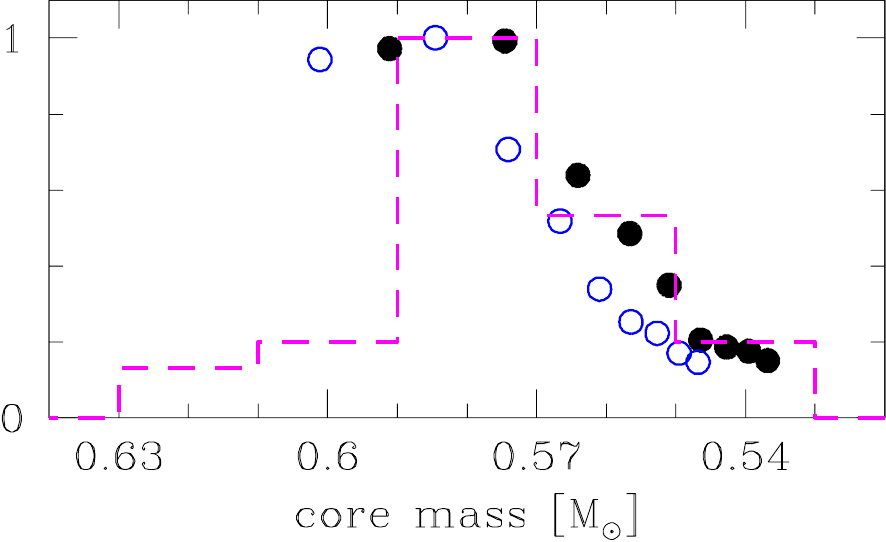} }
      \caption{
The open circles show the core-mass distribution (normalized to unity)
obtained from a constant SFH, which was backward-corrected for
visibility time and for the PNe birth rate at the points indicated in 
Fig.\,\ref{sfr2}. The mass-axis is drawn in reverse order to allow a
direct correspondence of the large dots from left to right in both
figures. For comparison the dashed line shows the final mass
distribution repeated from Fig.\,\ref{cm}. The filled circles
represent the same data points as open circles but obtained with
modified IFMR defined by Eq.\,\ref{mod_IFMR}. }
    \label{sfr2d1}
\end{figure}

\citet{Bensby2013} find that the age distributions for their stars
change with metallicity with the stars with solar and higher
metallicity being predominantly of intermediate age, and the
metal-poor populations peaking at ages older than 10\,Gyr. The
detailed results in the publication allows us to compare our histogram
with different subsets of \citet{Bensby2013}. Our result agrees best
with their ``regular-metallicity'' sample ($ -0.5 < [{\rm Fe/H}] <
+0.6 $), which we obtained by excluding the 15 most metal-poor ($
[{\rm Fe/H}] \le -0.5 $) stars from the whole sample. This
distribution (normalized to unity) is overplotted on our histogram in
Fig.\,\ref{sfr2} as the dashed (green) line. By considering all
uncertainties that concern post-AGB evolution and IFMR, the similarity
in shape of both dependencies seems remarkable.

The older, metal-poor population is much less apparent in our data.
This can be interpreted as a selection bias in that the lowest-mass
(oldest) stars may be lacking in the sample of compact PNe for reasons
discussed before. The abundance study of \citet{CGSB2009} shows that
bulge PNe have metallicities that range from solar-like to about
$-0.5$dex with few if any lower metallicity objects. Therefore, the
lower metallicity population in the bulge indeed appears to have
produced a few bright PNe. A deep survey for faint PNe may find a
missing low-metallicity, old-progenitor population. 

To verify the method, we tried to invert the correction routines and
apply them to the SFH that follows from \citet{Bensby2013}.
Figure\,\ref{sfr2} shows that the SFR can be crudely approximated as a
constant over the range 2--10\,Gyr. If everything is correct we should
obtain a distribution similar to the cspn masses shown in
Fig.\,\ref{cm}. Figure\,\ref{sfr2d1} shows this result. The open
circles correspond to those of Fig.\,\ref{sfr2}, and because equally
spaced total ages do not correspond to equally spaced cspn masses, we
made these dots denser than the histogram bins. The dashed (magenta)
line is copied from Fig.\,\ref{cm}. The similarity is obvious. We are
tempted to search for an IFMR that improves the fit. The result is
shown with filled circles. Equation\,\ref{fin_IFMR} needs only a
slight modification  by adding 0.01 \begin{equation} M_{\rm i} = (
M_{\rm f} - 0.472 + 0.01 ) / 0.076. \label{mod_IFMR} \end{equation}
This modification shifts cspn (final) masses towards lower values, and
if the new relation was drawn in Fig.\,\ref{m_if}, it would be shifted
downwards from the old dotted line, which shift is not much more than
the diameter of the big circles. This is much smaller than the scatter
between different IFMR. Apart from the IFMR, the mass-age relation is
also involved (as a linear approximation), and it is very likely that
both are not single relations of fixed metallicity. Applied here were
also the visibility corrections for which we have only estimations.
Therefore, we did not go further into fine-tuning of all these
relations. We can conclude that the proposed IFMR fits reasonably and
consistently to our data and to \citet{Bensby2013} within an accuracy
of about $0.01\,M_\sun$ for the final masses.

\section{Conclusions}

The purpose of the investigation is to use the masses and ages of the
central stars of planetary nebulae to constrain their evolution and to
derive a star-formation history of the Galactic bulge. Our conclusions
are as follows:

\begin{itemize}

\item We use HST images and VLT echelle spectra of 31 compact bulge
  planetary nebulae to derive expansion ages for the nebulae. The
  Torun models are used to calculate the mass-averaged expansion
  velocities. A  correction factor is derived from published
  hydrodynamical models \citep{Perinotto2004} to relate this to the
  expansion of the outer radius. The expansion age of the nebula and
  the temperature of the central star are fitted to post-AGB model
  tracks to derive masses of the central stars.

\item We use the set of Bl\"ocker post-AGB tracks. Two problems are
  identified with these tracks: the extremely slow evolution of the
  lowest mass tracks prohibits the formation of an ionized planetary
  nebula, and the final stellar masses are higher than white dwarf
  masses and asteroseismology masses. To force consistency, we exclude
  the lowest mass model from the model interpolations. We also make
  the ad-hoc adjustment that the envelope masses could be
  overestimated to accelerate the Bl\"ocker tracks by a  constant
  factor of three.  This produces consistency with white dwarf masses
  and allows for the lowest masses of PN central stars found from
  asteroseismology.  We note that the white dwarf mass distribution in
  the bulge is not known and that the acceleration proposed as
  constant can actually depend on stellar mass and envelope mass.

\item An IFMR is applied to the central star masses to derive the
  original (ZAMS) masses. An observational relation is derived from
  white dwarf masses in clusters. This gives a relation in very good
  agreement with \citet{Casewell2009} but with a shallowing at the
  lowest masses to allow for 0.53\,$M_\sun$ white dwarf masses in
  globular clusters.  A simple relation is proposed to estimate the
  total ages of the stars. 

\item  The derived ages are lower than seems plausible for bulge
  populations.  To improve the agreement between the derived ages and
  the published data on Galactic bulge ages, we use the IFMR as a free
  parameter and adjust it to a linear IFMR that relates our youngest
  and oldest object with corresponding youngest and oldest bulge stars
  in \citet{Bensby2013}. Such an IFMR appears steeper than other
  observational relations but is not as steep as some assumed for
  evolutionary calculations. With this IFMR, the total ages of our
  bulge objects cover the desired range. A further acceleration of the
  Bl\"ocker tracks is a possible alternative to adjusting IFMR.

\item To convert the age distribution of the PN central stars to a
  star-formation history, the inherent bias needs to be quantified. We
  estimate the total number of PNe present in the bulge as 2000.  We
  calculate the visibility time as a compact PN as a function of final
  mass and the stellar death rate (or PN birth rate) as function of
  age of the stellar population. The age distribution is corrected for
  these functions.  After the correction an apparent peak at 3\,Gyr
  disappears, which confirms the role of observational bias.

\item  The age distribution obtained by us is consistent with the
  subsample of \citet{Bensby2013} with metallicities close to solar
  values. The derived star-formation history shows a constant rate
  over time in general. 

\item  The set of proposed procedures was eventually verified by
applying it in an inverted version to the final, approximately
constant SFH. This resulted (as expected) in a core-mass distribution
similar to the one obtained from PNe with an accuracy of about
$0.01\,M_\sun$. The derived masses, ages and relations compose a
rather consistent picture.

\item Our results agree with an extended star formation in the
  Galactic bulge, which was approximately constant between  3 and
  10\,Gyr ago. This is more consistent with a secular development of
  the bulge, as expected from interaction with a bar. There may still
  be an older component of the bulge with a different origin, as the
  metal-poor population in the bulge region seems to be poorly
  represented among the compact PNe.

\end{itemize}

\begin{acknowledgements}
K.G. acknowledges financial support by Uniwersytet Mikolaja Kopernika
w Toruniu and A.A.Z. by the STFC. We gratefully acknowledge the
permission by D.\,Sch\"onberner and M.\,Steffen to use their
hydrodynamic models in this paper. We thank D.\,Sch\"onberner for
extensive comments. The final version benefited from helpful
suggestions of the referee M.\,Richer.
\end{acknowledgements}

\Online

\begin{appendix} 
\label{onl_app}
\section{The observations and models of planetary nebulae}

The model fitting procedure was described in brief in
Sect.\,\ref{phi_mod}. A full description can be found in
\citet{GZAGGW2006}, where we applied a genetic algorithm for the first
time to locate the best fit parameters. 

We searched for the density and velocity distributions as simply as
possible. The density was assumed similar to a reversed parabola with
both edges modified to fit the images. The best fit was judged by
visual inspection of the superposed images shown below in the boxes
marked ``surface brightness''. The stellar temperature, luminosity,
and the ionized mass was derived by PIKAIA as a compromise to a
simultaneously satisfying fit to all available emission strengths (see
Table\,\ref{pne_ratios}). The density values at the innermost and
outermost edges were additionally modified to reproduce the strengths
of emissions originated there, but the velocity field was solely
derived by the PIKAIA routine, which judges the results by the quality
of emission profile fitting (least squares method). The best velocity
field was searched among linear or parabola shapes. More complicated
velocity fields were not considered. Wherever the parabola provided no
improvement of the fit, we adopted the linear velocity as the
simplest.

This method was applied identically for all 31 objects presented
below. The largest nebulae and those barely resolved are treated in
the same way to allow us to discuss the whole data set consistently.

\begin{figure*}
\centering
 \includegraphics[width=5.5cm]{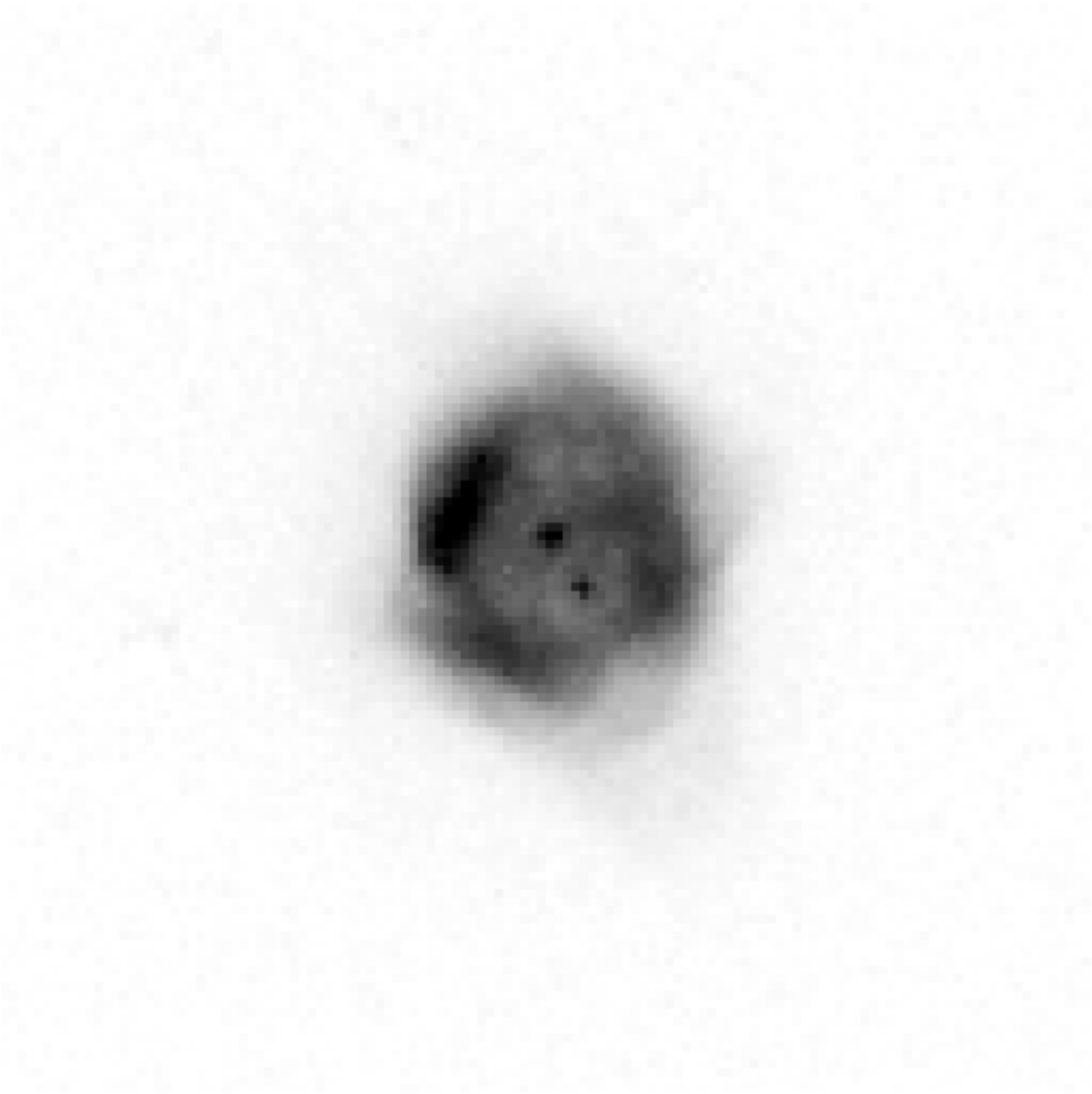}
 \includegraphics[width=10cm]{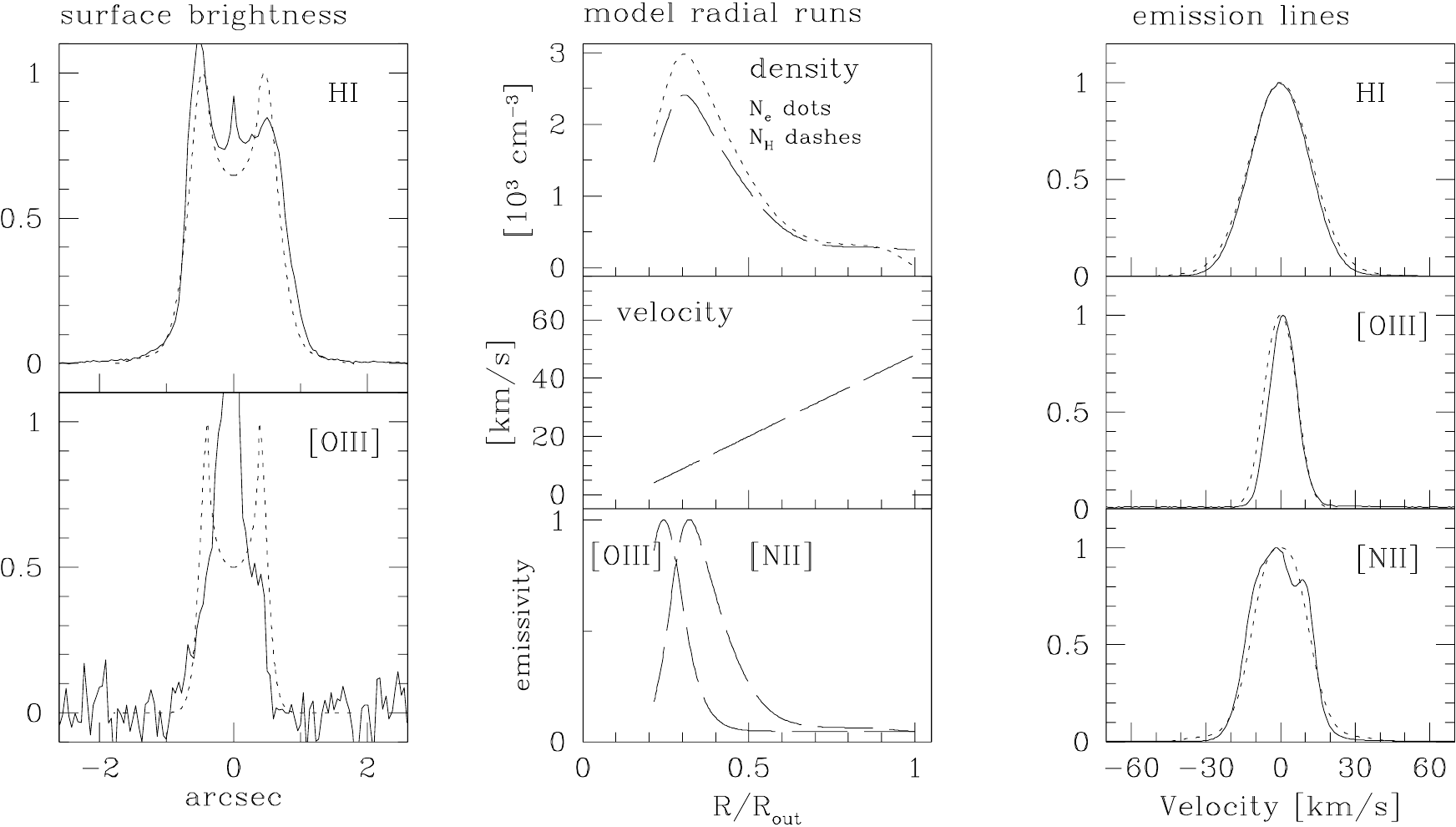}
 \caption{The nebula H\,1-55 (PN\,G\,001.7$-$04.4).  The left panel
   shows the HST H$\alpha$ image in linear grey scale, which is
   rotated so that the VLT spectrograph slit runs horizontally through
   the centre. In the small panels, the observed data are drawn in
   solid lines (as a rule), while model fits and model parameters are
   in dotted and dashed lines. The ``surface brightness'' panel shows
   the horizontal slices through the H$\alpha$ and [\ion{O}{iii}]
   image centre (averaged over nine pixels) with the model brightness
   profile superposed. Next, the radial distributions of the best-fit
   model for selected parameters are shown. The right-most panel shows
   the emission lines observed at the VLT with the superposed modelled
   line profiles corrected for instrumental broadening of 0.1\AA. }
    \label{mod_fit_1}
\end{figure*}

The series of figures (Figs.\,\ref{mod_fit_1}--\ref{mod_fit_last}, one
per object) present the HST observations  with the derived model
dependencies. The HST image are shown rotated, such that the VLT
echelle spectroscopic slit is horizontal. In most cases, this puts the
slit along the minor axis of the PN. The echelle spectrum was
extracted from the central arcsecond only. For extended objects, outer
areas of the slit were not used. Wherever no observational data was
available, we left the corresponding box empty.

\begin{longtab}
\centering
\begin{longtable}{llllllllll}
\caption{\label{pne_ratios}Galactic bulge PNe observed H$\beta$ fluxes and line strengths together with the corresponding photoionization model results. The line strengths are normalized to the strengths of H$\beta$ equal 1. For each object, the first row contains the observed parameters while the corresponding model values are typed below (in the next line).}\\
\hline\hline
\noalign{\smallskip}
PN\,G & name & $\log F({\rm H}\beta)$ & \ion{He}{ii} & [\ion{N}{ii}] & [\ion{O}{i}] & [\ion{O}{ii}] & [\ion{O}{iii}] & [\ion{S}{ii}] & [\ion{S}{iii}] \\
 &  &  & 4686\AA & 6583\AA & 6302\AA & 3729\AA & 5007\AA & 6731\AA & 6312\AA \\
\hline
\noalign{\smallskip}
\endfirsthead
\caption{continued.} \\
\hline\hline
\noalign{\smallskip}
PN\,G & name & $\log F({\rm H}\beta)$ & \ion{He}{ii} & [\ion{N}{ii}] & [\ion{O}{i}] & [\ion{O}{ii}] & [\ion{O}{iii}] & [\ion{S}{ii}] & [\ion{S}{iii}] \\
 &  &  & 4686\AA & 6583\AA & 6302\AA & 3729\AA & 5007\AA & 6731\AA & 6312\AA \\
\hline
\noalign{\smallskip}
\endhead
%
\hline
\endfoot
%
\hline
\endlastfoot
001.7-04.4   &  H 1-55       & -11.5 &  --      & 2.77E+00 & 2.00E-02 & 2.30E-01 & 1.10E-01 & 1.90E-01 & 2.90E-03 \\
             &               & -12.2 & 2.15E-05 & 3.49E+00 & 4.97E-03 & 2.72E-01 & 1.56E-01 & 1.77E-01 & 6.33E-03 \\
\noalign{\smallskip}
002.3-03.4   &  H 2-37       & -11.9 & --       & 1.11E+00 & 1.59E-02 & --       & 8.07E+00 & 8.00E-02 & 2.00E-02 \\
             &               & -11.5 & 2.78E-01 & 1.29E+00 & 1.35E-03 & 3.07E-01 & 6.51E+00 & 2.05E-02 & 1.82E-02 \\
\noalign{\smallskip}
002.8+01.7   &  H 2-20       & -11.0 & --       & 2.81E+00 & 2.00E-02 & 9.90E-01 & 2.70E-01 & 1.10E-01 & --       \\
             &               & -11.2 & 8.39E-06 & 3.12E+00 & 8.42E-03 & 5.72E-01 & 3.38E-01 & 5.30E-02 & 5.30E-02 \\
\noalign{\smallskip}
002.9-03.9   &  H 2-39       & -11.6 & 3.80E-01 & 5.80E-02 & --       & --       & 1.30E+01 & 1.29E-02 & 8.70E-03 \\
             &               & -11.5 & 4.43E-01 & 6.14E-02 & 3.01E-05 & 5.80E-02 & 1.09E+01 & 6.67E-04 & 2.48E-03 \\
\noalign{\smallskip}
003.1+03.4   &  H 2-17       & -11.3 & --       & 1.91E+00 & --       & 3.20E-01 & 1.00E-03 & 1.60E-01 & --       \\
             &               & -11.7 & 5.49E-06 & 1.41E+00 & 3.23E-04 & 1.12E-01 & 4.12E-02 & 1.69E-01 & 1.86E-02 \\
\noalign{\smallskip}
003.6+03.1   &  M 2-14       & -11.4 & 2.40E-02 & 2.94E+00 & 2.20E-02 & 7.00E-01 & 2.36E+00 & 6.70E-02 & 1.30E-02 \\
             &               & -10.7 & 4.68E-04 & 2.49E+00 & 2.09E-03 & 1.61E-01 & 2.77E+00 & 2.71E-02 & 1.45E-02 \\
\noalign{\smallskip}
003.8+05.3   &  H 2-15       & -12.3 & --       & 5.88E+00 & 1.20E-01 & --       & 9.24E+00 & 4.00E-01 & --       \\
             &               & -11.8 & 4.03E-02 & 2.96E+00 & 5.12E-02 & 1.11E+00 & 7.12E+00 & 1.27E-01 & 1.91E-02 \\
\noalign{\smallskip}
004.1-03.8   &  KFL 11       & -12.2 & --       & 4.20E-01 & --       & --       & 8.65E+00 & 4.80E-02 & --       \\ 
             &               & -11.0 & 5.59E-03 & 8.24E-01 & 1.07E-02 & 3.99E-01 & 6.34E+00 & 2.05E-02 & 7.82E-03 \\
\noalign{\smallskip}
004.8+02.0   &  H 2-25       & -11.2 & --       & 9.50E-01 & 1.00E-02 & 1.50E+00 & 6.40E-01 & 3.00E-02 & 1.00E-02 \\
             &               & -11.1 & 1.13E-05 & 7.28E-01 & 5.85E-04 & 7.12E-01 & 5.90E-01 & 4.07E-02 & 1.08E-02 \\
\noalign{\smallskip}
006.1+08.3   &  M 1-20       & -10.8 & --       & 4.50E-01 & 4.70E-02 & 5.50E-01 & 1.00E+01 & 2.30E-02 & --       \\
             &               & -10.6 & 1.27E-02 & 3.67E-01 & 1.17E-02 & 1.55E-01 & 9.58E+00 & 1.17E-02 & 8.57E-03 \\
\noalign{\smallskip}
006.3+04.4   &  H 2-18       & -11.5 & 5.00E-02 & 7.80E-02 & 1.00E-02 & 1.60E-01 & 1.30E+01 & 1.80E-02 & 1.00E-02 \\
             &               & -11.0 & 7.80E-02 & 2.57E-03 & 5.58E-08 & 5.65E-03 & 7.80E+00 & 4.78E-06 & 2.22E-04 \\
\noalign{\smallskip}
006.4+02.0   &  M 1-31       & -10.0 & --       & 1.39E+00 & 6.40E-02 & 3.30E-01 & 7.66E+00 & 6.90E-02 & 1.70E-02 \\
             &               & -9.51 & 1.84E-03 & 1.65E+00 & 6.33E-04 & 2.57E-02 & 4.20E+00 & 5.76E-03 & 1.16E-02 \\
\noalign{\smallskip}
008.2+06.8   &  He 2-260     & -11.7 & --       & 9.10E-01 & 2.00E-02 & 1.54E+00 & 4.80E-02 & 4.00E-02 & 7.00E-03 \\
             &               & -11.0 & 8.80E-07 & 4.17E-01 & 1.90E-03 & 1.27E-01 & 3.75E-01 & 1.08E-02 & 7.46E-03 \\
\noalign{\smallskip}
008.6-02.6   &  MaC 1-11     & -11.2 & --       & 8.10E-02 & --       & --       & 8.48E+00 & --       & --       \\
             &               & -11.1 & 3.39E-02 & 7.65E-02 & 5.64E-06 & 2.92E-02 & 8.75E+00 & 6.37E-04 & 2.60E-03 \\
\noalign{\smallskip}
351.1+04.8   &  M 1-19       & -10.8 & --       & 8.20E-01 & --       & --       & 5.02E+00 & 3.90E-02 & --       \\
             &               & -11.0 & 2.55E-03 & 8.92E-01 & 2.72E-03 & 3.64E-01 & 5.09E+00 & 3.48E-02 & 1.61E-02 \\
\noalign{\smallskip}
351.9-01.9   &  Wray 16-286  & -10.5 & --       & 5.90E-01 & --       & --       & 1.06E+01 & 5.00E-02 & --       \\
             &               & -10.5 & 5.37E-02 & 6.92E-01 & 2.53E-02 & 1.80E-01 & 1.27E+01 & 3.80E-02 & 1.16E-02 \\
\noalign{\smallskip}
352.6+03.0   &  H 1-8        & -10.5 & --       & 2.50E+00 & --       & --       & 8.81E+00 & 9.70E-02 & --       \\
             &               & -10.6 & 1.85E-02 & 1.55E+00 & 8.52E-03 & 1.49E-01 & 8.93E+00 & 3.19E-02 & 1.08E-02 \\
\noalign{\smallskip}
354.5+03.3   &  Th 3-4       & -10.6 & --       & 9.20E-01 & --       & --       & 1.65E+01 &  --      &  --      \\ 
             &               & -10.5 & 1.41E-03 & 7.23E-01 & 1.39E-03 & 3.21E-02 & 4.56E+00 & 2.73E-03 & 1.04E-02 \\
\noalign{\smallskip}
354.9+03.5   &  Th 3-6       & -11.2 & --       & 2.03E+00 & --       &  --      & 2.09E+00 & 1.20E-01 &  --      \\
             &               & -11.0 & 3.85E-04 & 2.42E+00 & 3.16E-03 & 4.48E-01 & 2.18E+00 & 5.58E-02 & 9.04E-03 \\
\noalign{\smallskip}
355.9+03.6   &  H 1-9        & -10.6 & --       & 6.30E-01 & 1.70E-02 &  --      & 1.89E+00 & 1.00E-02 & 2.30E-02 \\
             &               & -10.3 & 3.95E-05 & 5.74E-01 & 2.96E-03 & 2.50E-01 & 1.92E+00 & 1.55E-02 & 1.40E-02 \\
\noalign{\smallskip}
356.1-03.3   &  H 2-26       & -12.7 & 1.80E-01 & 6.21E+00 & 2.40E-01 &  --      & 6.28E+00 & 6.00E-01 &  --      \\
             &               & -12.0 & 4.14E-01 & 7.57E+00 & 1.24E-01 & 1.60E+00 & 6.31E+00 & 3.89E-01 & 3.89E-01 \\
\noalign{\smallskip}
356.5-03.6   &  H 2-27       & -11.9 & --       & 1.58E+00 & --       &  --      & 7.01E+00 & 1.90E-01 &  --      \\
             &               & -11.8 & 1.30E-02 & 3.07E+00 & 4.34E-02 & 1.11E+00 & 4.43E+00 & 1.29E-01 & 1.47E-02 \\
\noalign{\smallskip}
356.8+03.3   &  Th 3-12      & -11.8 & --       & 3.05E+00 & --       &  --      & 4.10E-01 & 1.20E-01 &  --      \\
             &               & -11.9 & 1.16E-05 & 2.96E+00 & 9.43E-03 & 3.32E-01 & 4.89E-01 & 1.03E-01 & 5.94E-03 \\
\noalign{\smallskip}
356.9+04.4   &  M 3-38       & -11.5 & 3.00E-01 & 1.47E+00 & 1.28E-01 &  --      & 1.53E+01 & 1.10E-01 & 4.70E-02 \\
             &               & -11.8 & 2.60E-01 & 1.68E+00 & 2.24E-03 & 1.04E-01 & 1.11E+01 & 1.70E-02 & 2.10E-02 \\
\noalign{\smallskip}
357.1-04.7   &  H 1-43       & -11.5 & --       & 1.94E+00 & 1.40E-02 &  --      & --       & 9.60E-02 &  --      \\
             &               & -12.1 & 2.18E-10 & 1.06E+00 & 3.31E-05 & 1.70E-02 & 1.38E-04 & 2.18E-01 & 1.54E-03 \\
\noalign{\smallskip}
357.2+02.0   &  H 2-13       & -10.9 & --       & 6.60E-01 & 5.00E-02 & --       & 1.55E+01 & 1.30E-01 & 1.70E-02 \\
             &               & -10.6 & 8.05E-02 & 1.31E+00 & 2.26E-02 & 3.52E-01 & 1.31E+01 & 4.39E-02 & 2.05E-02 \\
\noalign{\smallskip}
358.5-04.2   &  H 1-46       & -10.4 & --       & 4.72E-01 & 2.00E-02 & --       & 4.78E+00 & 2.41E-02 & 1.58E-02 \\
             &               & -10.2 & 5.11E-03 & 2.59E-01 & 4.07E-03 & 5.26E-02 & 6.63E+00 & 1.10E-02 & 1.37E-02 \\
\noalign{\smallskip}
358.5+02.9   &  Al 2-F       & -12.0 & --       & 3.90E-02 & --       & --       & 1.39E+01 &  --      & --       \\
             &               & -12.0 & 7.20E-01 & 3.88E-02 & 2.95E-06 & 2.65E-02 & 1.09E+01 & 1.77E-04 & 1.30E-03 \\
\noalign{\smallskip}
358.7+05.2   &  M 3-40       & -11.0 & --       & 2.02E+00 & --       & --       & --       & 1.30E-01 & --       \\
             &               & -11.2 & 3.75E-03 & 2.36E+00 & 1.05E-02 & 4.58E-01 & 4.45E+00 & 6.18E-02 & 1.46E-02 \\
\noalign{\smallskip}
358.9+03.4   &  H 1-19       & -10.7 & --       & 2.24E+00 & --       & --       & 3.98E+00 & 6.00E-02 & --       \\
             &               & -10.9 & 8.99E-04 & 1.92E+00 & 2.50E-03 & 1.51E-01 & 2.93E+00 & 2.02E-02 & 9.40E-03 \\
\noalign{\smallskip}
359.2+04.7   &  Th 3-14      & -11.4 & --       & 1.12E+00 & --       & 6.30E-01 & 1.00E-02 & 1.60E-01 & --       \\
             &               & -11.2 & 3.59E-09 & 2.15E+00 & 2.39E-03 & 1.12E-01 & 7.55E-04 & 1.04E-01 & 1.58E-03 \\
\end{longtable}
\end{longtab}

\begin{figure*}
\centering
 \includegraphics[width=5.5cm]{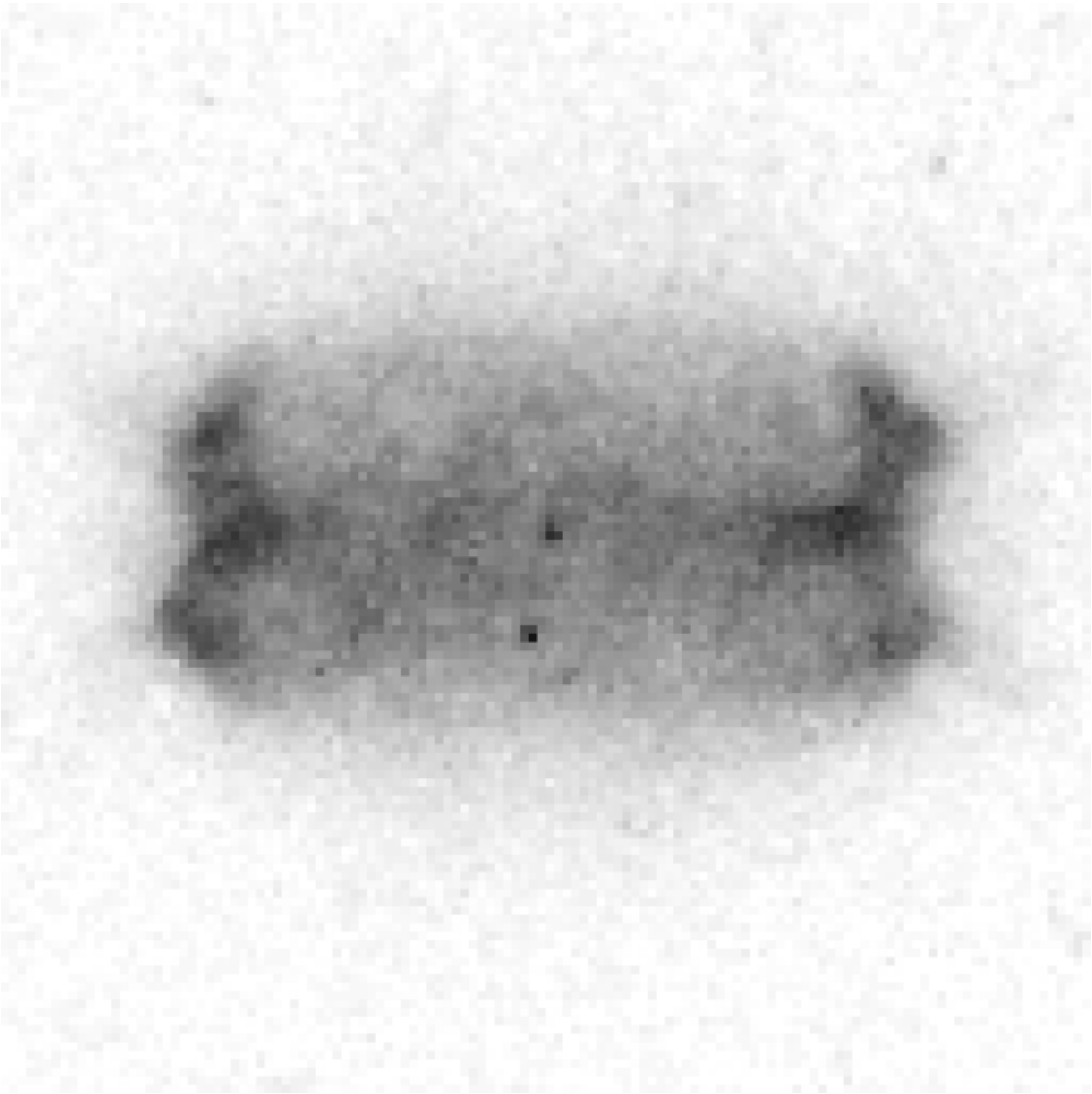} 
 \includegraphics[width=10cm]{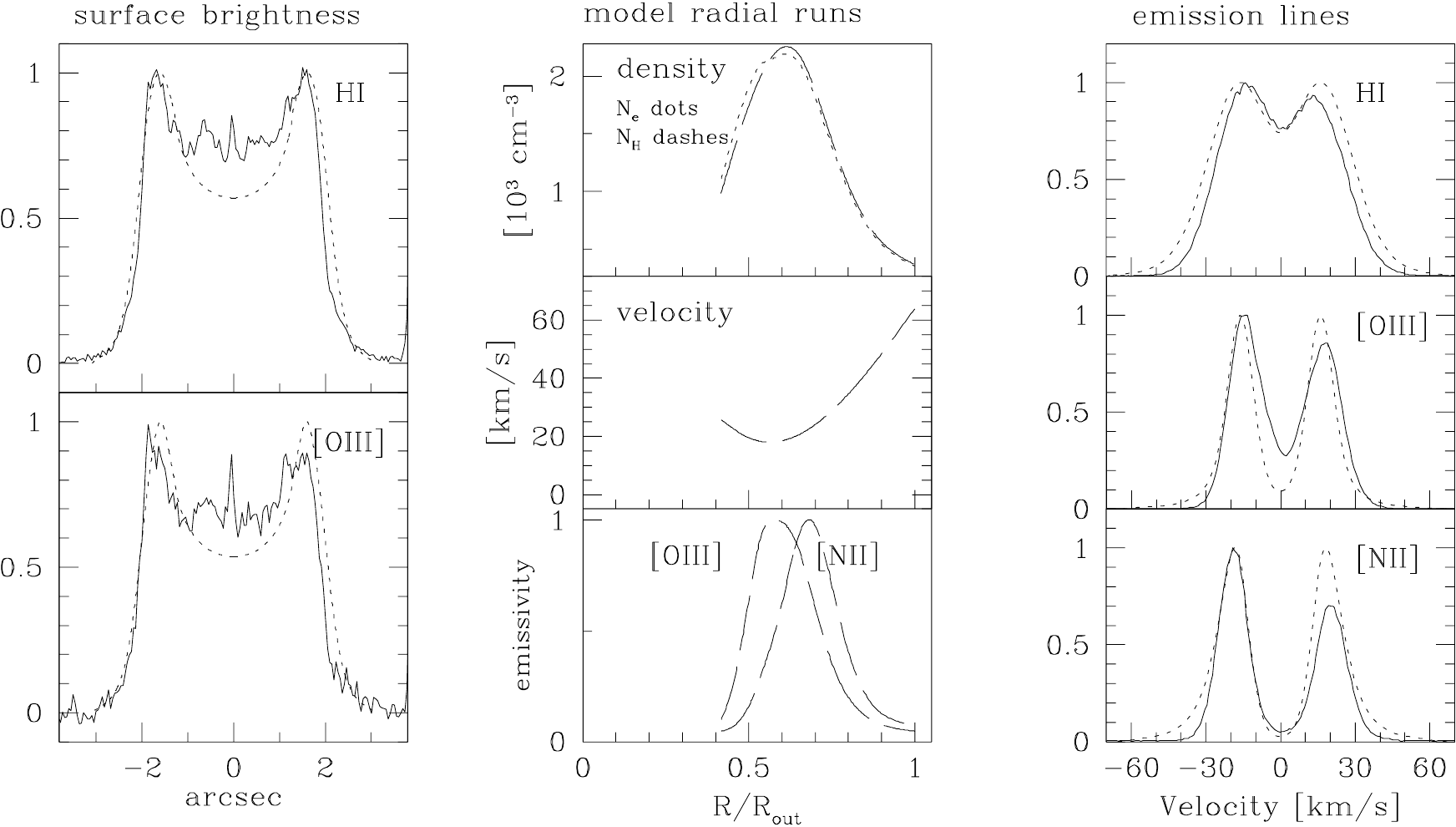} 
 \caption{The nebula H\,2-37 (PN\,G\,002.3$-$03.4). The data are presented as
   in Fig.\,\ref{mod_fit_1}.  }
\end{figure*}

\begin{figure*}
\centering
 \includegraphics[width=5.5cm]{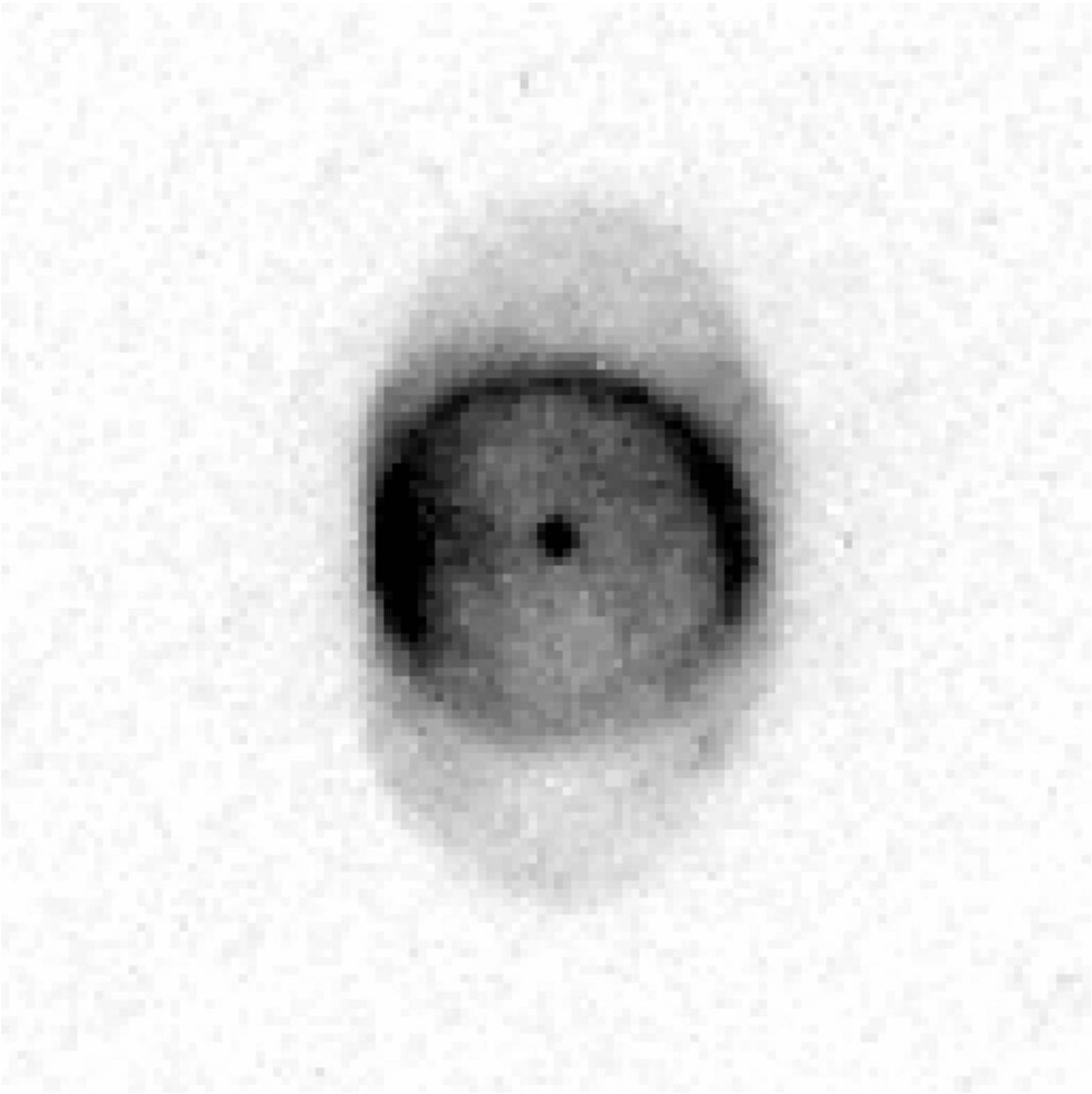} 
 \includegraphics[width=10cm]{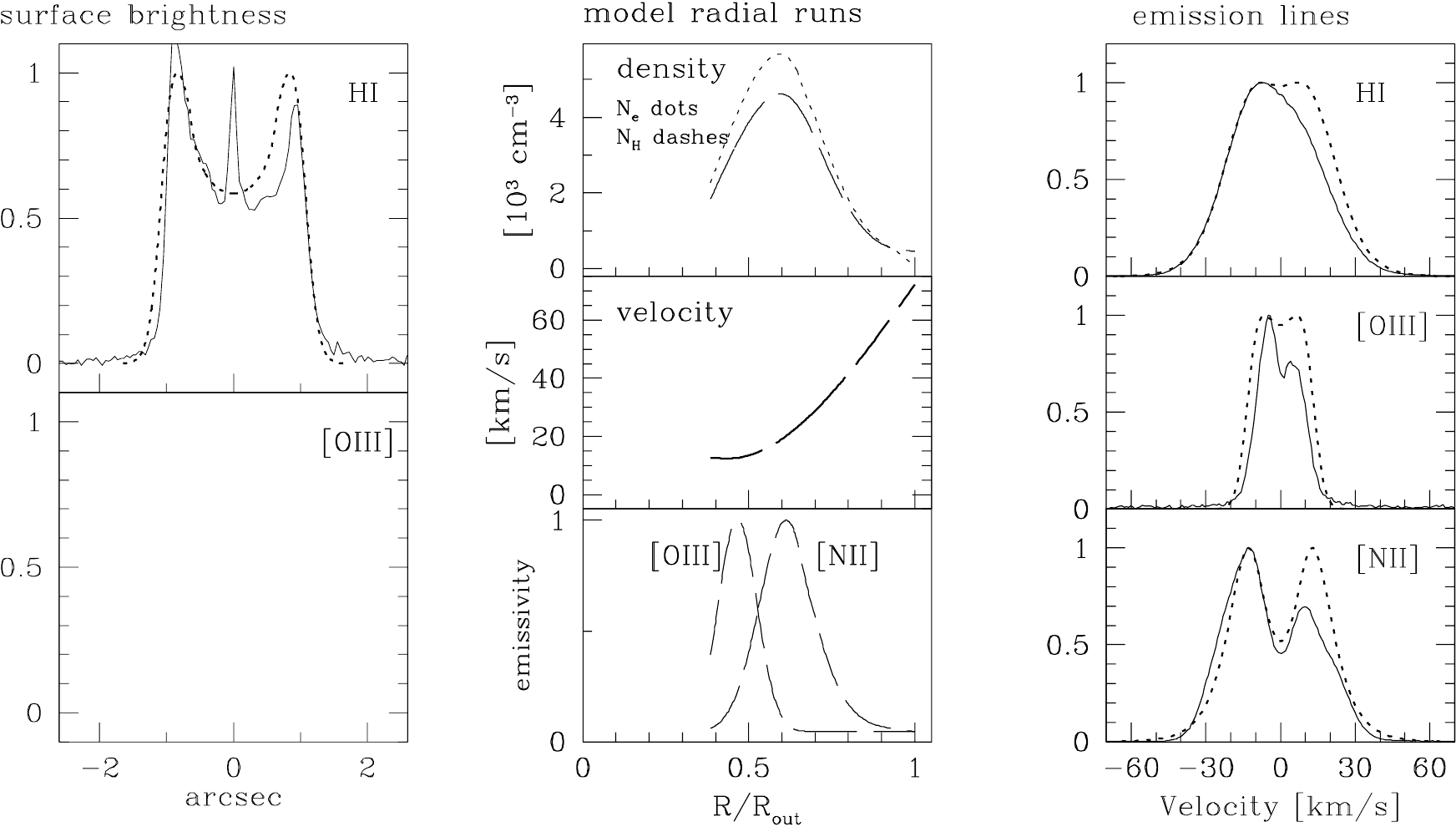}
 \caption{The nebula H\,2-20 (PN\,G\,002.8+01.7). The data are presented as in
   Fig.\,\ref{mod_fit_1}.    }
\end{figure*}

\begin{figure*}
\centering
 \includegraphics[width=5.5cm]{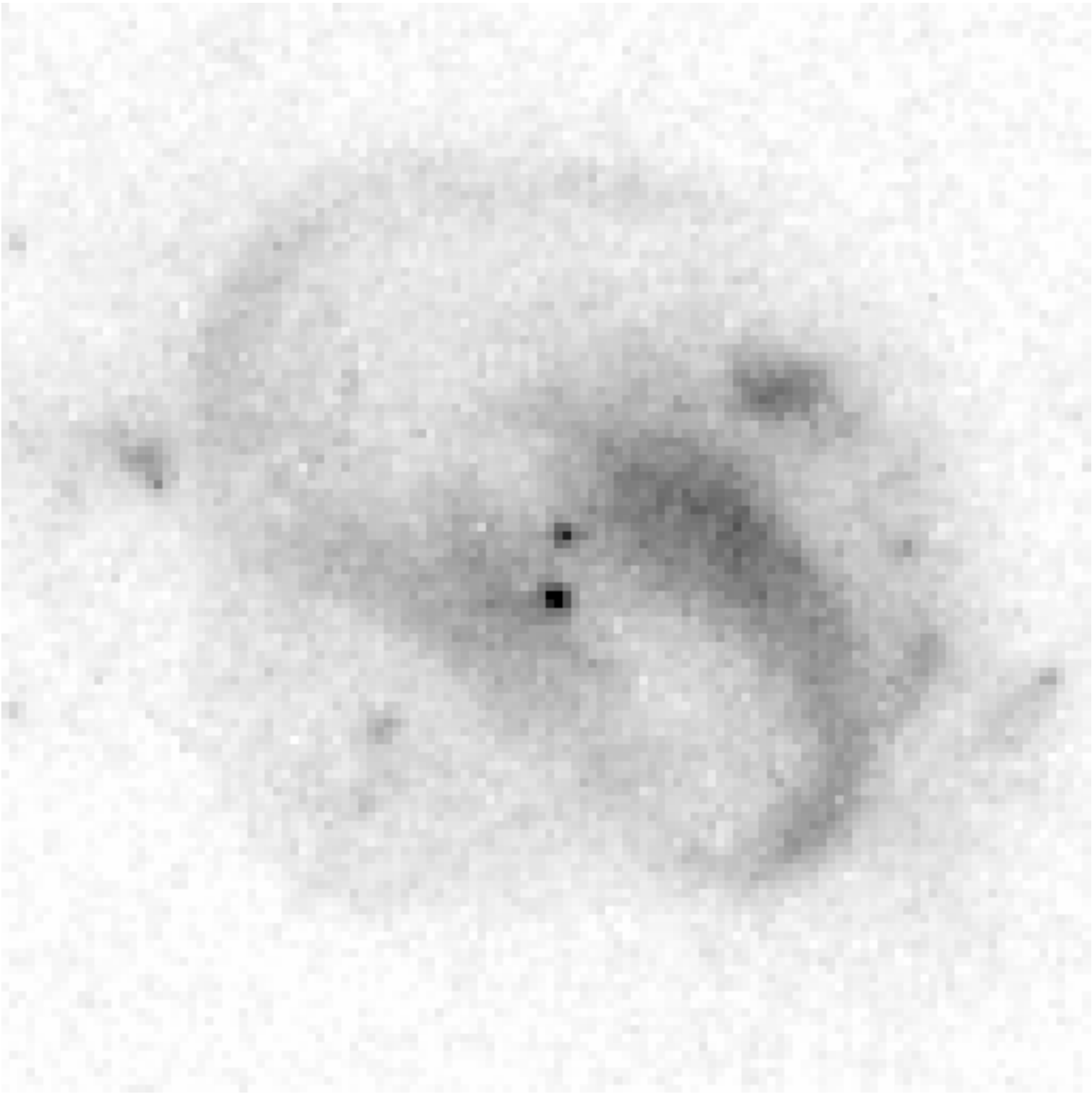} 
 \includegraphics[width=10cm]{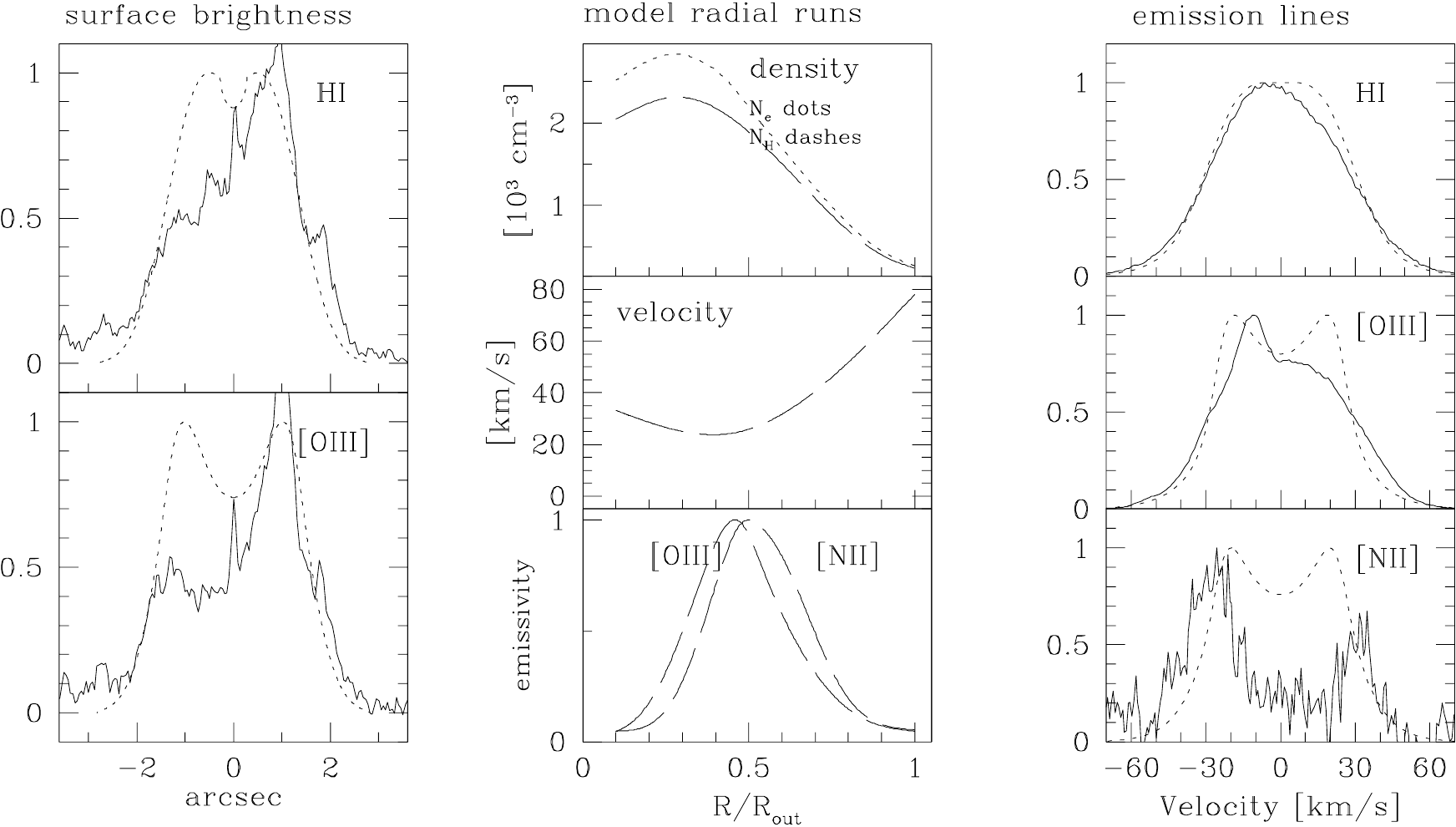}
 \caption{The nebula H\,2-39 (PN\,G\,002.9-03.9). The data are presented as in
   Fig.\,\ref{mod_fit_1}. In spite of the complex structure, the velocity
   profile is well reproduced by the 1-D model.  }
\end{figure*}

\begin{figure*}
\centering
 \includegraphics[width=5.5cm]{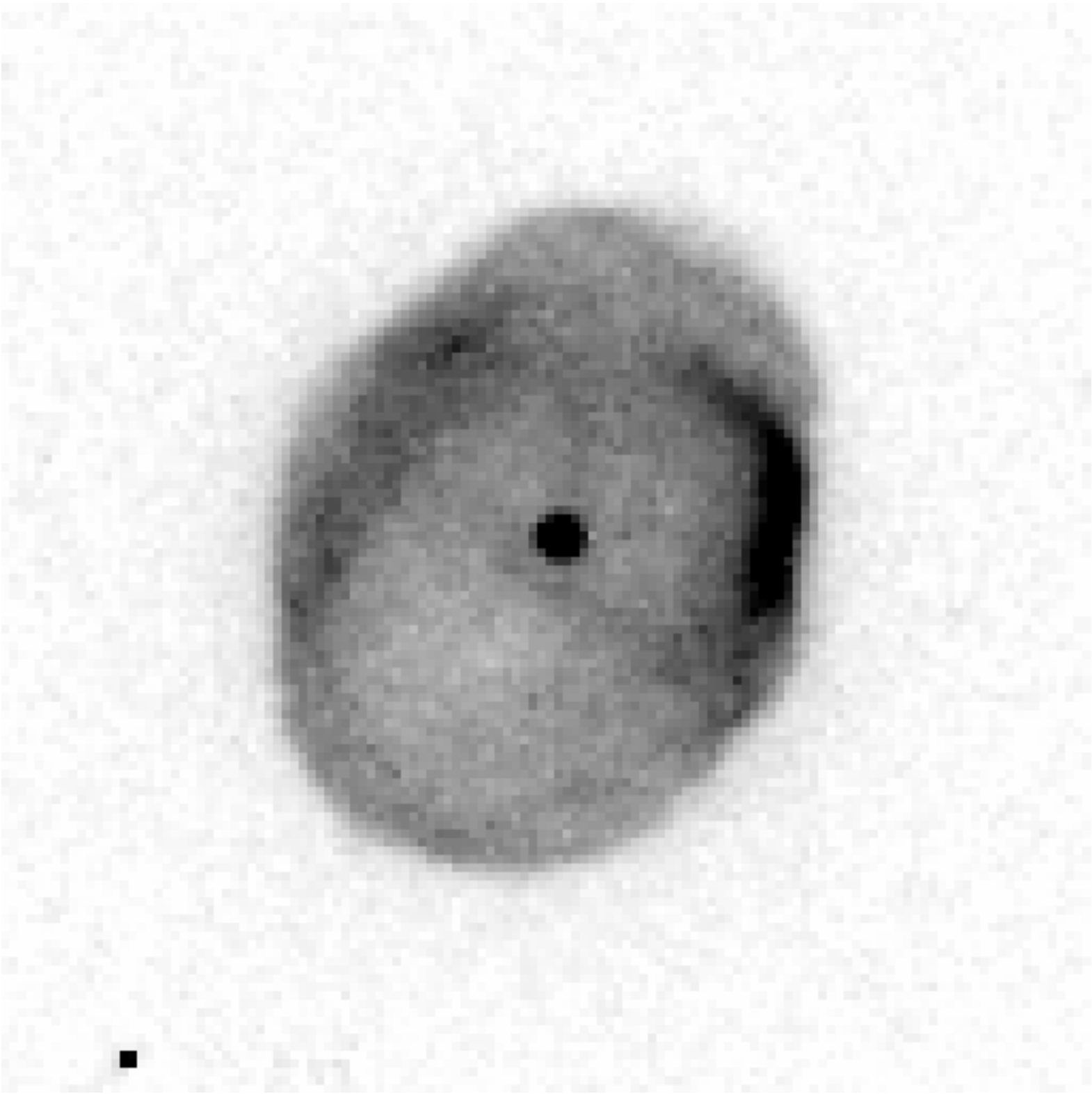} 
 \includegraphics[width=10cm]{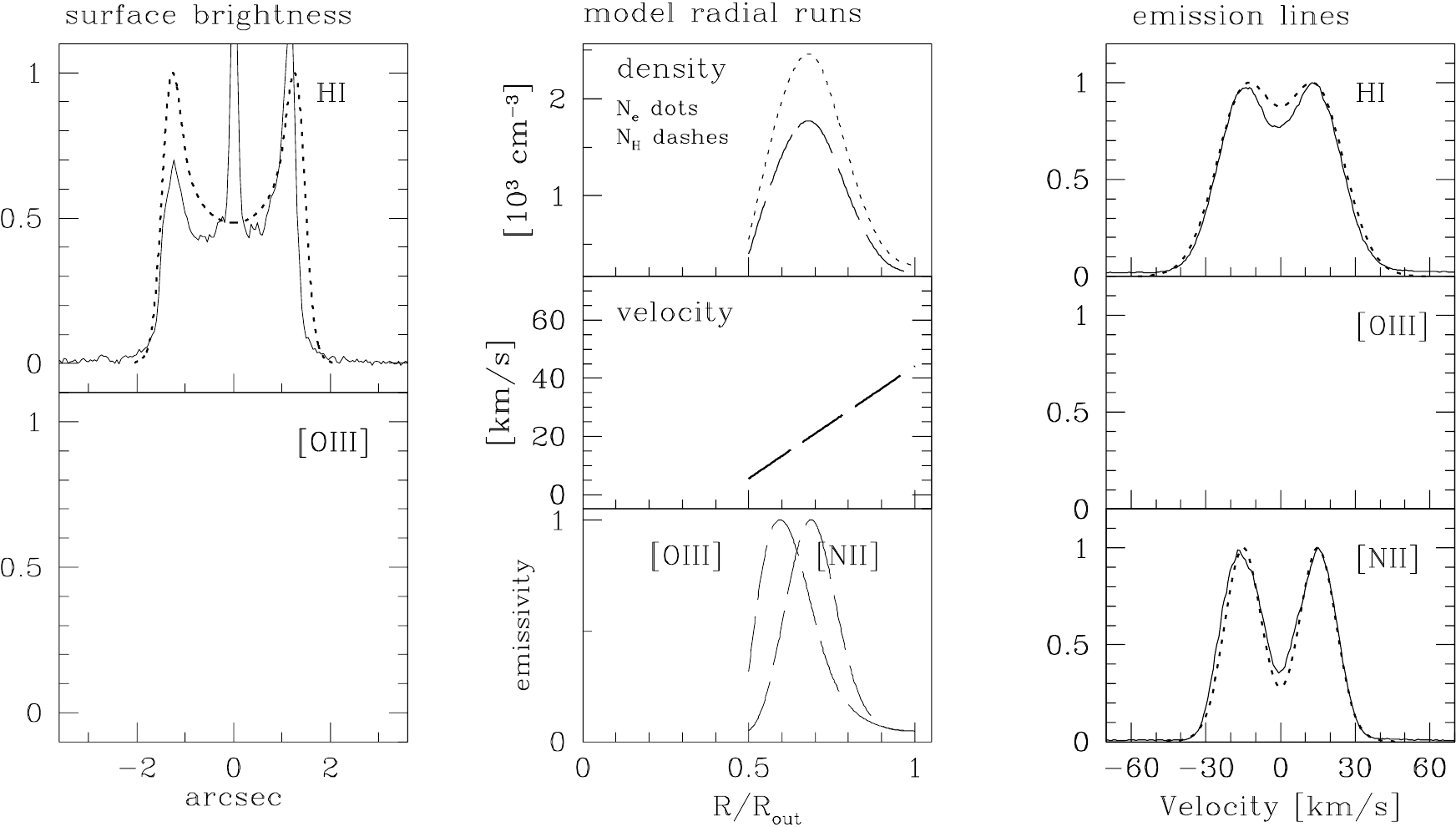}
 \caption{The nebula H 2-17 (PN\,G\,003.1+03.4). The data are presented as in
   Fig.\,\ref{mod_fit_1}.    }
\end{figure*}

\begin{figure*}
\centering
 \includegraphics[width=5.5cm]{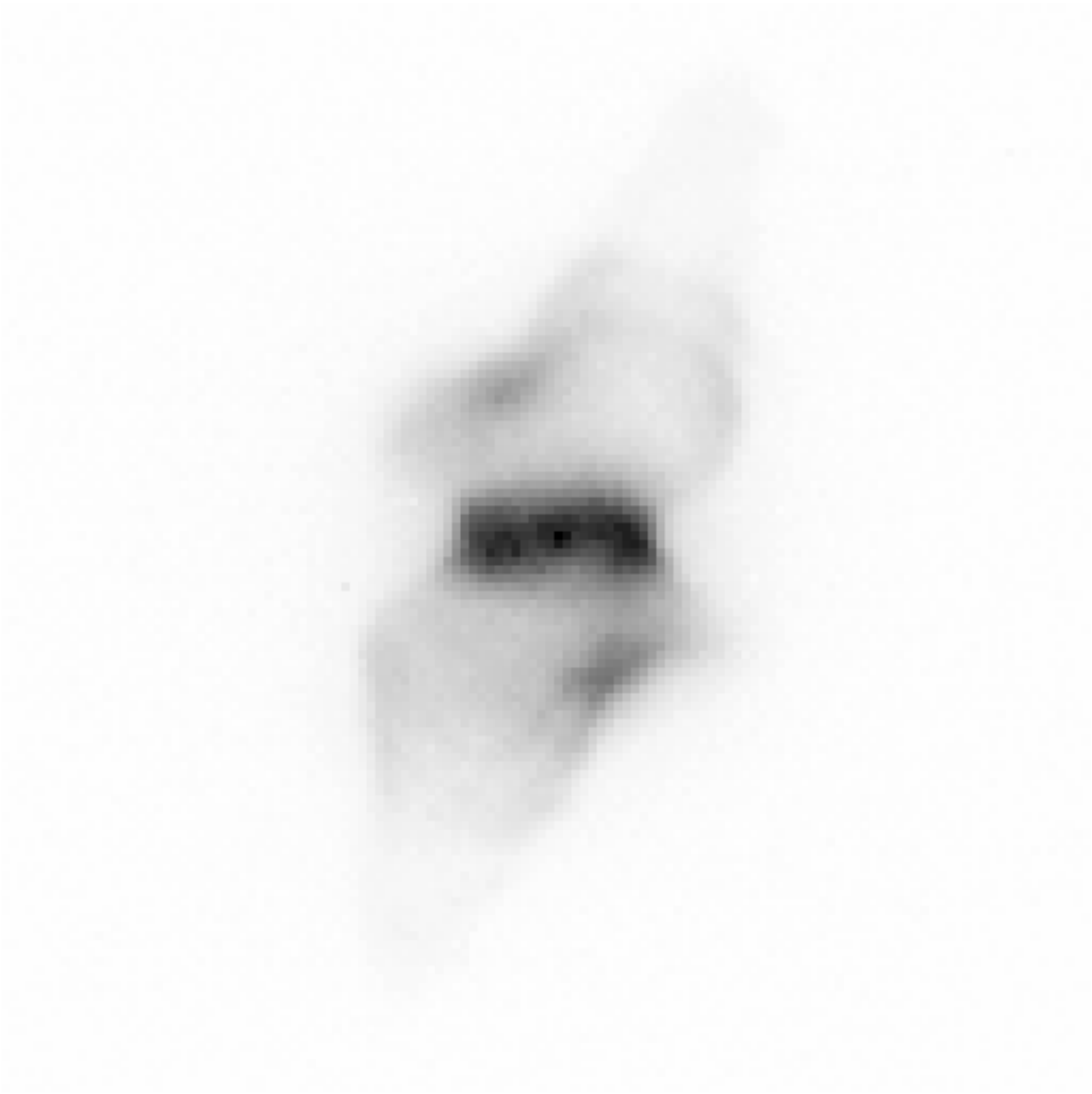} 
 \includegraphics[width=10cm]{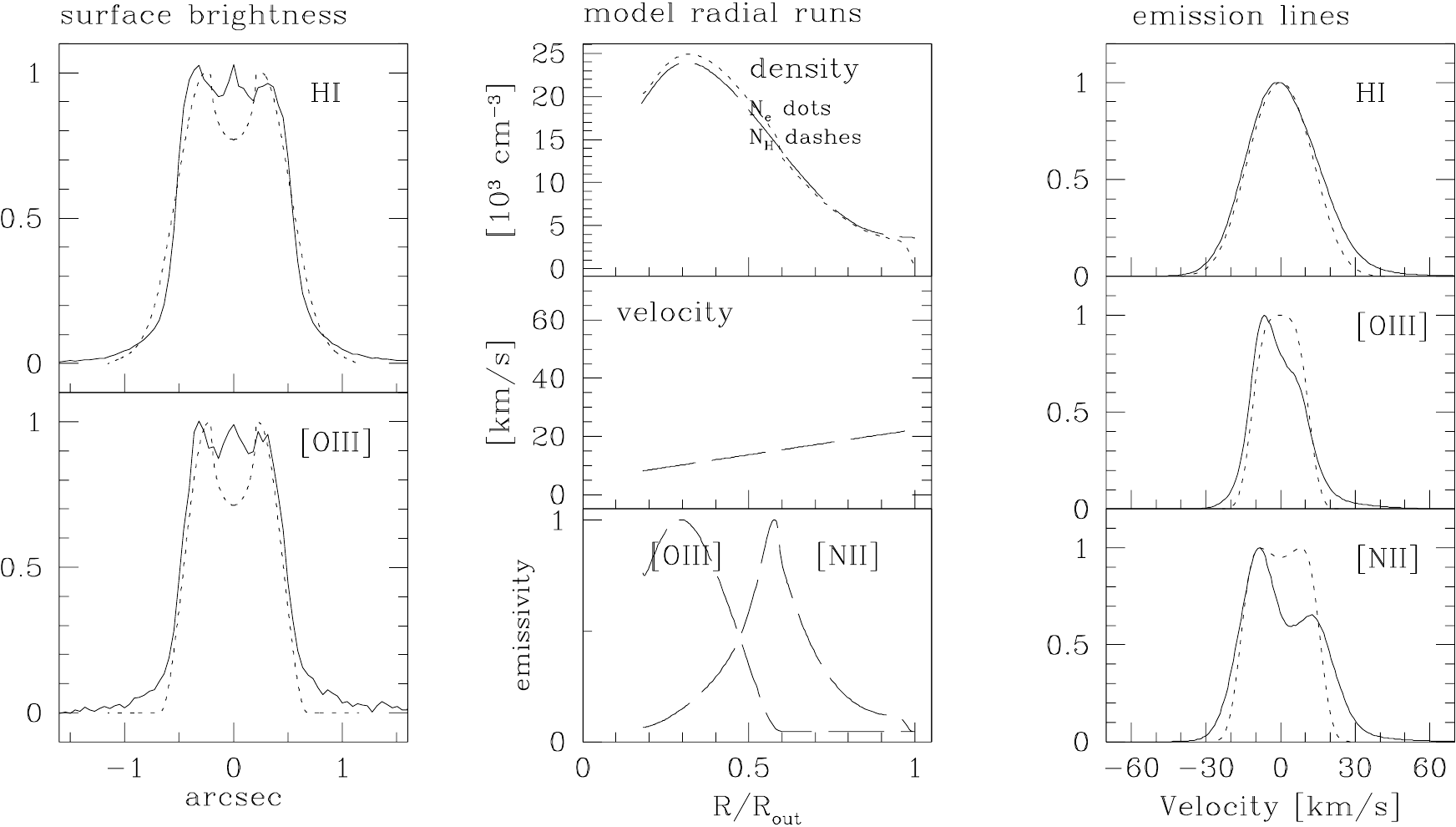}
 \caption{The nebula M\,2-14 (PN\,G\,003.6+03.1). The data are presented as in
   Fig.\,\ref{mod_fit_1}.  The slit is along the minor axis. The high velocity
   wing of the [\ion{N}{ii}] line is not in the model (which assumes a linear velocity 
   gradient). It may arise from the base of the polar flow.   }
\end{figure*}

\begin{figure*}
\centering
 \includegraphics[width=5.5cm]{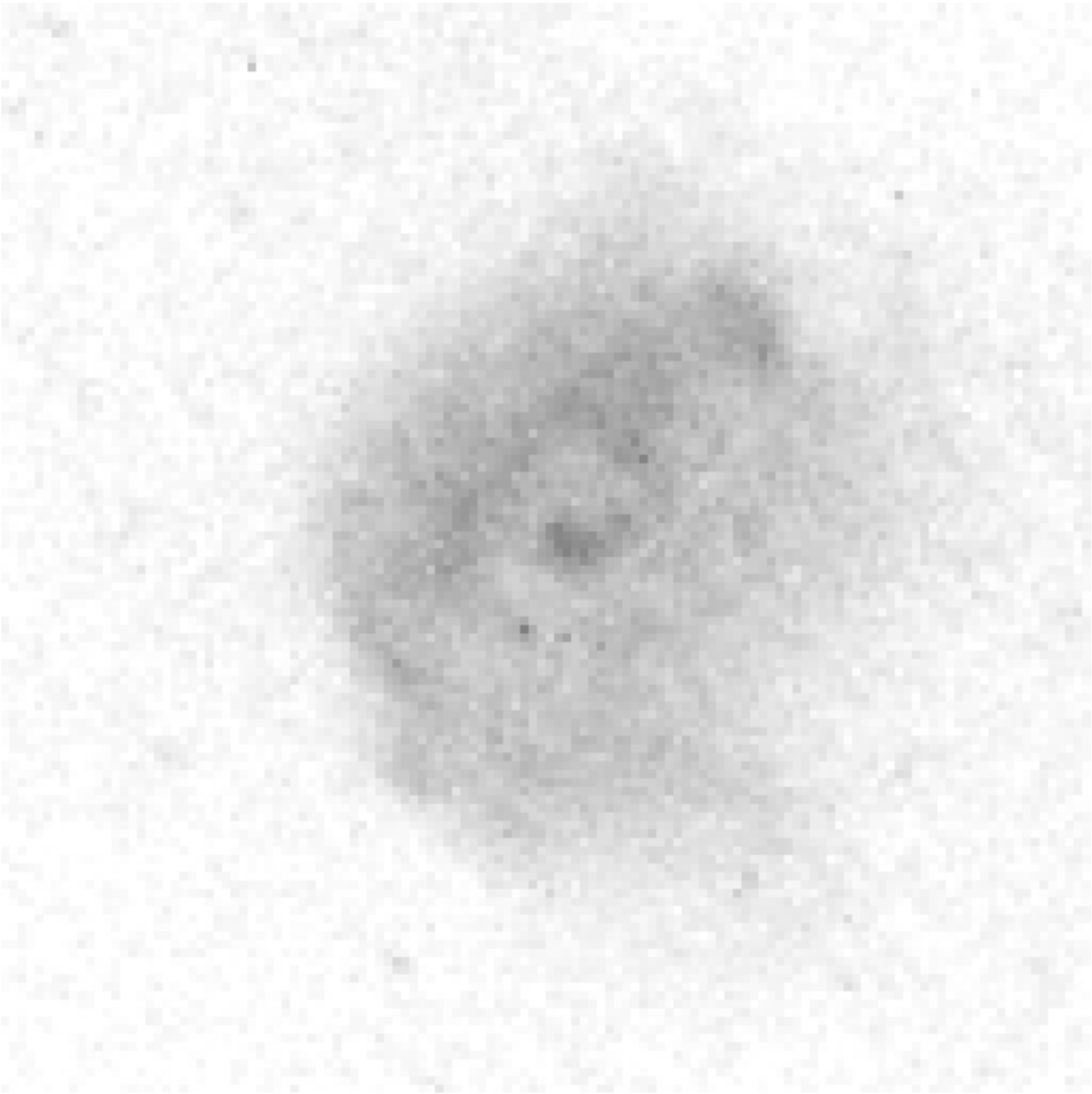} 
 \includegraphics[width=10cm]{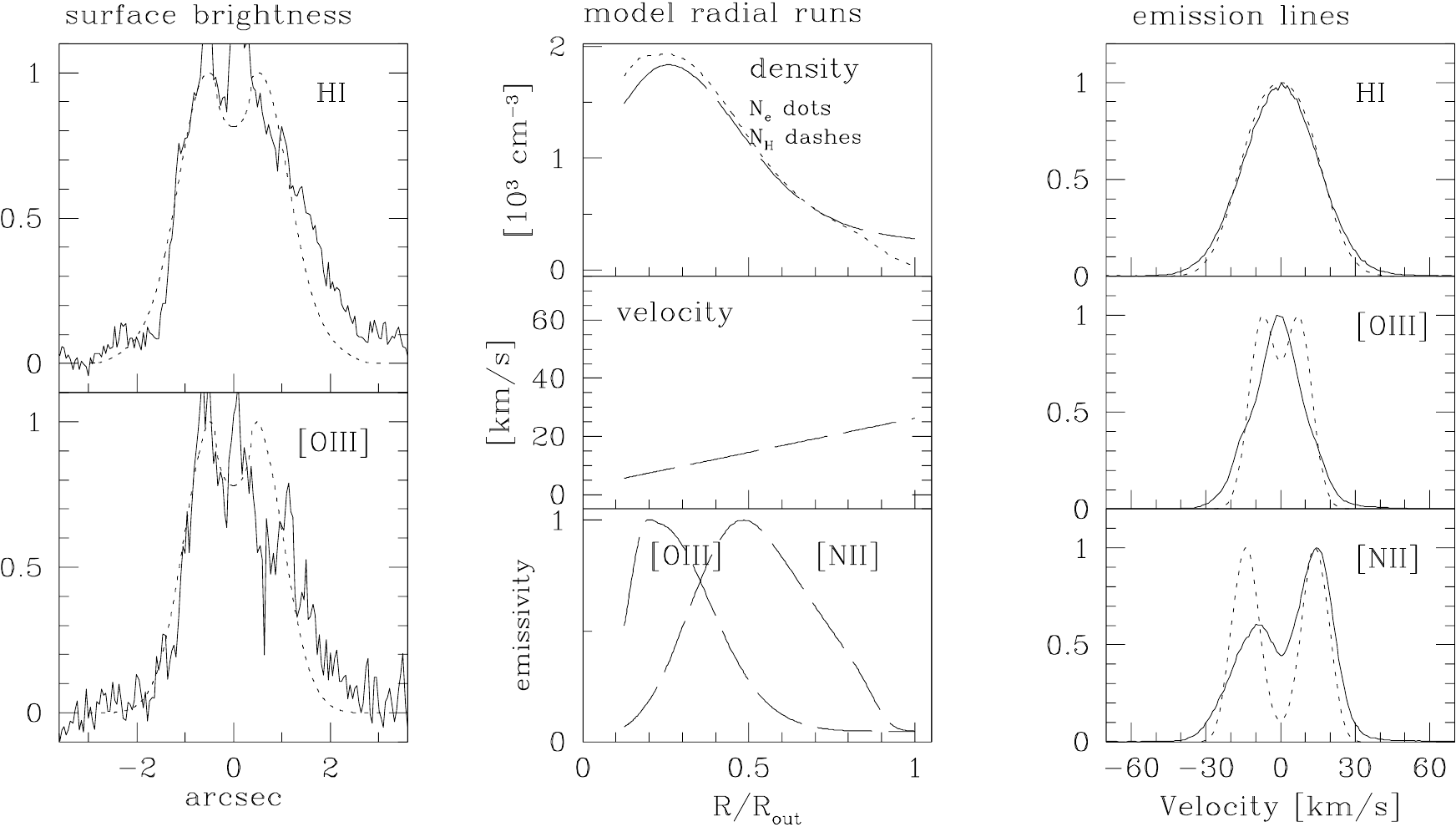}
 \caption{The nebula H\,2-15 (PN\,G\,003.8+05.3). The data are presented as in
   Fig.\,\ref{mod_fit_1}.    }
\end{figure*}

\begin{figure*}
\centering
 \includegraphics[width=5.5cm]{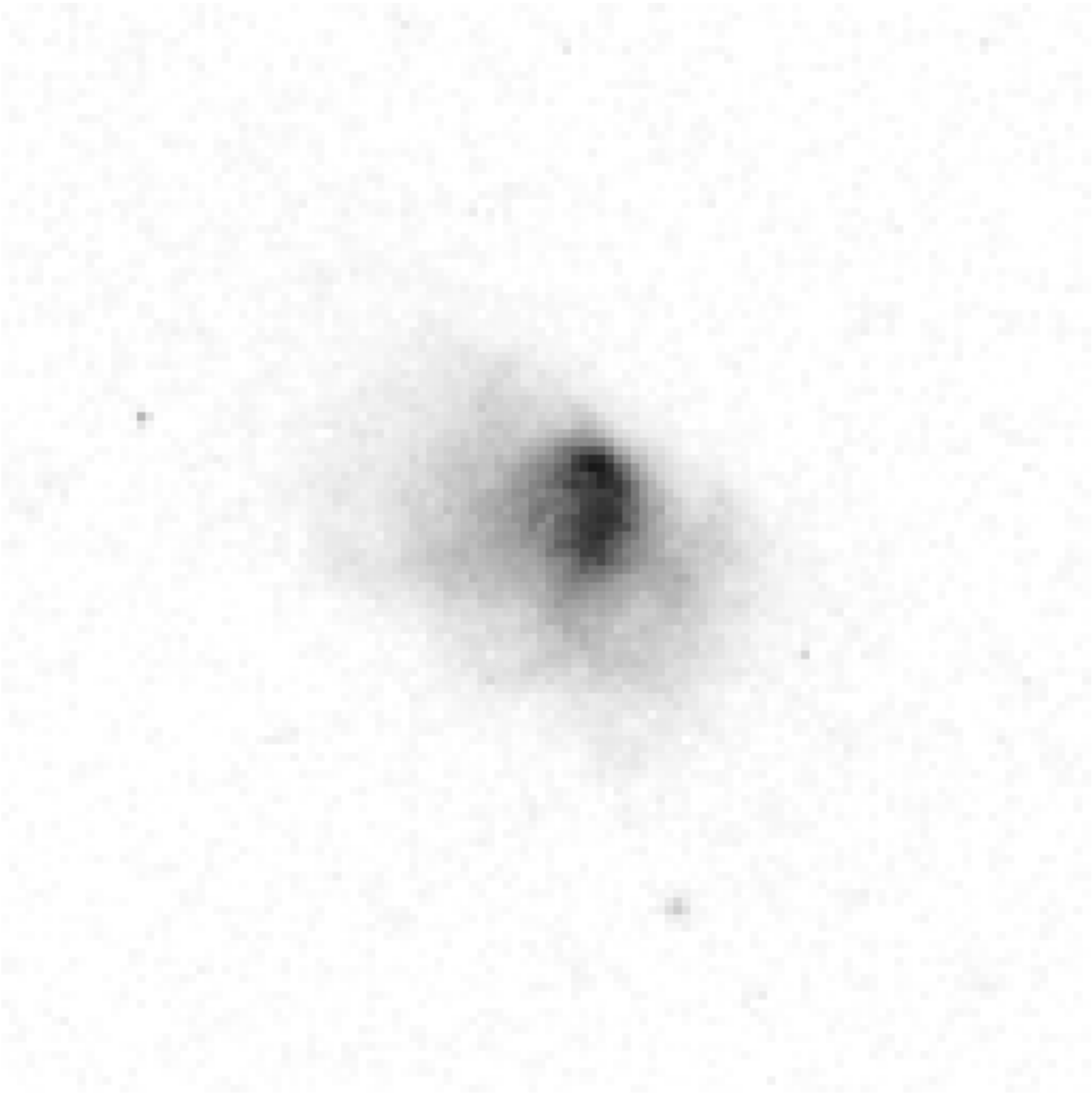} 
 \includegraphics[width=10cm]{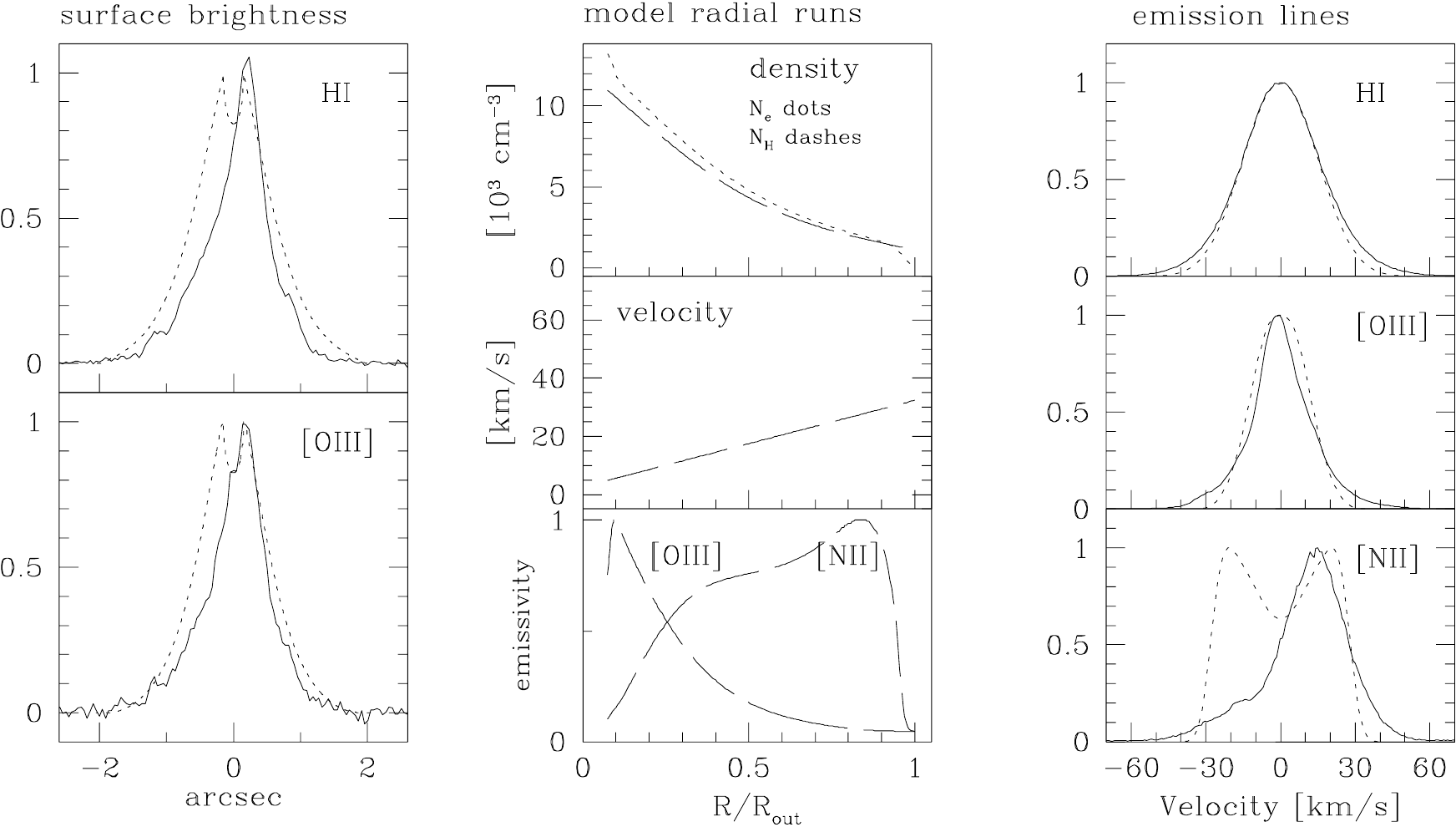}
 \caption{The nebula KFL\,11 (PN\,G\,004.1$-$03.8). The data are presented as
   in Fig.\,\ref{mod_fit_1}. The [\ion{N}{ii}] profile is one sided, indicating an
   asymmetry in the nebula, which is also evident from the image. However, the
   velocity field is well fitted.  }
\end{figure*}

\begin{figure*}
\centering
 \includegraphics[width=5.5cm]{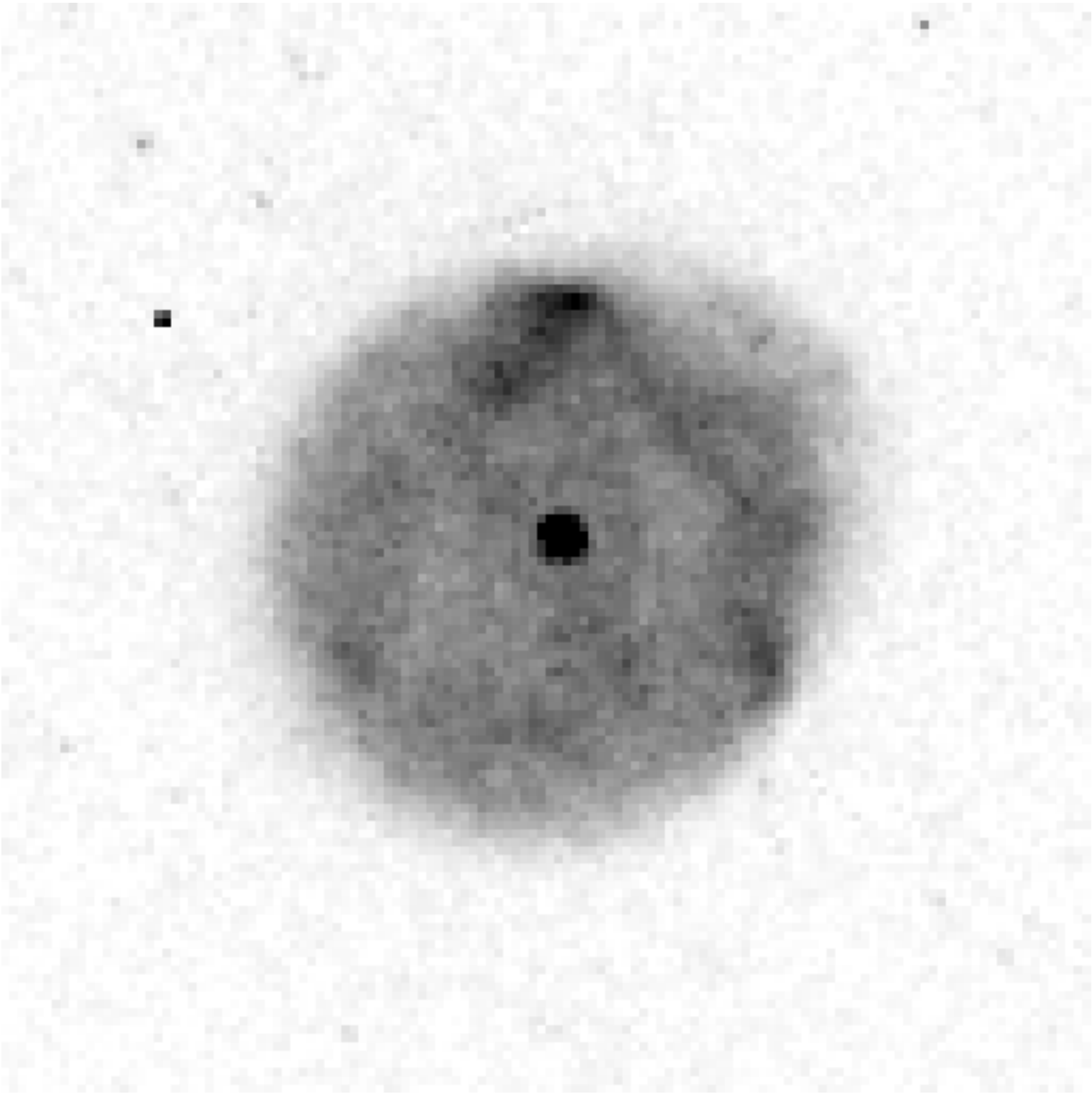} 
 \includegraphics[width=10cm]{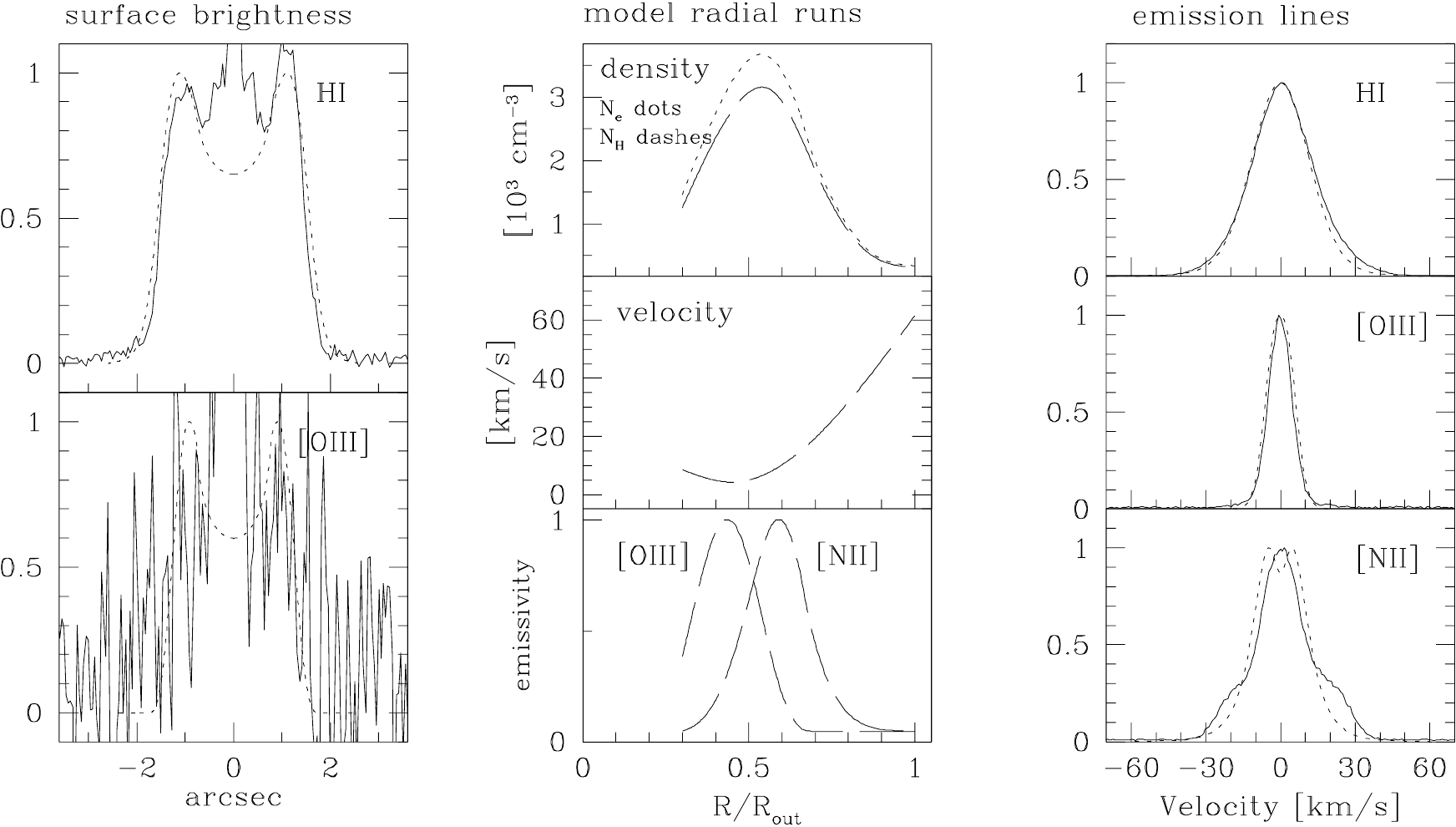}
 \caption{The nebula H\,2-25 (PN\,G\,004.8+02.0). The data are presented as in
   Fig.\,\ref{mod_fit_1}.    }
\end{figure*}

\begin{figure*}
\centering
 \includegraphics[width=5.5cm]{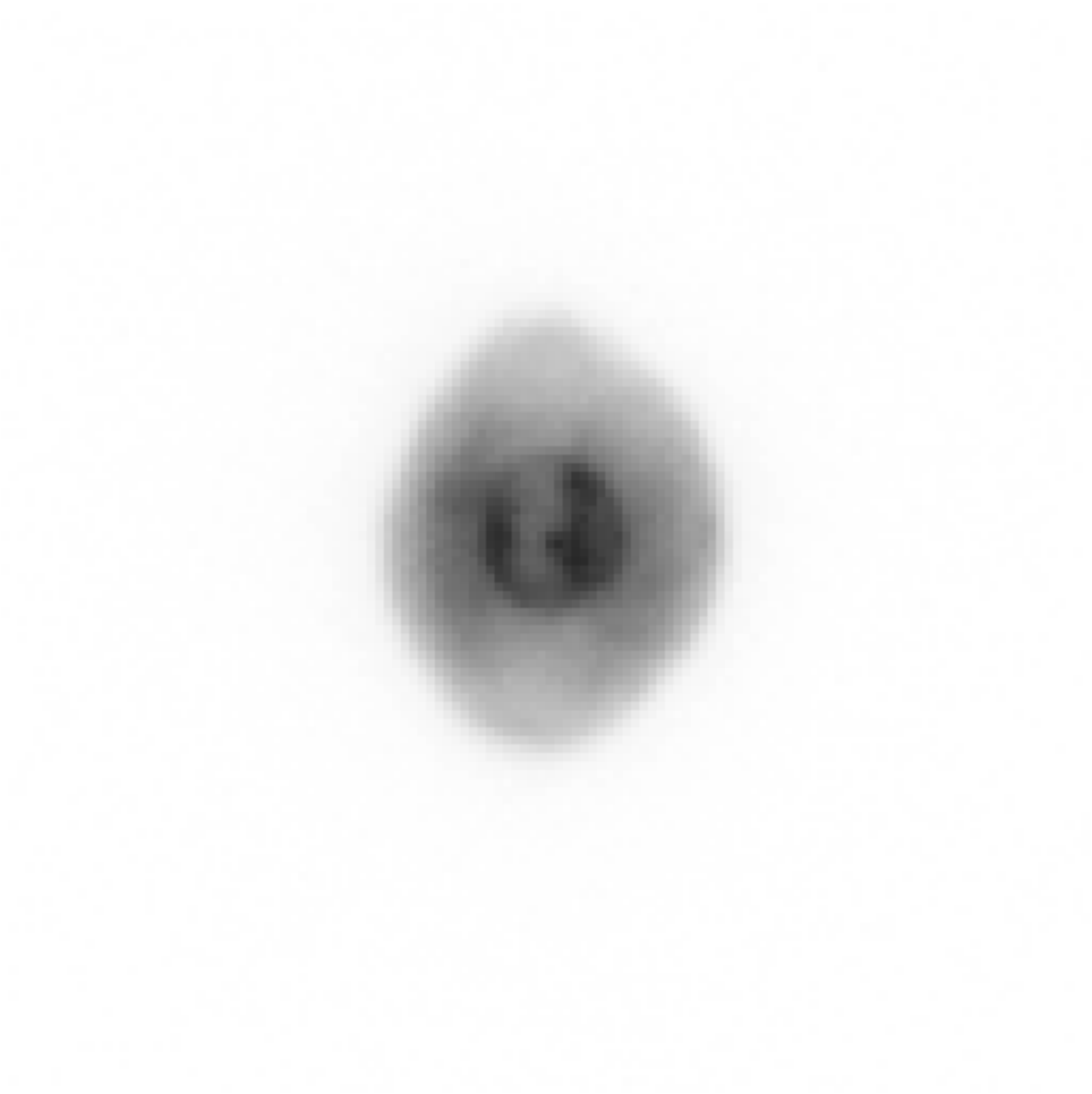} 
 \includegraphics[width=10cm]{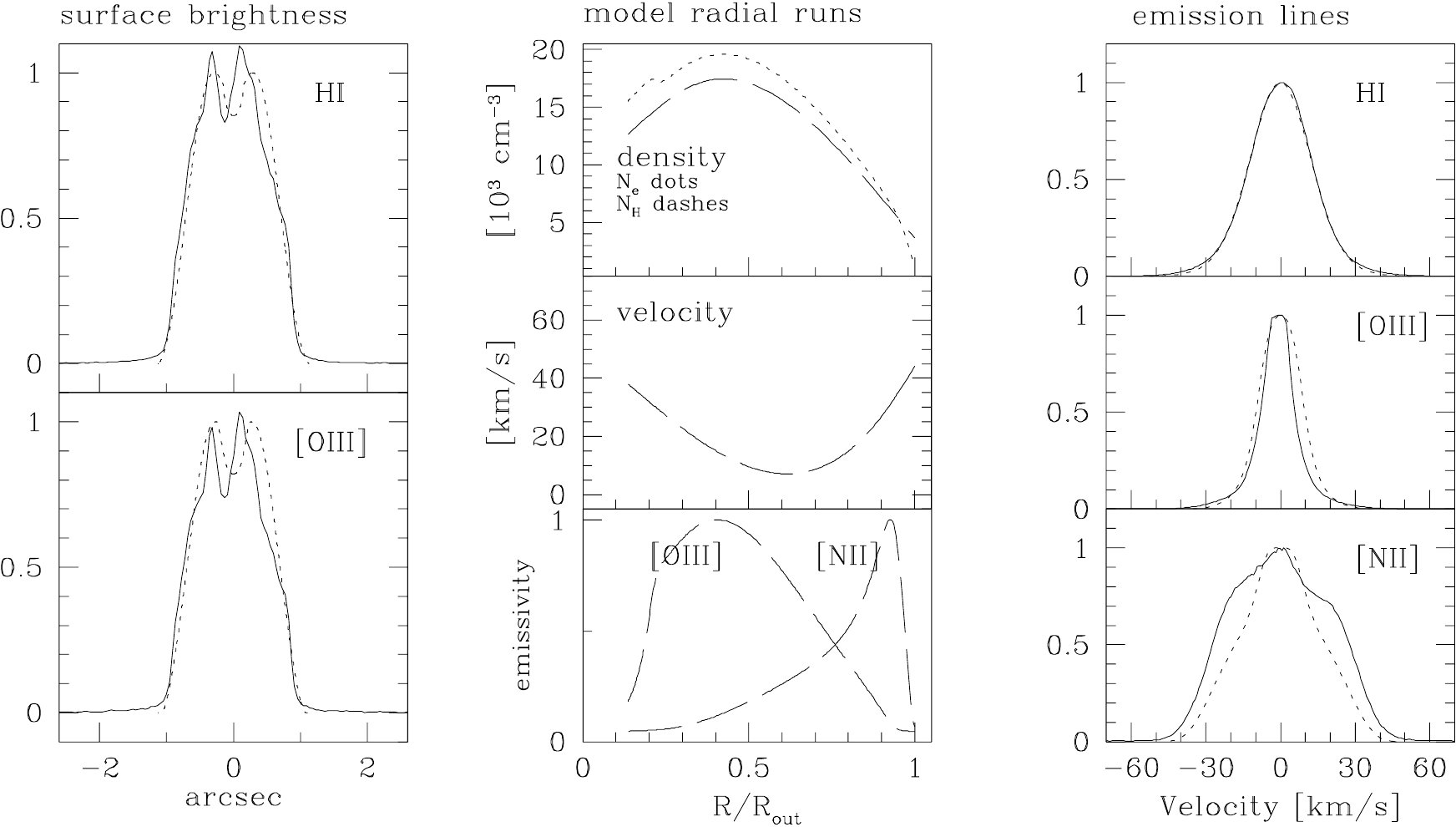}
 \caption{The nebula M\,1-20 (PN\,G\,006.1+08.3). The data are presented as in
   Fig.\,\ref{mod_fit_1}.    }
\end{figure*}

\begin{figure*}
\centering
 \includegraphics[width=5.5cm]{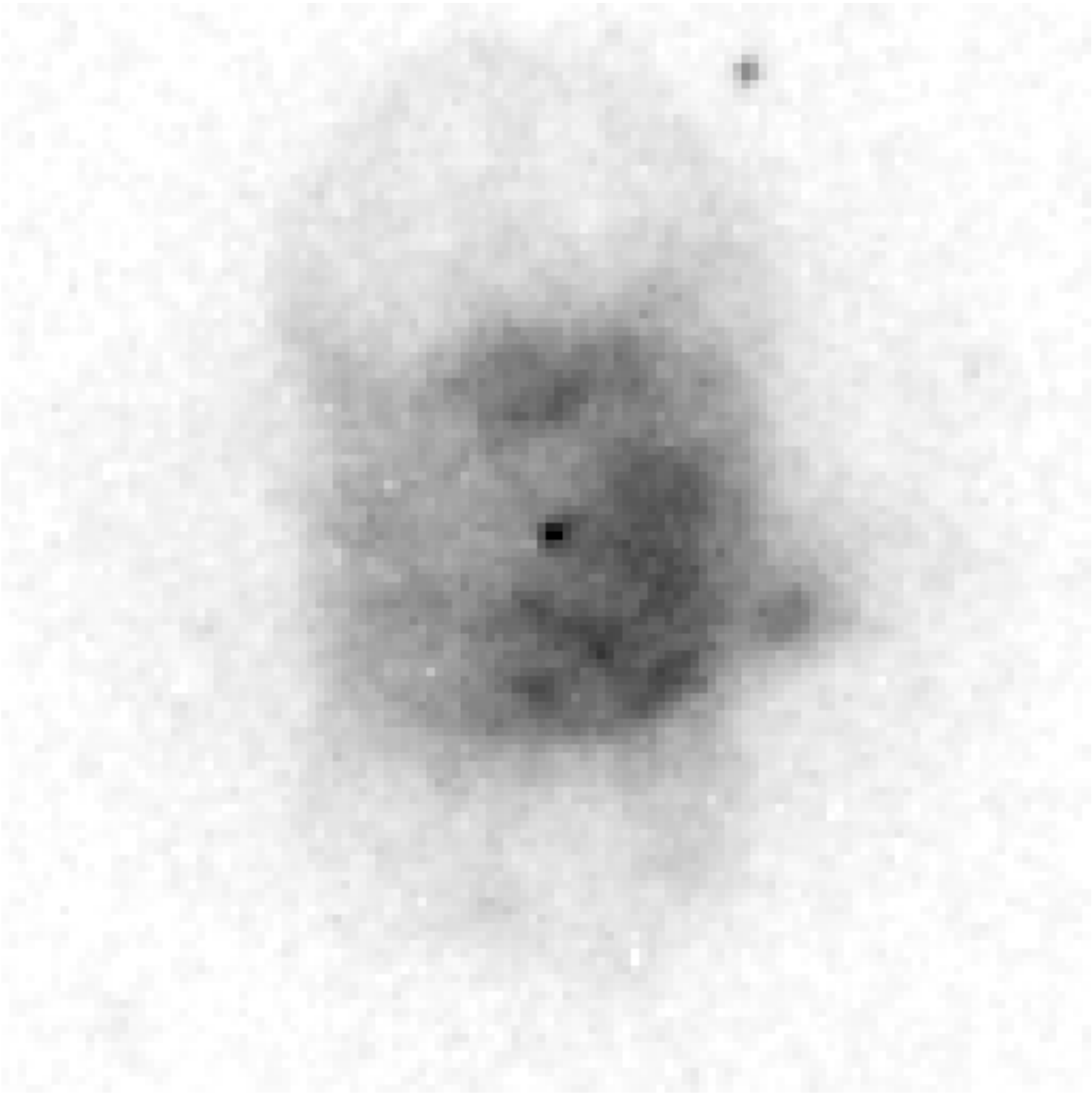} 
 \includegraphics[width=10cm]{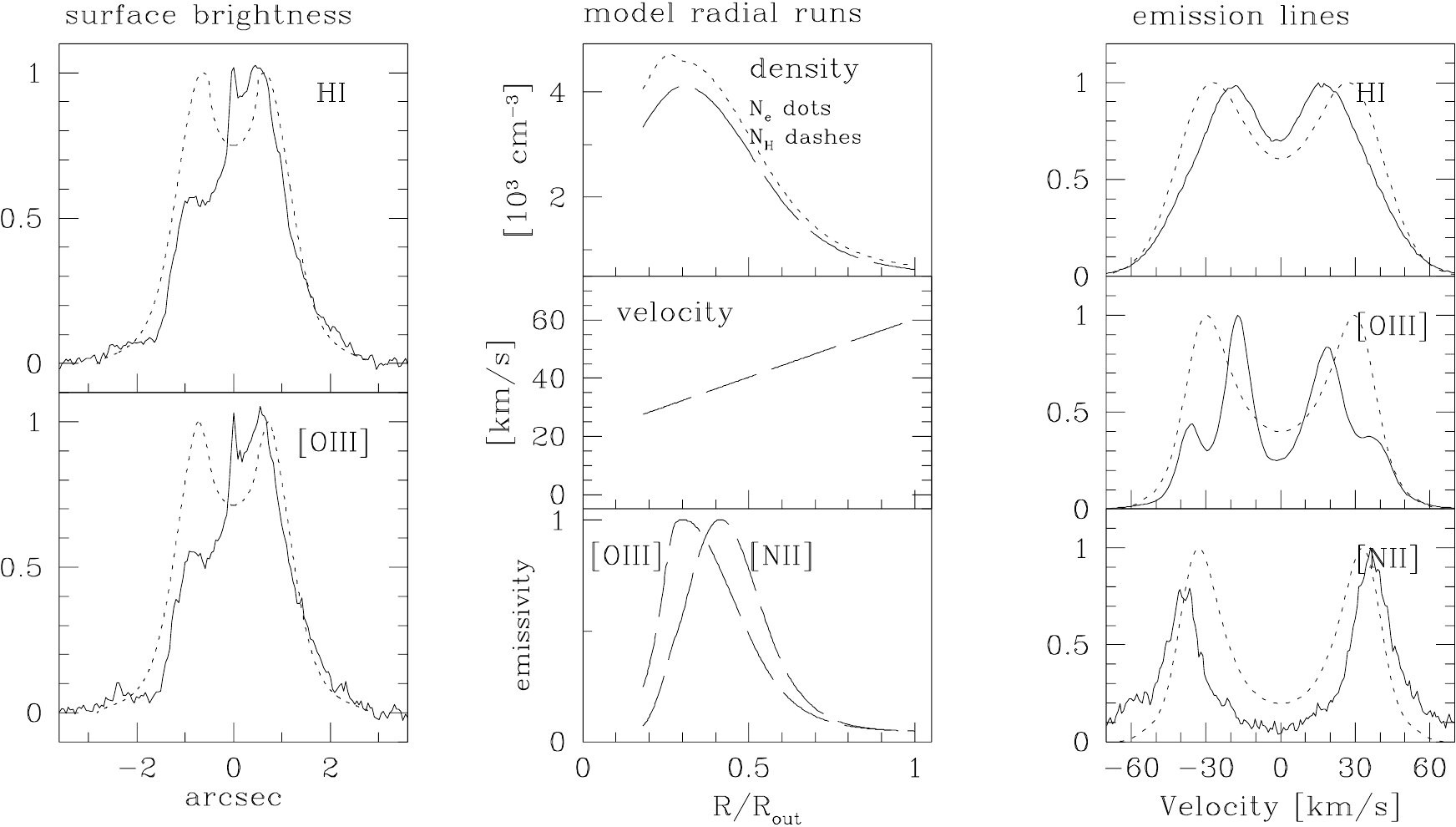}
 \caption{The nebula H\,2-18 (PN\,G\,006.3+04.4). The data are presented as in
   Fig.\,\ref{mod_fit_1}.    }
\end{figure*}

\begin{figure*}
\centering
 \includegraphics[width=5.5cm]{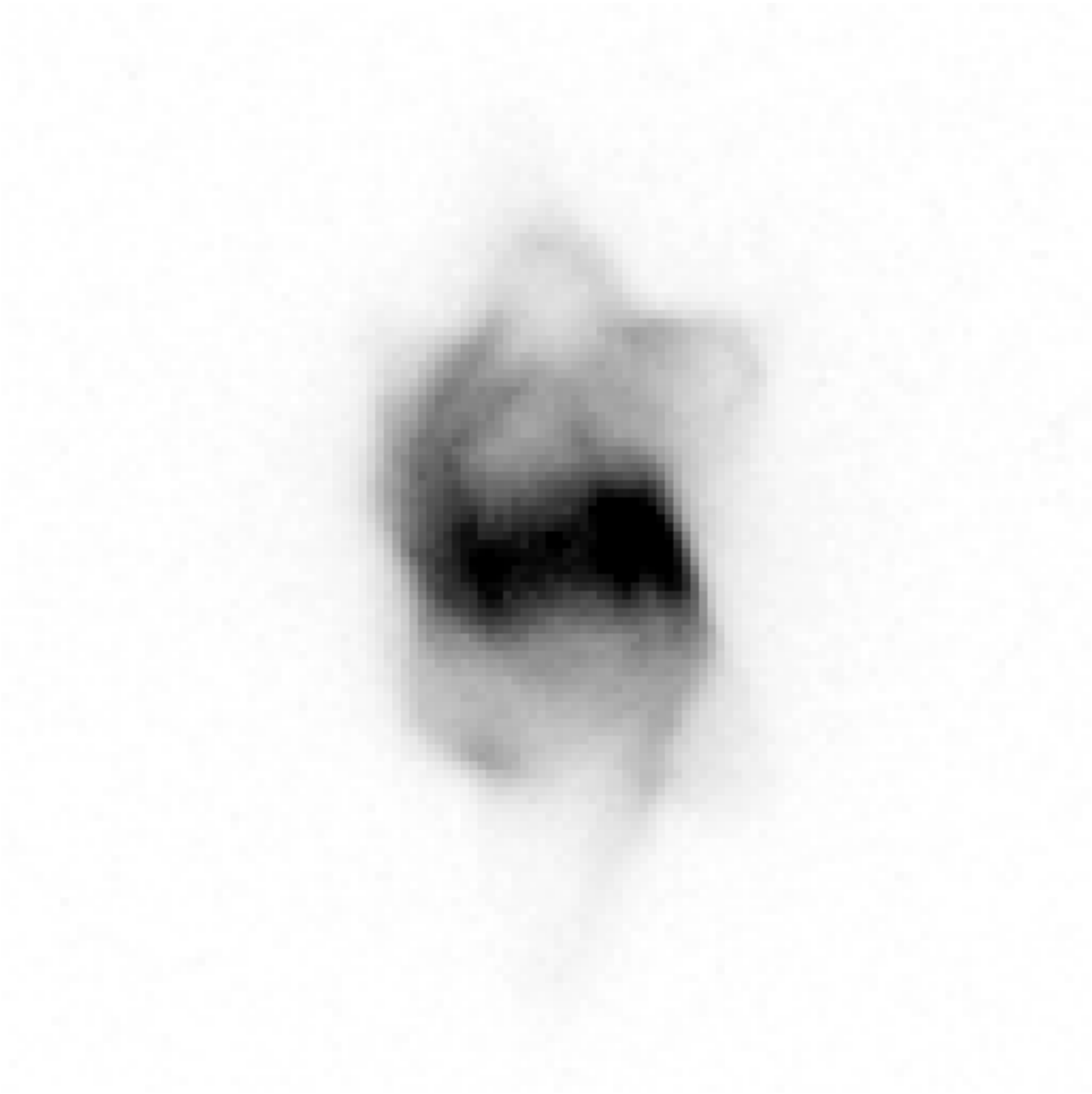} 
 \includegraphics[width=10cm]{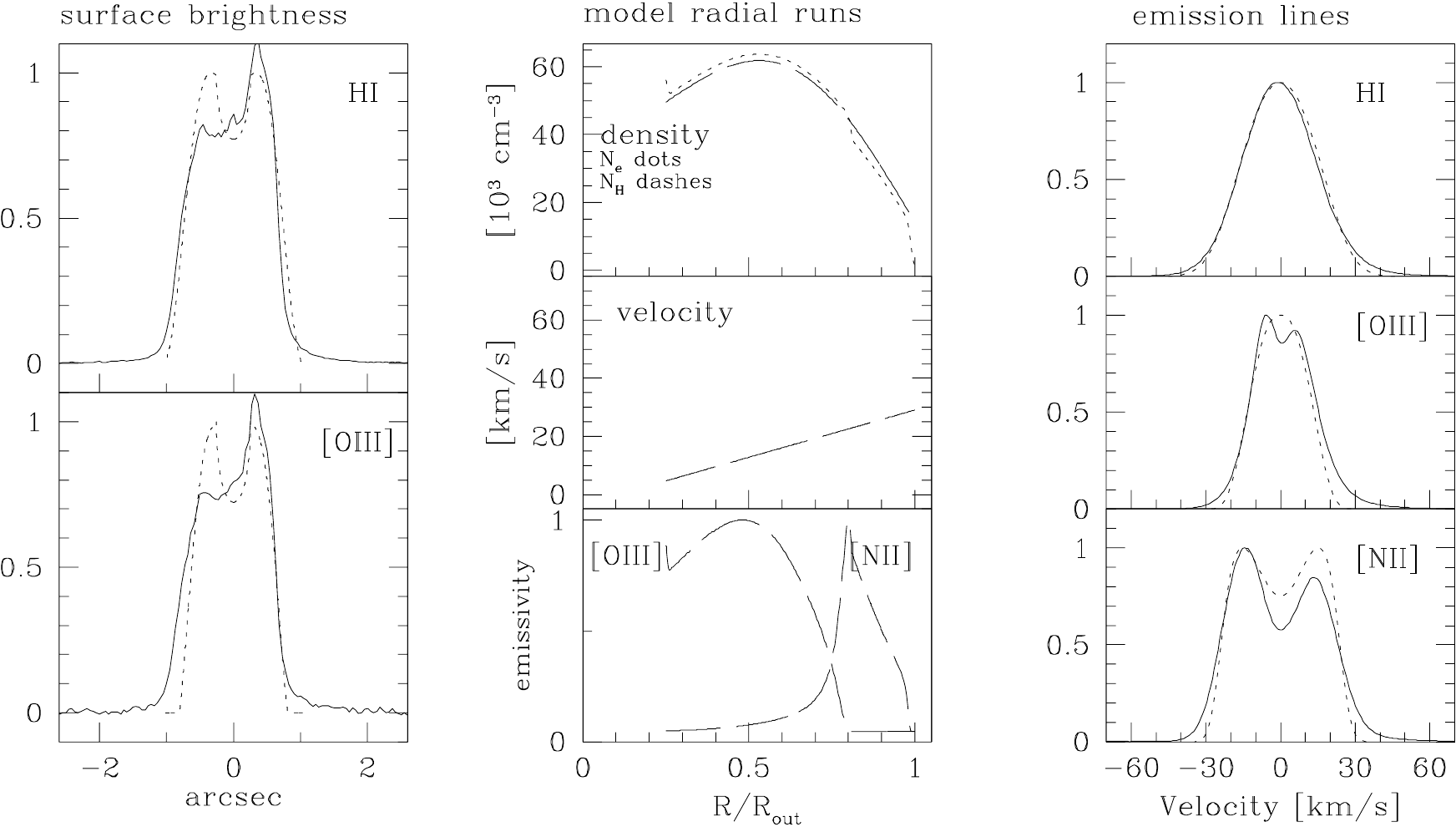}
 \caption{The nebula M\,1-31 (PN\,G\,006.4+02.0). The data are presented as in
   Fig.\,\ref{mod_fit_1}.    }
\end{figure*}

\begin{figure*}
\centering
 \includegraphics[width=5.5cm]{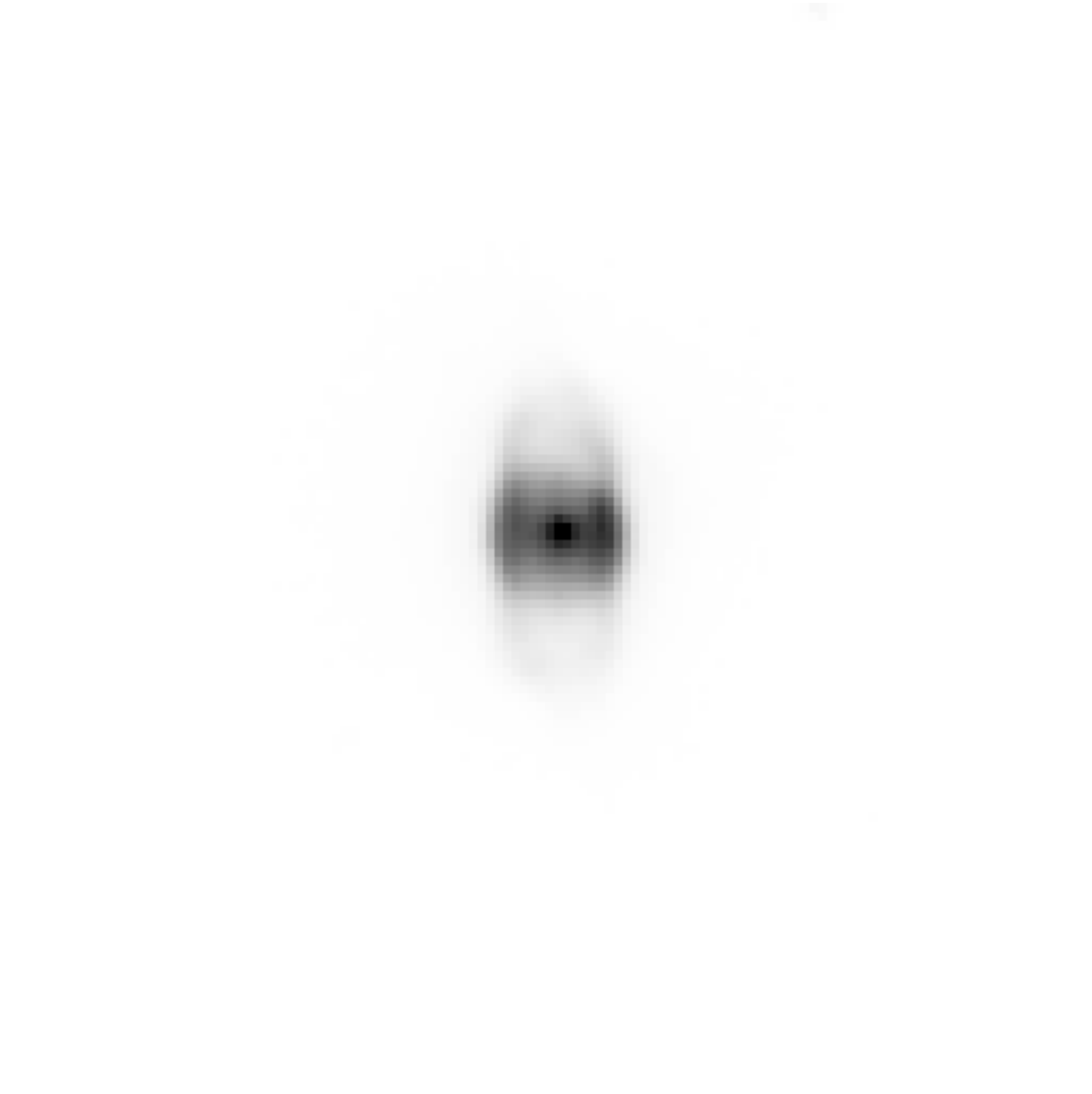} 
 \includegraphics[width=10cm]{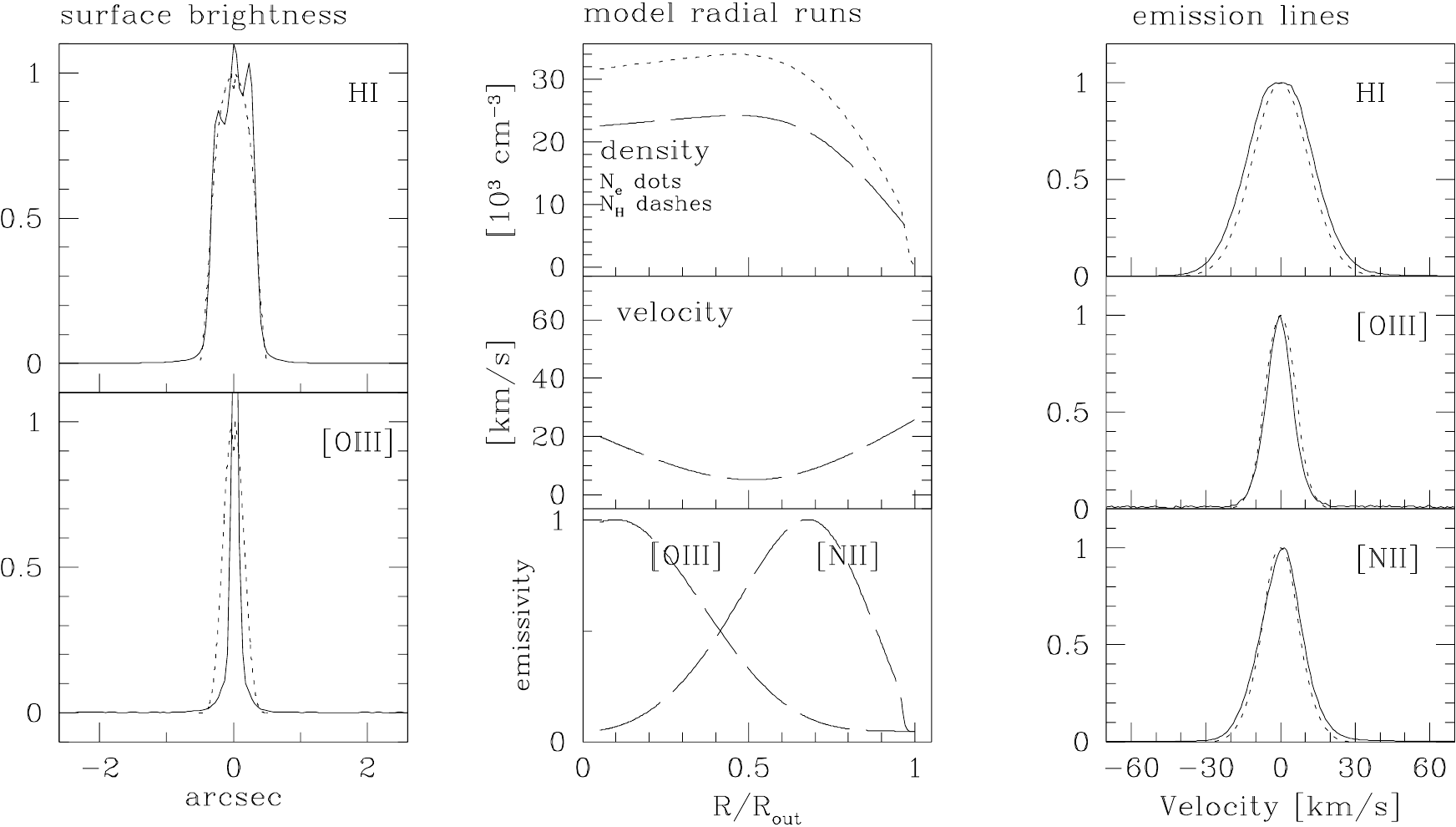}
 \caption{The nebula He\,2-260 (PN\,G\,008.2+06.8). The data are presented as
   in Fig.\,\ref{mod_fit_1}.    }
\end{figure*}

\begin{figure*}
\centering
 \includegraphics[width=5.5cm]{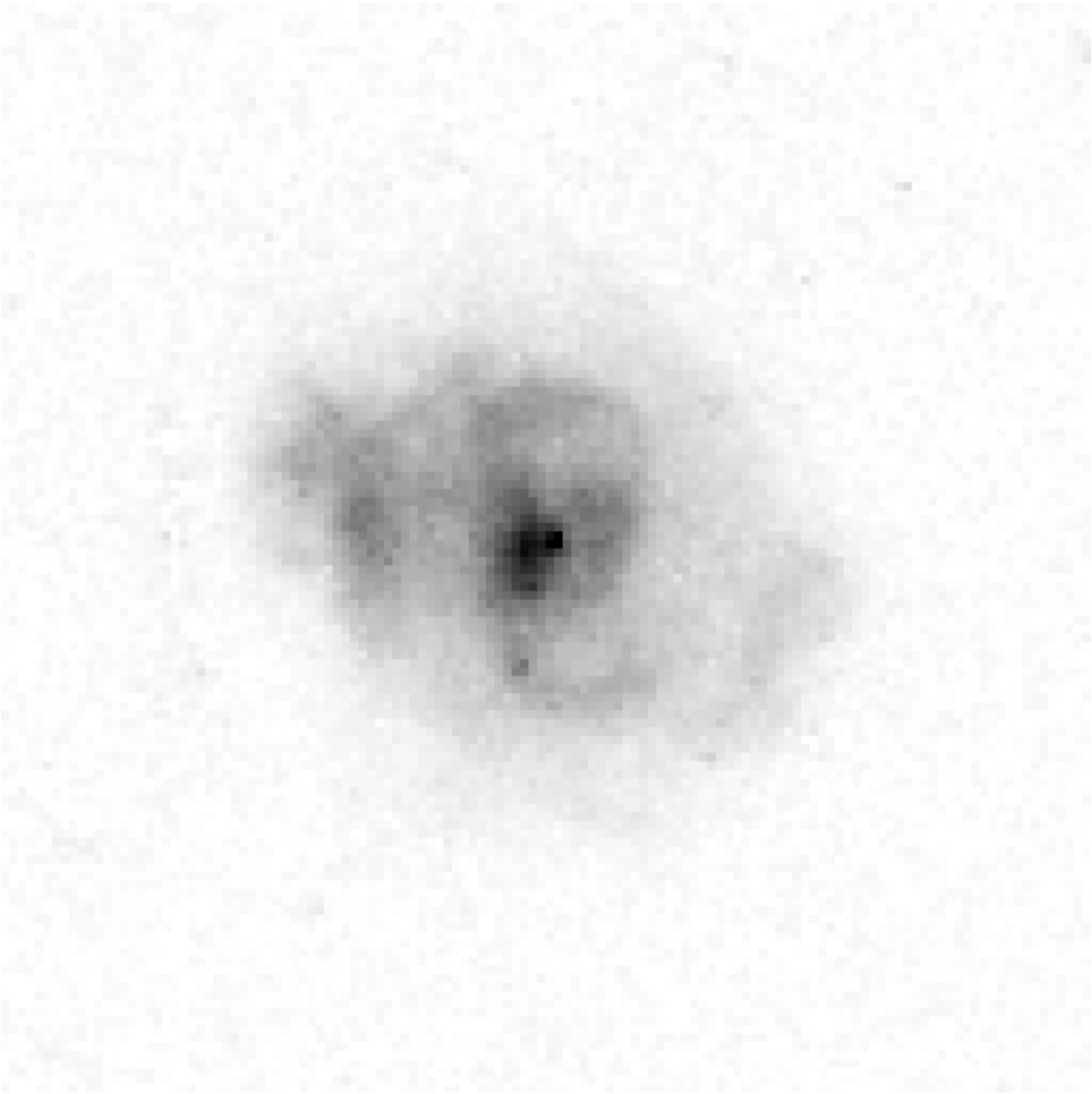} 
 \includegraphics[width=10cm]{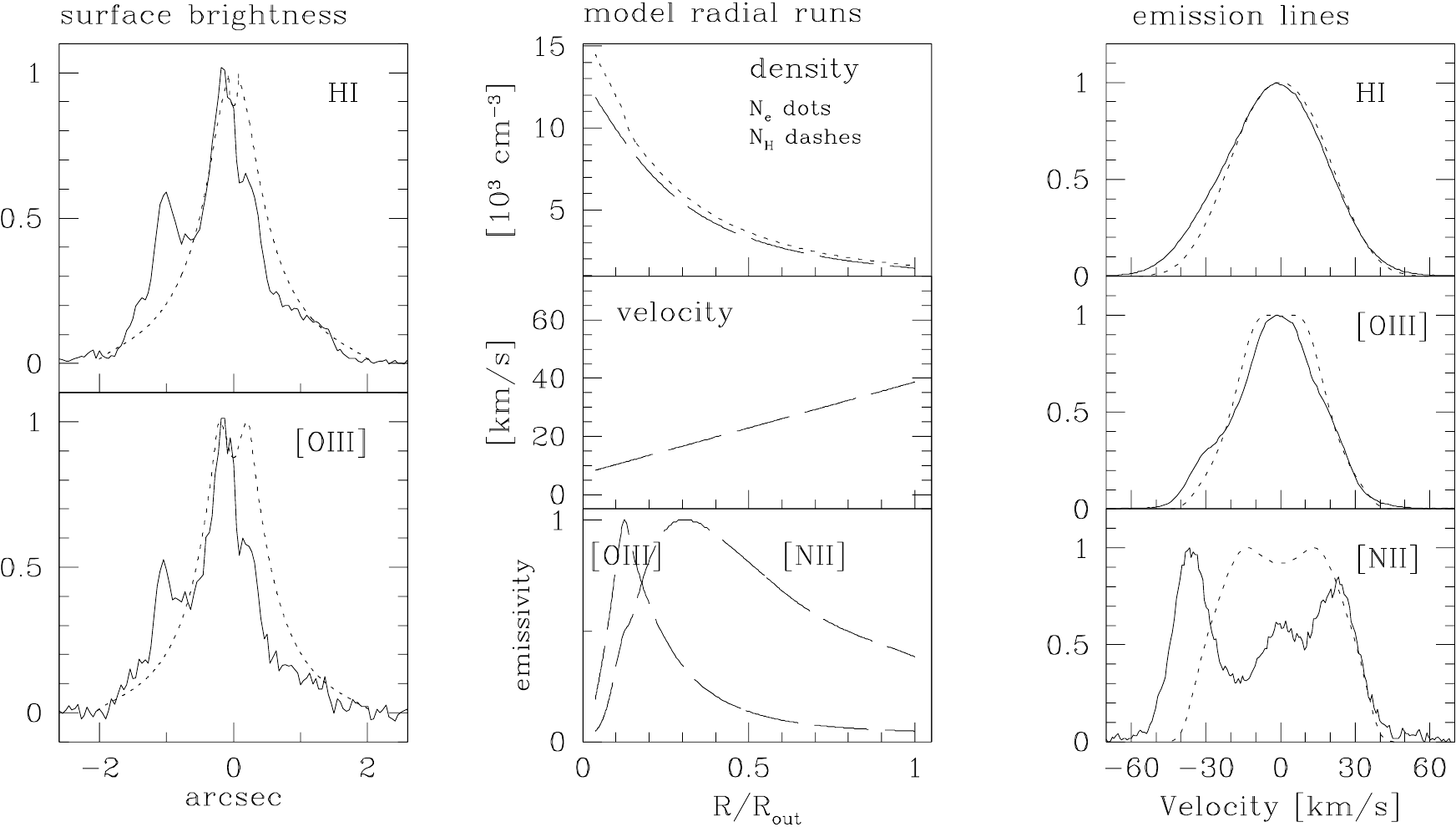}
 \caption{The nebula MaC\,1-11 (PN\,G\,008.6$-$02.6). The data are presented
   as in Fig.\,\ref{mod_fit_1}. The [\ion{N}{ii}] line shows a velocity asymmetry,
   which may also be just visible in [\ion{O}{iii}]. The model fits the red-shifted
   velocities.      }
\end{figure*}

\begin{figure*}
\centering
 \includegraphics[width=5.5cm]{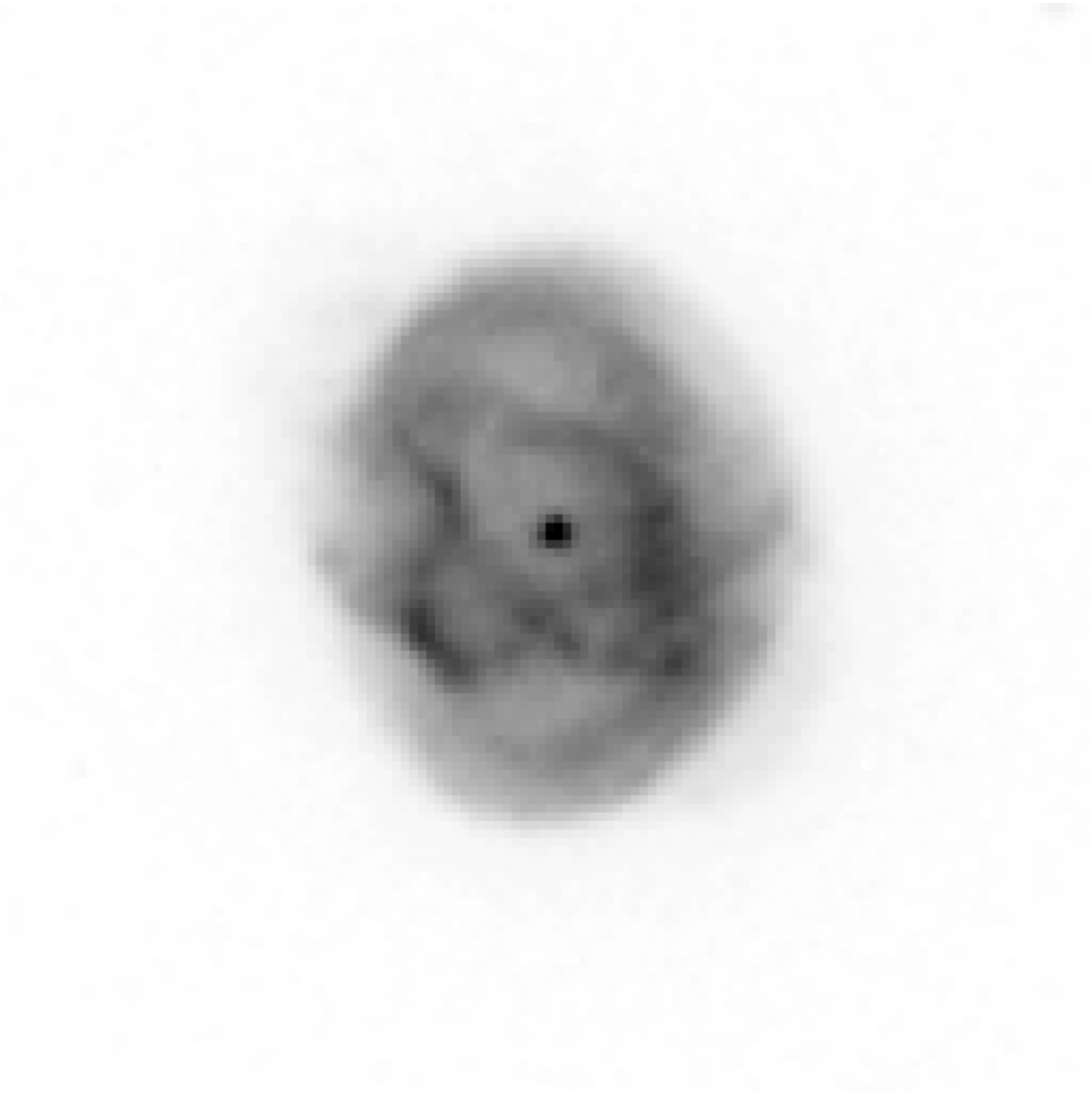} 
 \includegraphics[width=10cm]{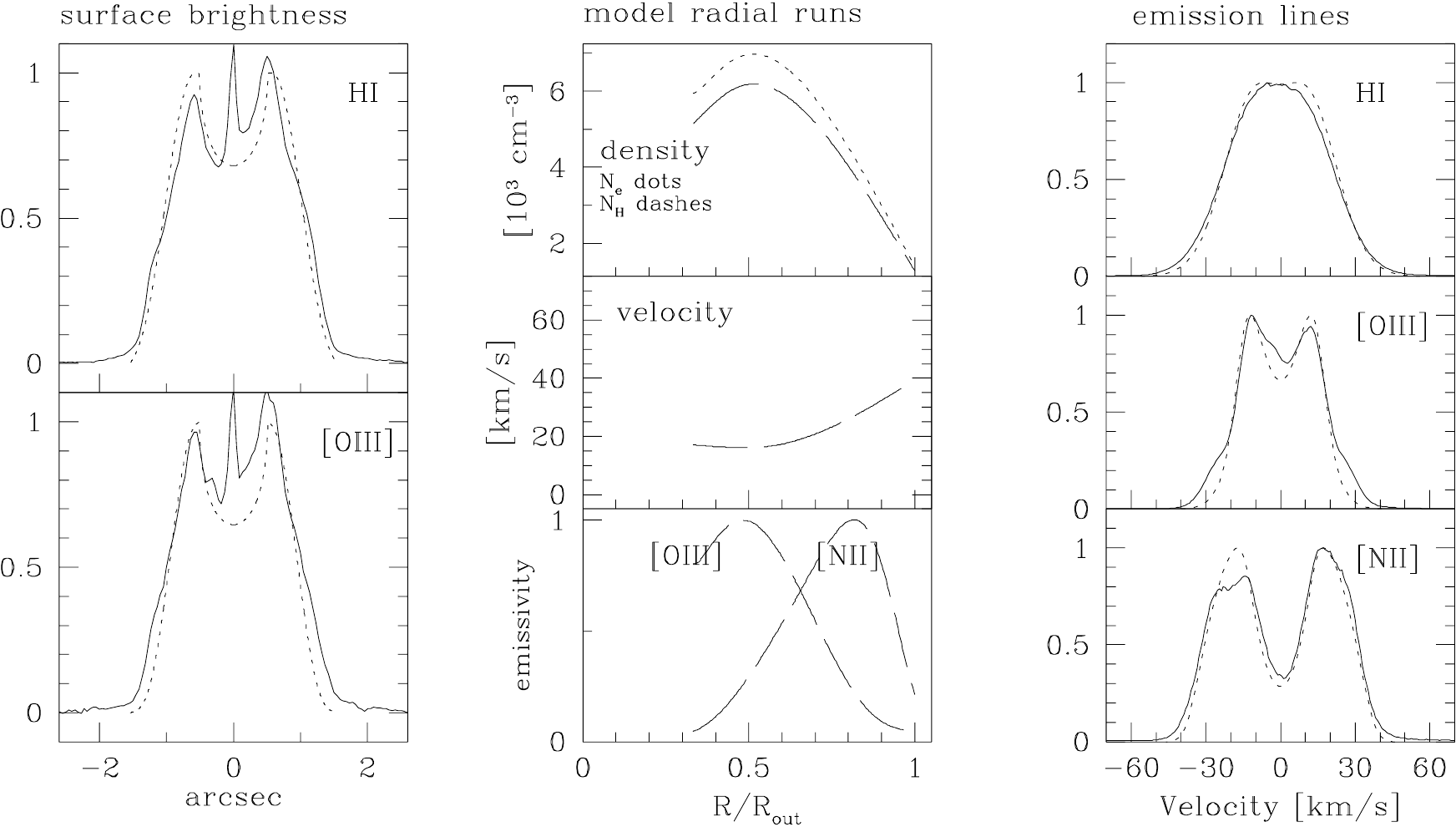}
 \caption{The nebula M\,1-19 (PN\,G\,351.1+04.8). The data are presented as in
   Fig.\,\ref{mod_fit_1}.    }
\end{figure*}

\begin{figure*}
\centering
 \includegraphics[width=5.5cm]{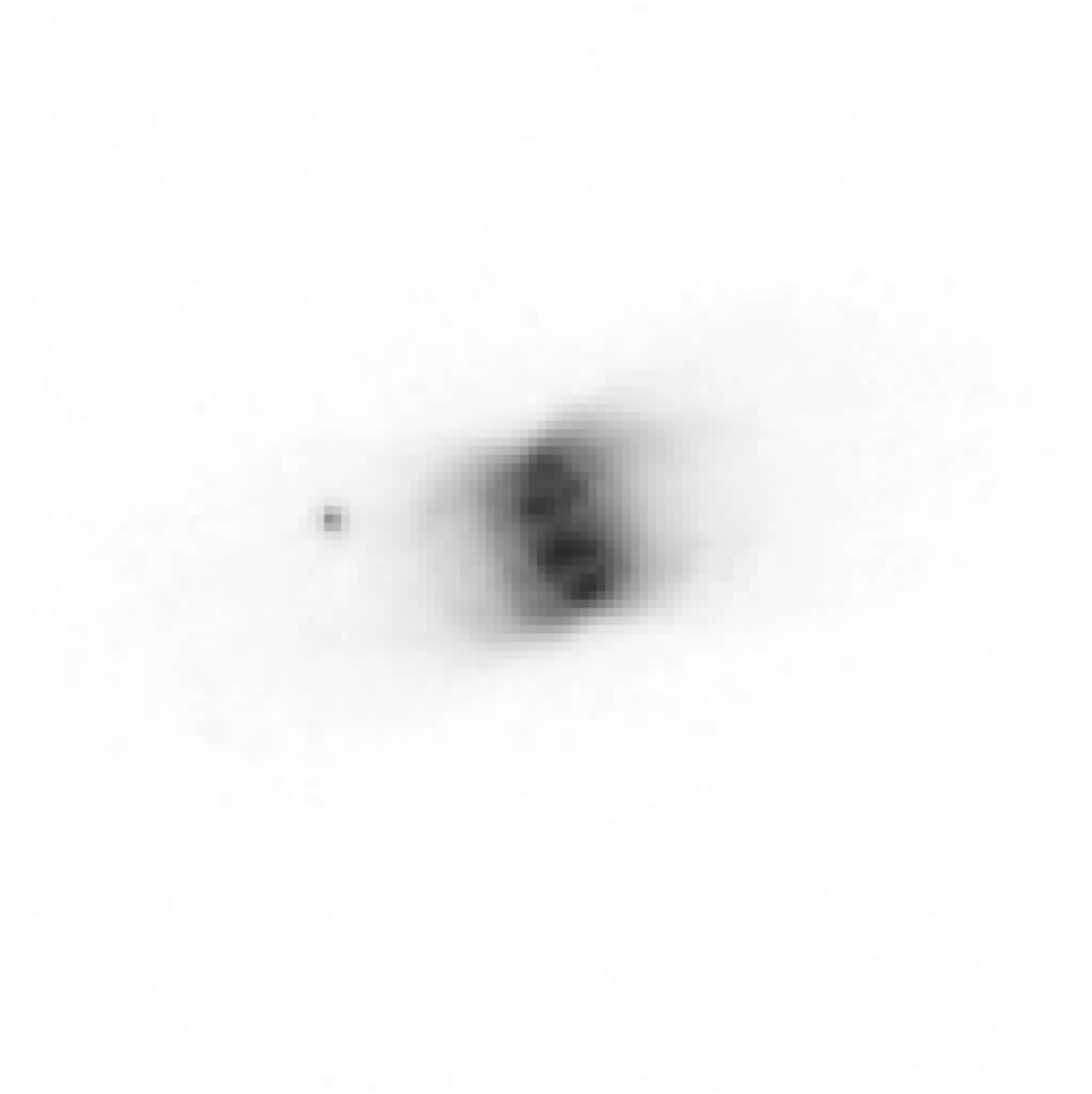} 
 \includegraphics[width=10cm]{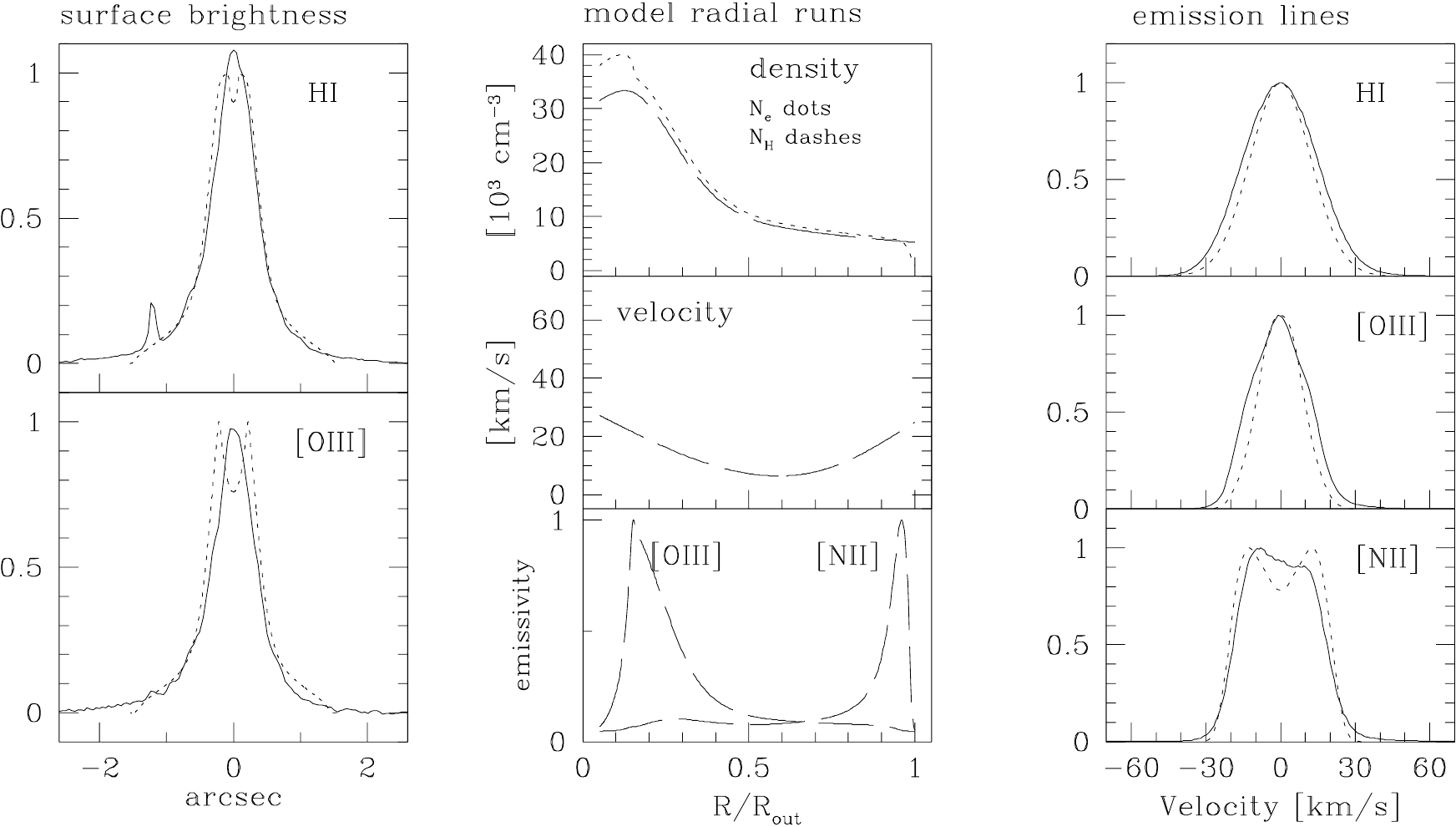}
 \caption{The nebula Wray\,16-286 (PN\,G\,351.9-01.9). The data are presented
   as in Fig.\,\ref{mod_fit_1}.  The slit was oriented along the major axis.
 }
\end{figure*}

\begin{figure*}
\centering
 \includegraphics[width=5.5cm]{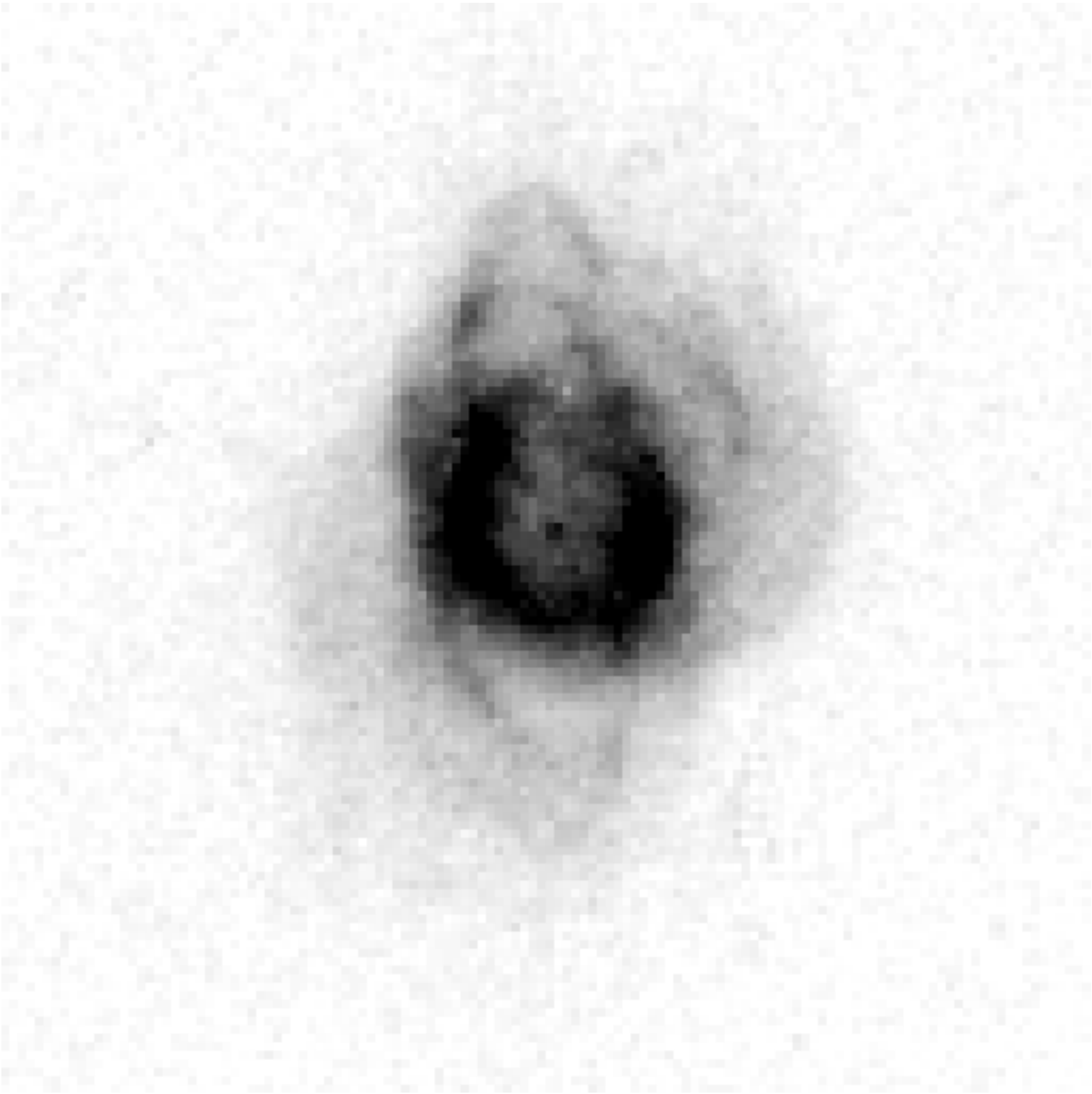} 
 \includegraphics[width=10cm]{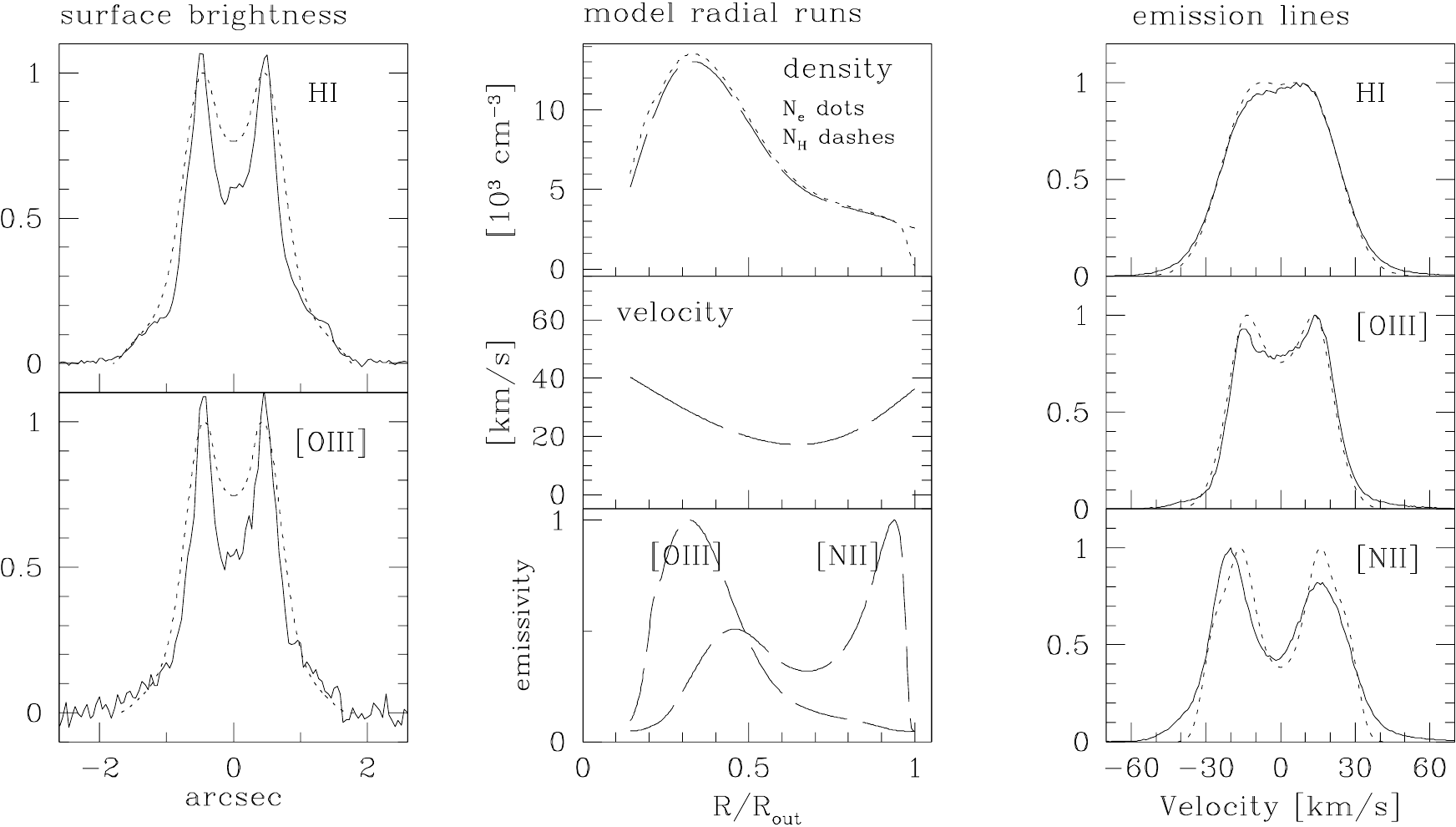}
 \caption{The nebula H\,1-8 (PN\,G\,352.6+03.0). The data are presented as in
   Fig.\,\ref{mod_fit_1}.    }
\end{figure*}

\begin{figure*}
\centering
 \includegraphics[width=5.5cm]{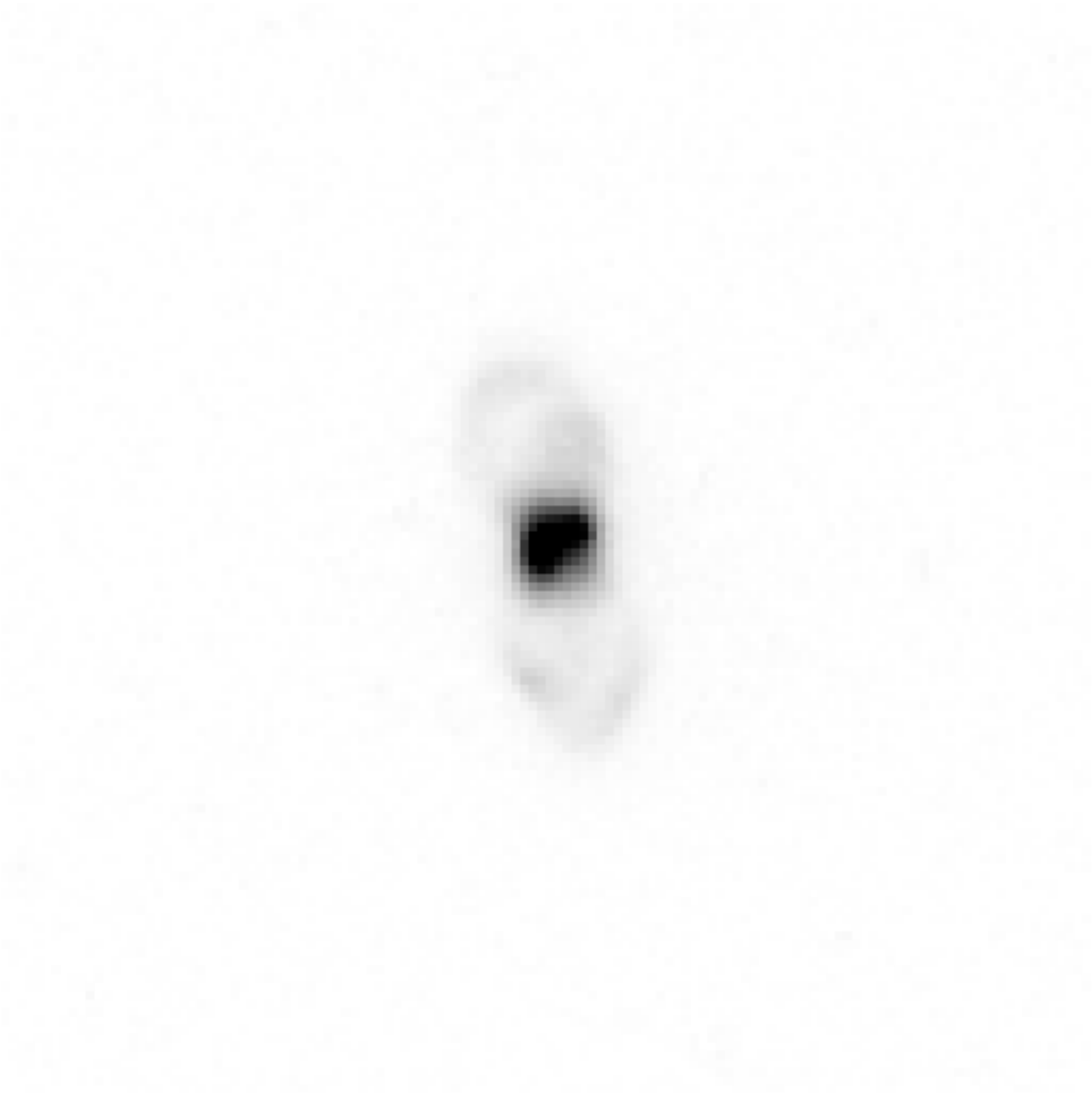} 
 \includegraphics[width=10cm]{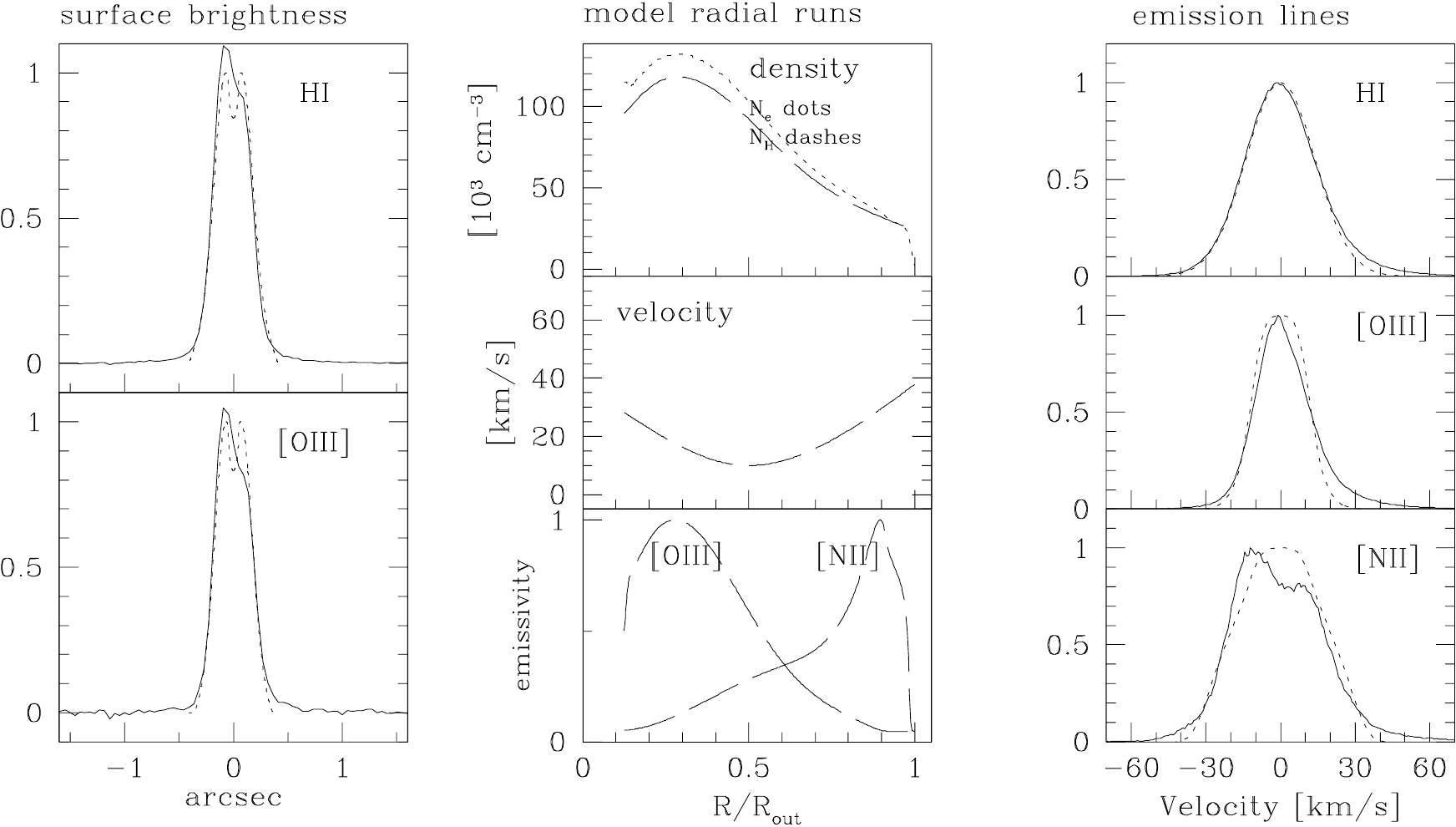}
 \caption{The nebula Th\,3-4 (PN\,G\,354.5+03.3). The data are presented as in
   Fig.\,\ref{mod_fit_1}.    }
\end{figure*}

\begin{figure*}
\centering
 \includegraphics[width=5.5cm]{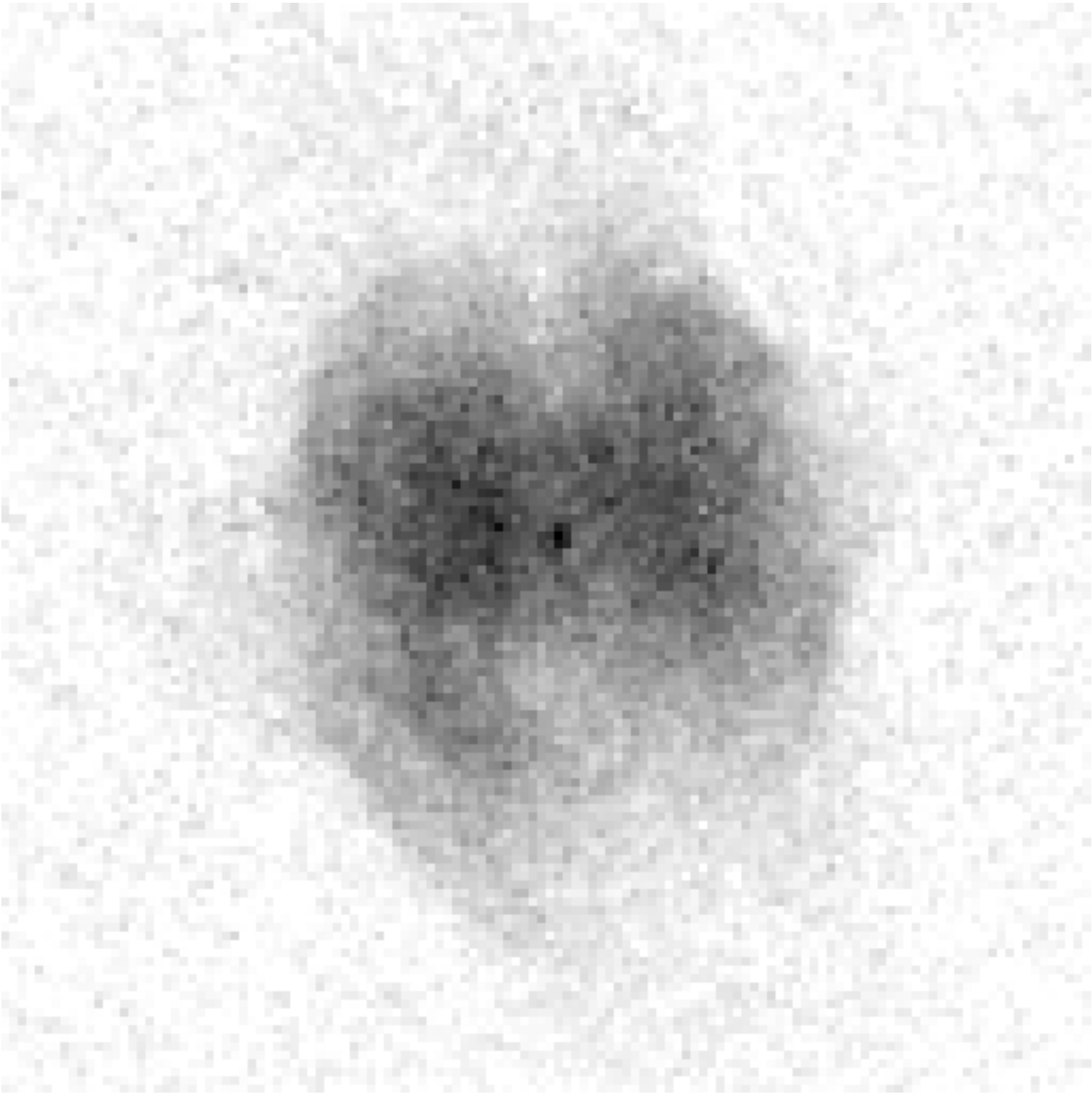} 
 \includegraphics[width=10cm]{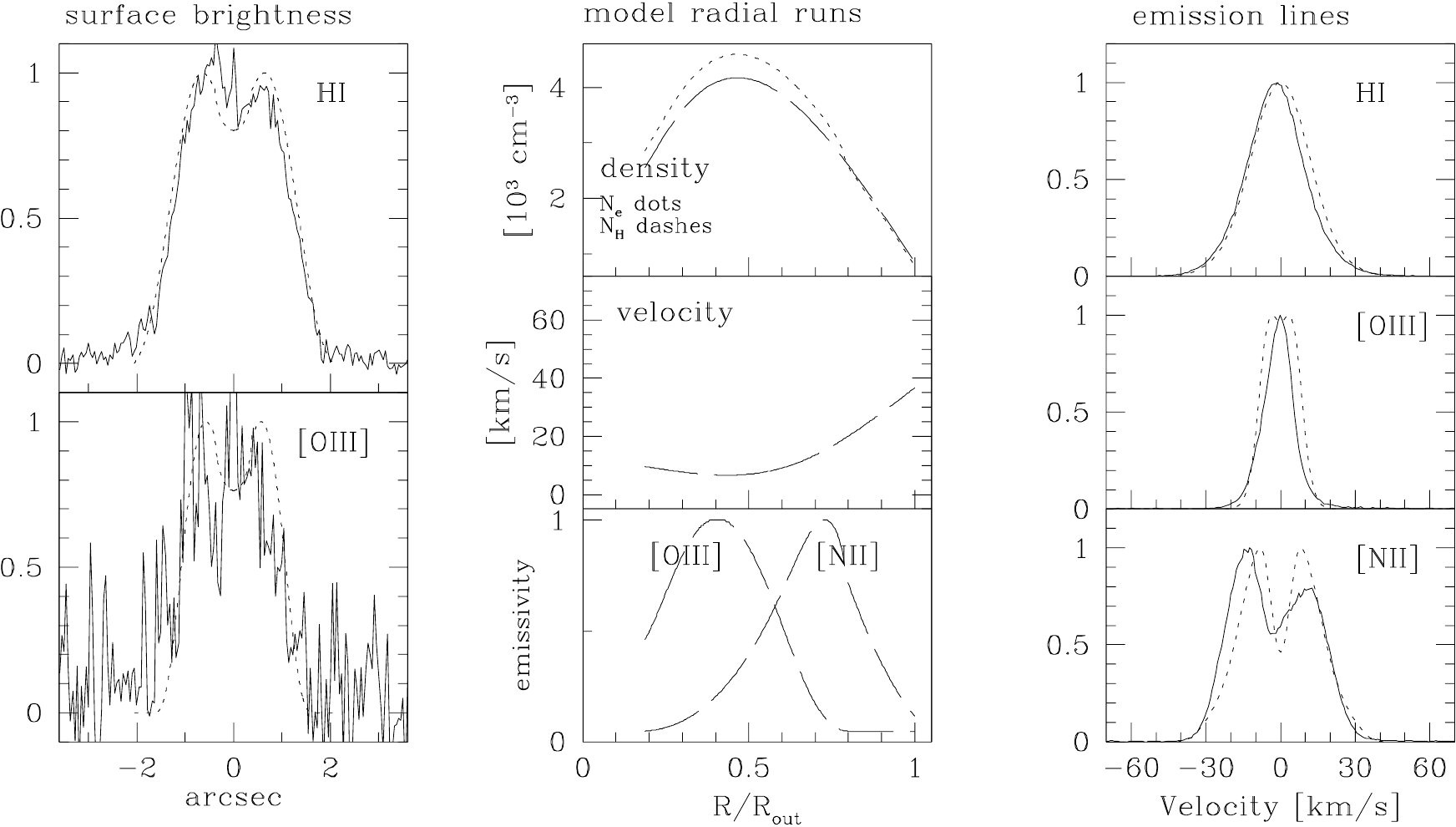}
 \caption{The nebula Th\,3-6 (PN\,G\,354.9+03.5). The data are presented as in
   Fig.\,\ref{mod_fit_1}.    }
\end{figure*}

\begin{figure*}
\centering
 \includegraphics[width=5.5cm]{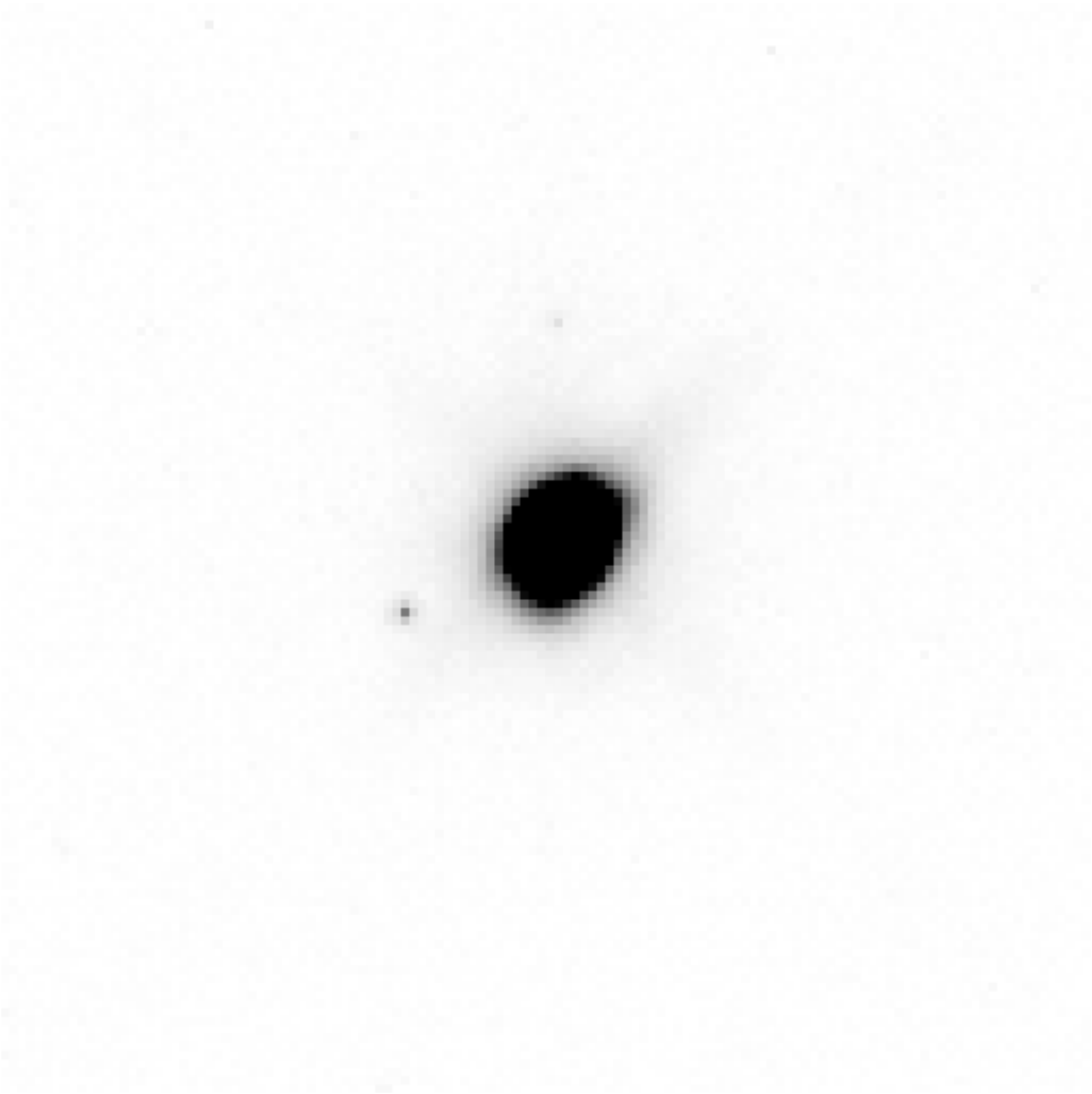} 
 \includegraphics[width=10cm]{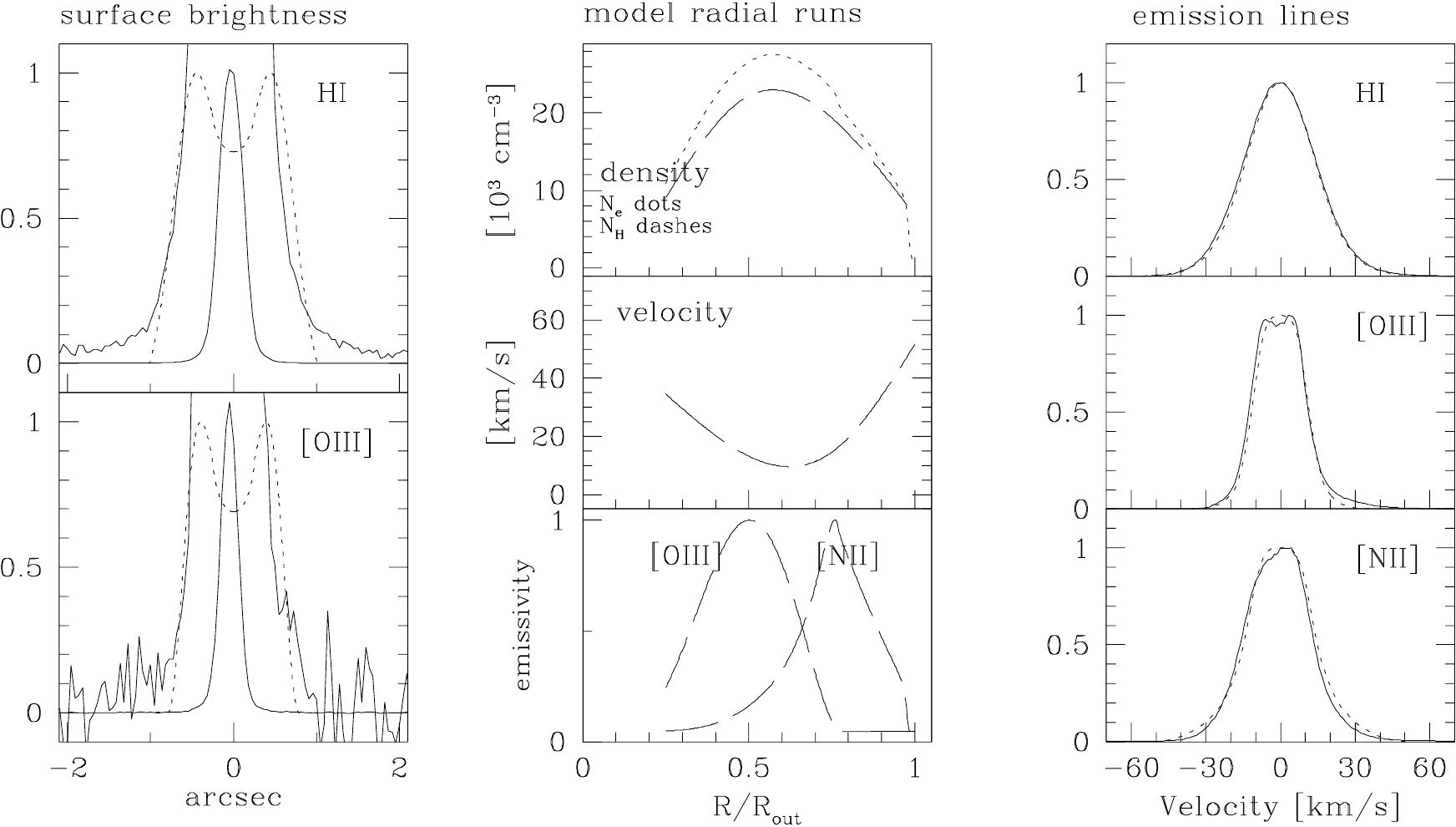}
 \caption{The nebula H\,1-9 (PN\,G\,355.9+03.6). The data are presented as in
   Fig.\,\ref{mod_fit_1}. The radial profile is shown scaled by a factor of
   100. The nebula is very compact and the model does not fit the radial
   distribution well, but the outer radius of the  model closely aligns with
   the edge of the image, as seen in the figure.  }
\end{figure*}

\begin{figure*}
\centering
 \includegraphics[width=5.5cm]{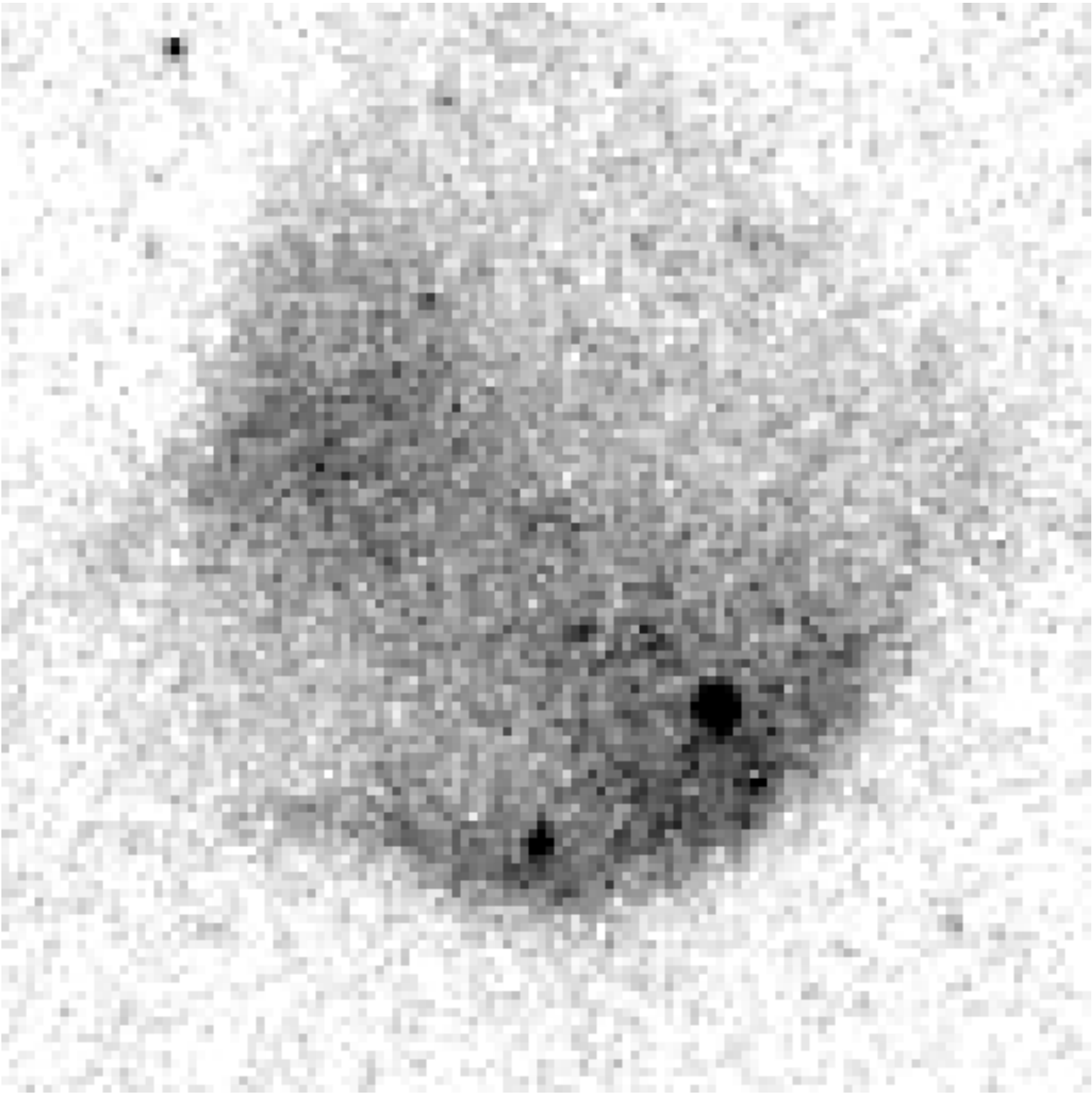} 
 \includegraphics[width=10cm]{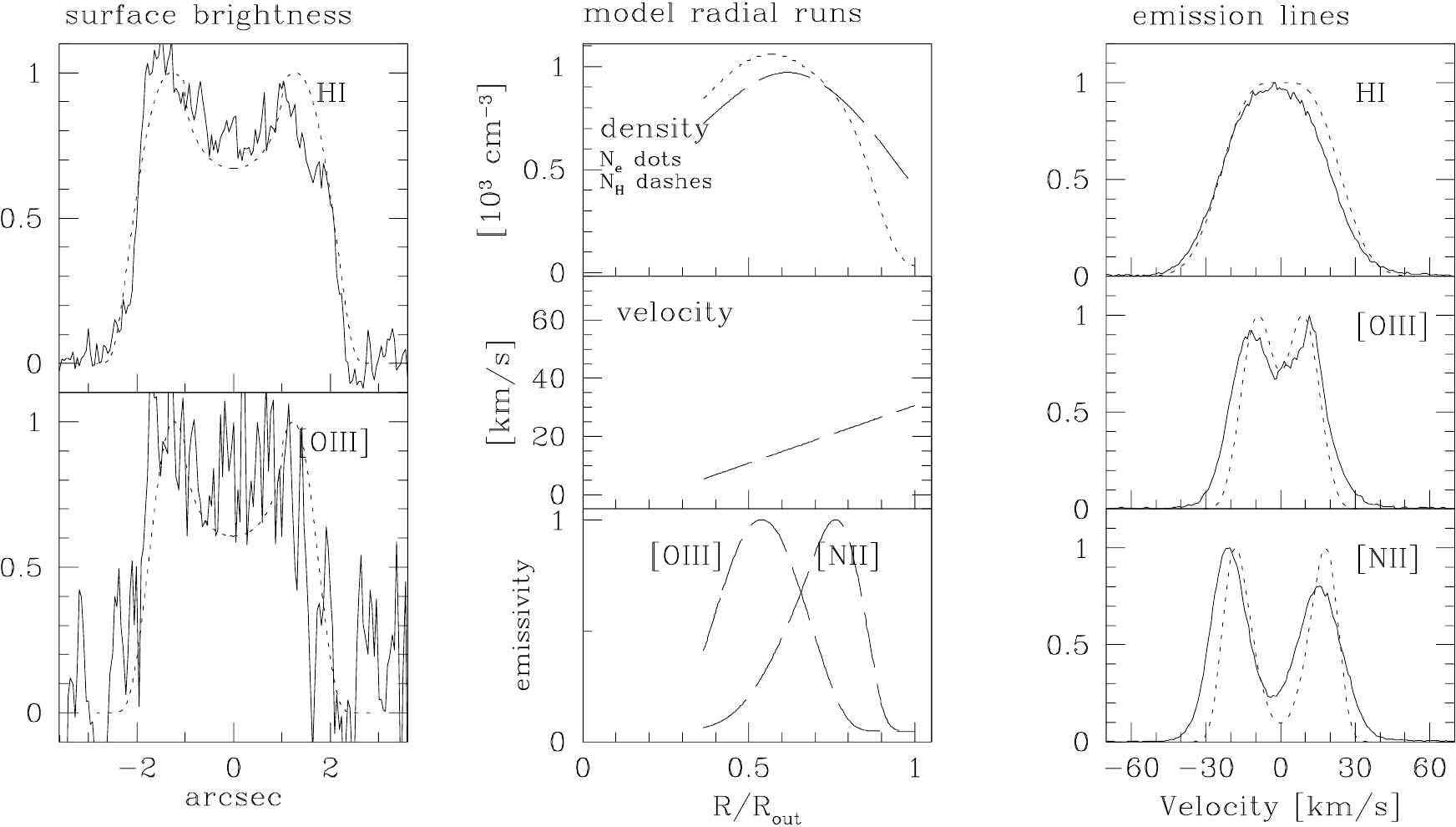}
 \caption{The nebula H\,2-26 (PN\,G\,356.1-03.3). The data are presented as in
   Fig.\,\ref{mod_fit_1}.    }
\end{figure*}

\begin{figure*}
\centering
 \includegraphics[width=5.5cm]{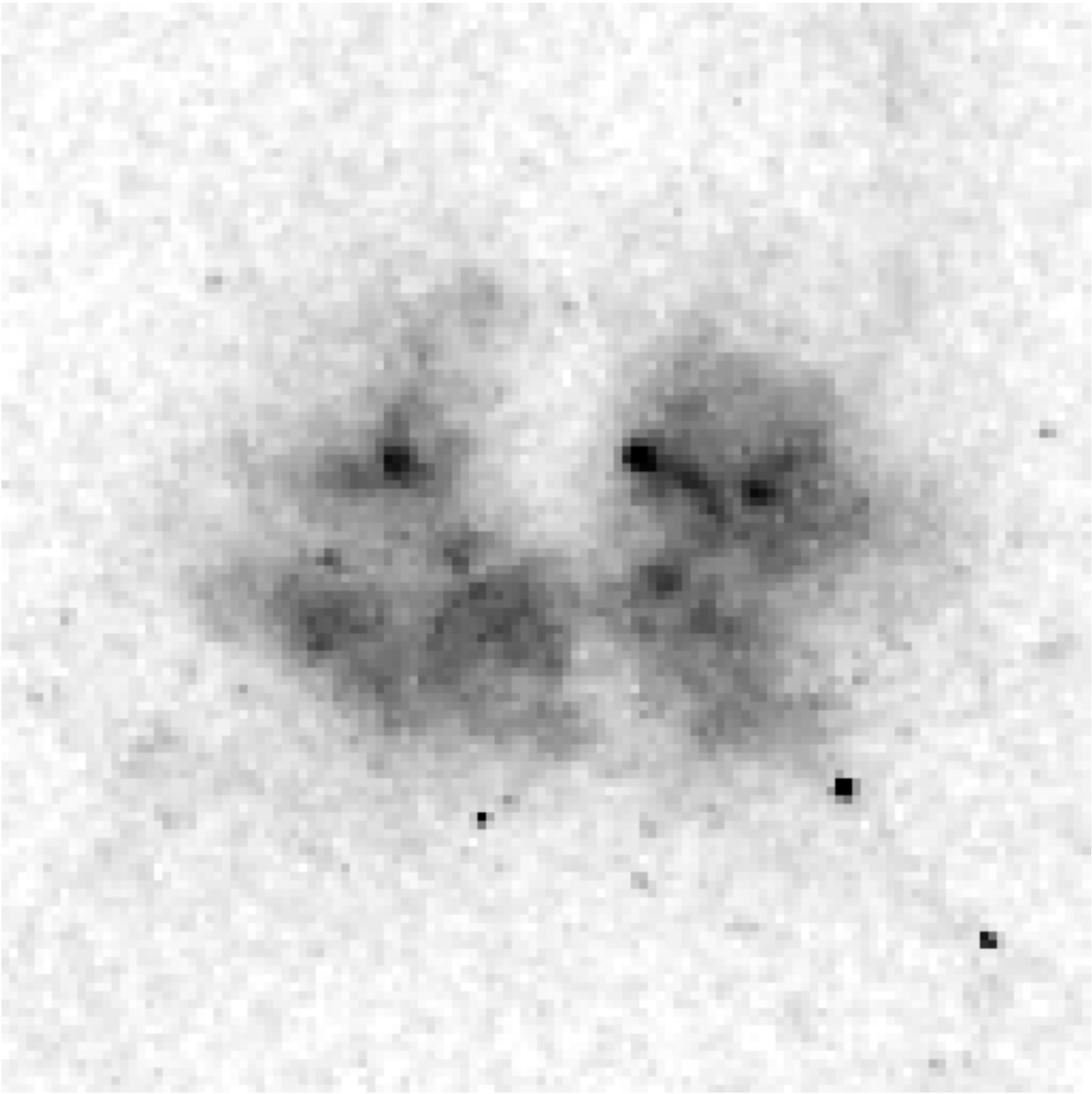} 
 \includegraphics[width=10cm]{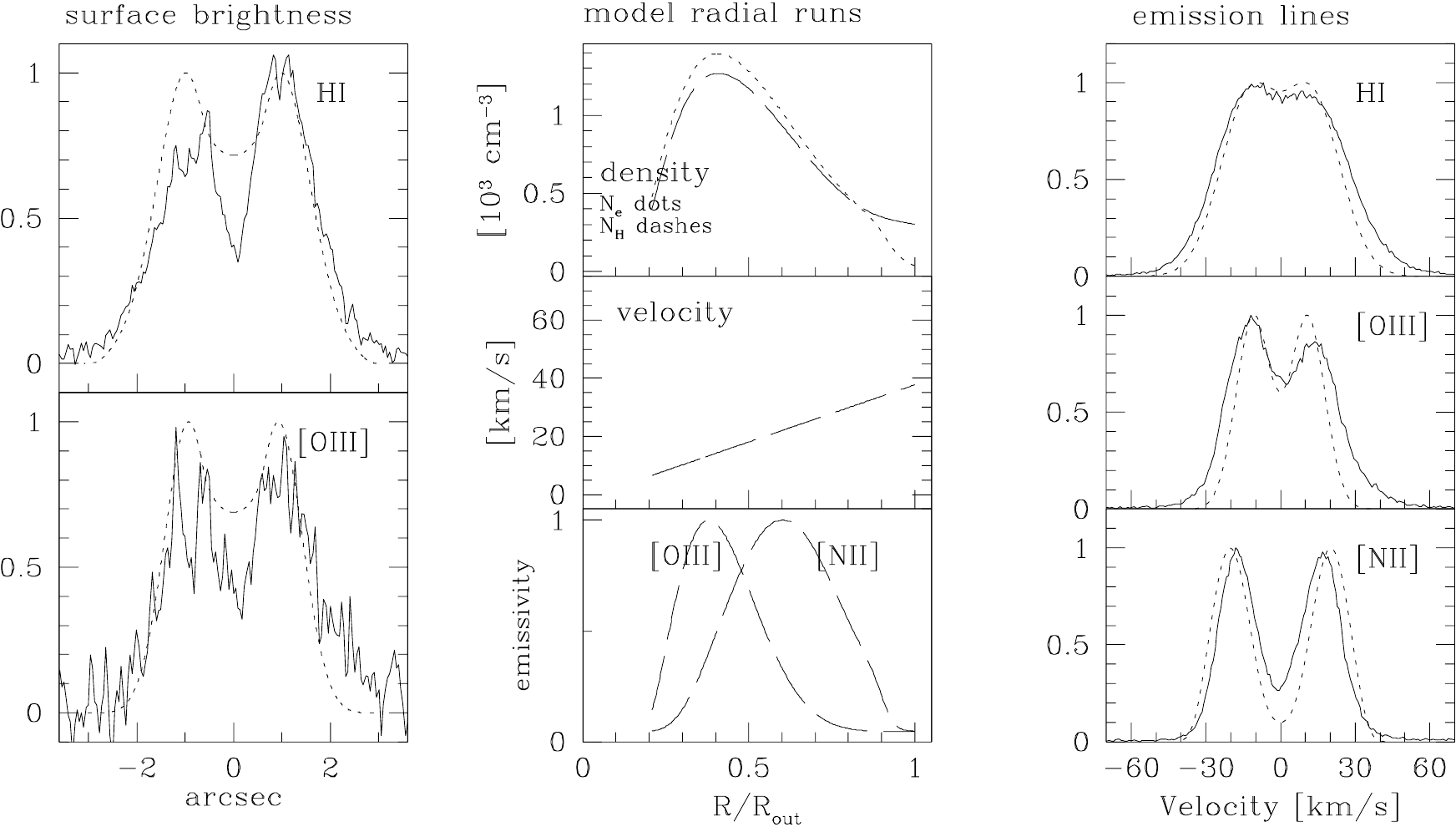}
 \caption{The nebula H\,2-27 (PN\,G\,356.5-03.6). The data are presented as in
   Fig.\,\ref{mod_fit_1}.    }
\end{figure*}

\begin{figure*}
\centering
 \includegraphics[width=5.5cm]{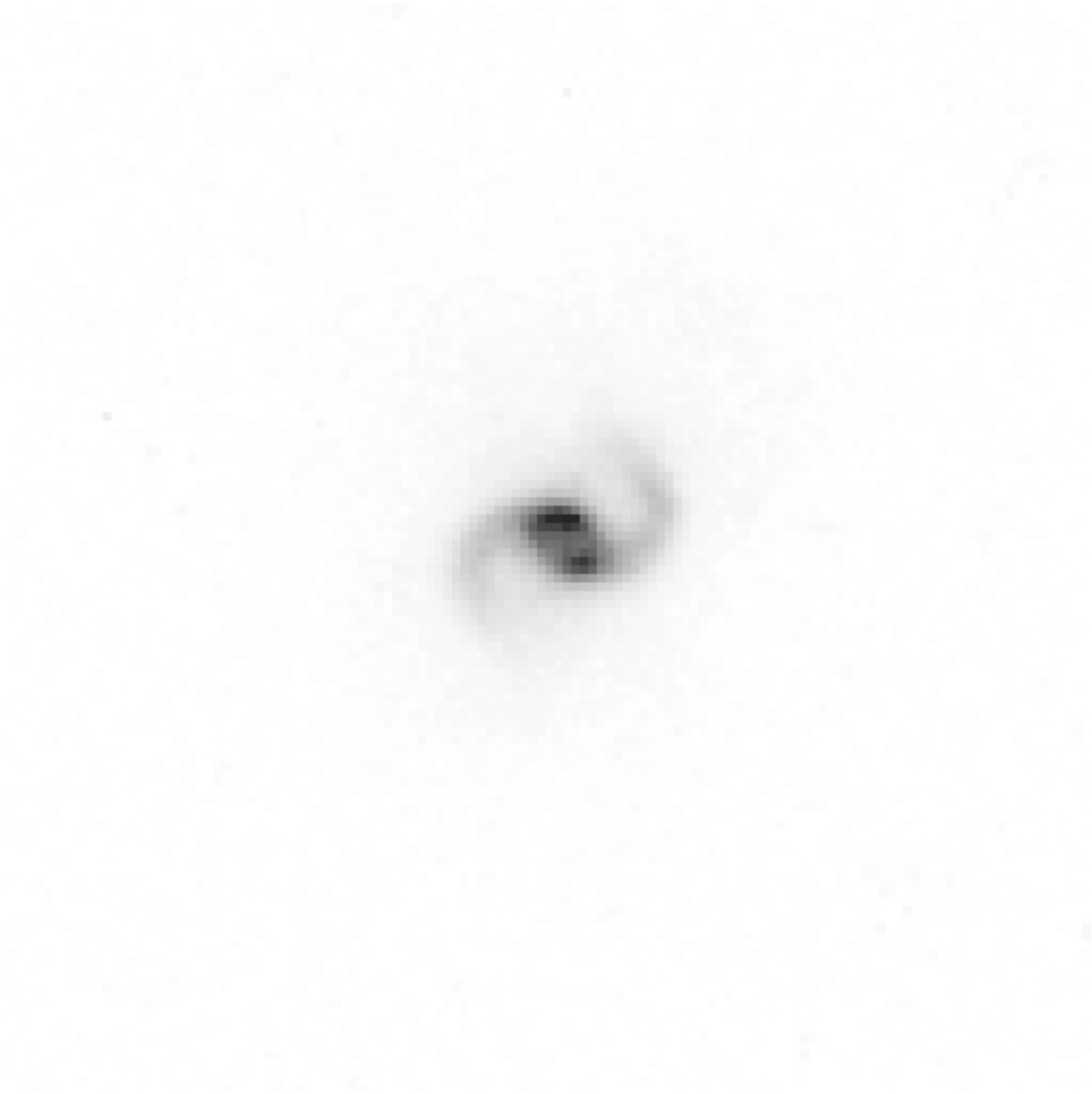} 
 \includegraphics[width=10cm]{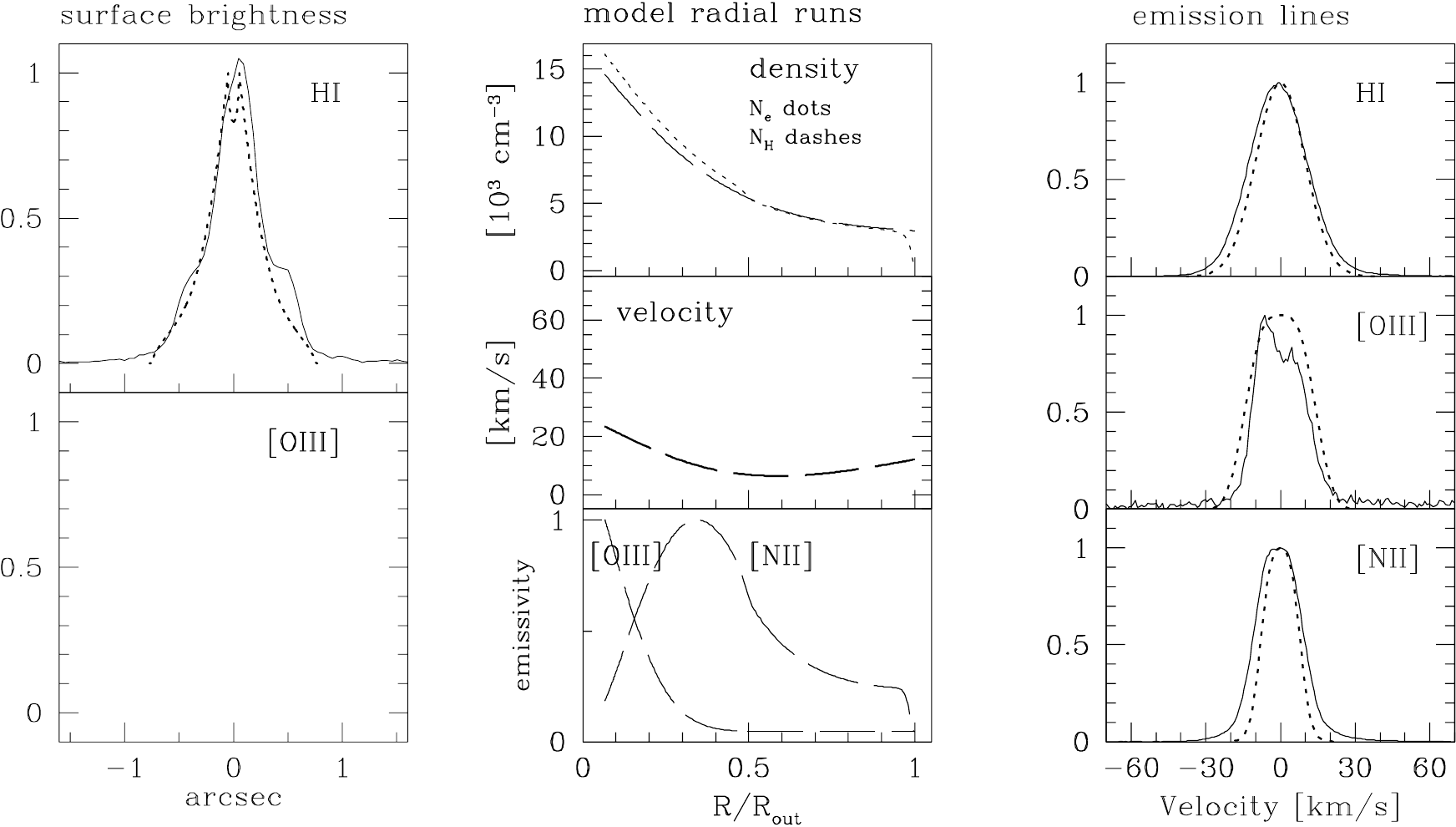}
 \caption{The nebula Th\,3-12 (PN\,G\,356.8+03.3). The data are presented as
   in Fig.\,\ref{mod_fit_1}. The very compact nebula shows a spiral-like
   structure, likely tracing a bipolar flow.  The profiles do not show
   evidence for high-velocity flows and are well fitted.   }
\end{figure*}

\begin{figure*}
\centering
 \includegraphics[width=5.5cm]{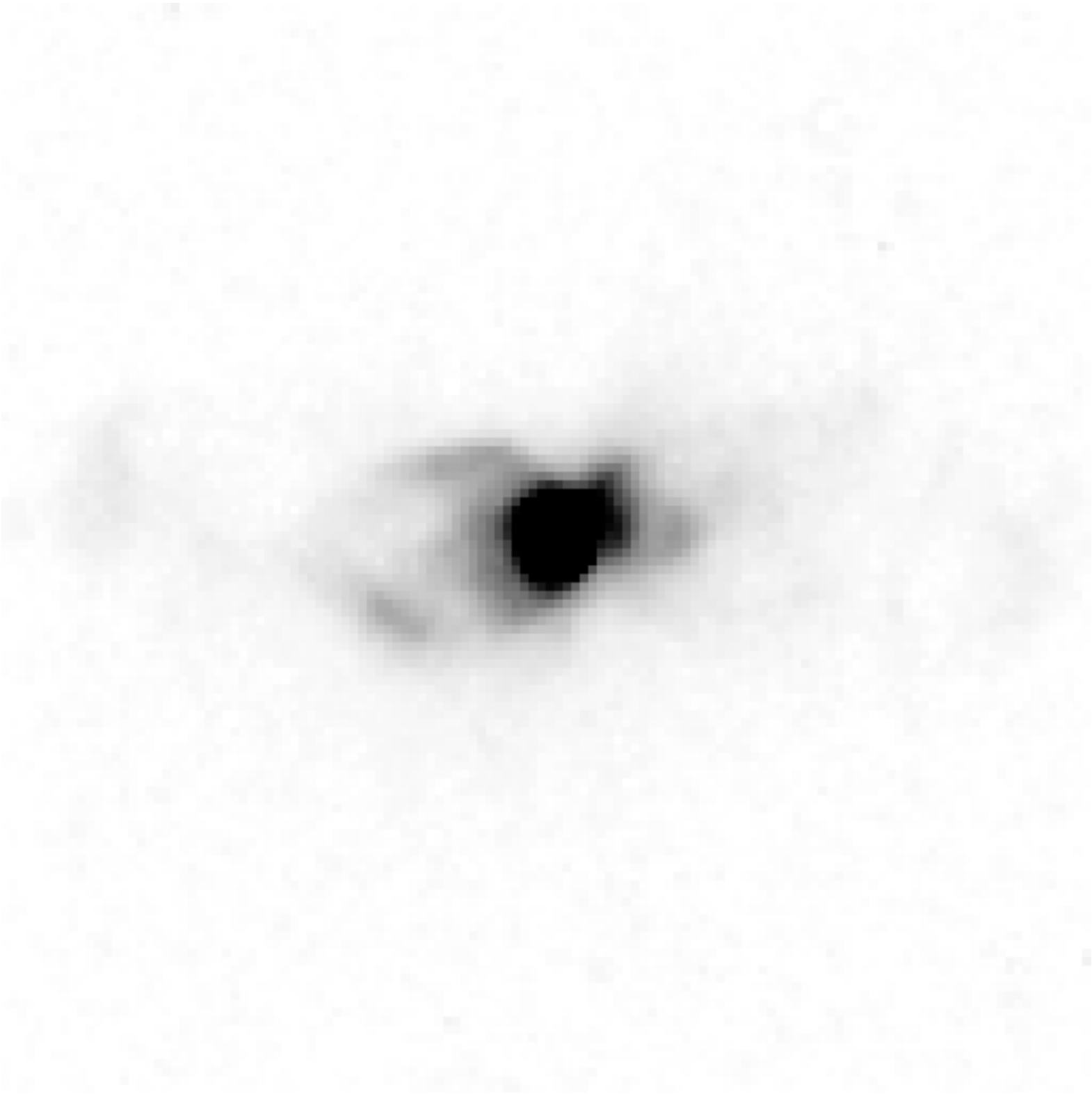} 
 \includegraphics[width=10cm]{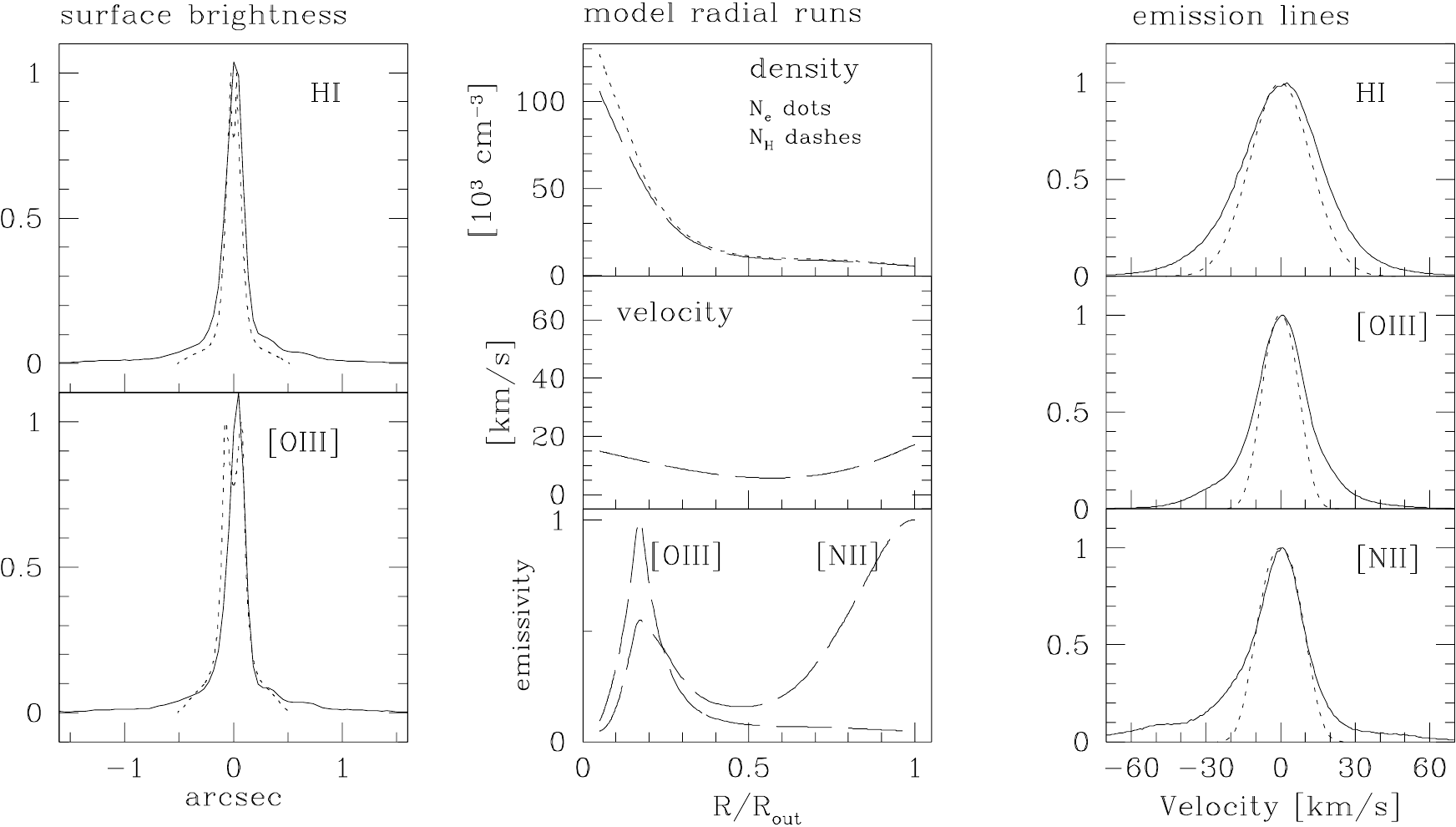}
 \caption{The nebula M\,3-38 (PN\,G\,356.9+04.4). The data are presented as in
   Fig.\,\ref{mod_fit_1}. The slit is oriented along the major axis.   }
\end{figure*}

\begin{figure*}
\centering
 \includegraphics[width=5.5cm]{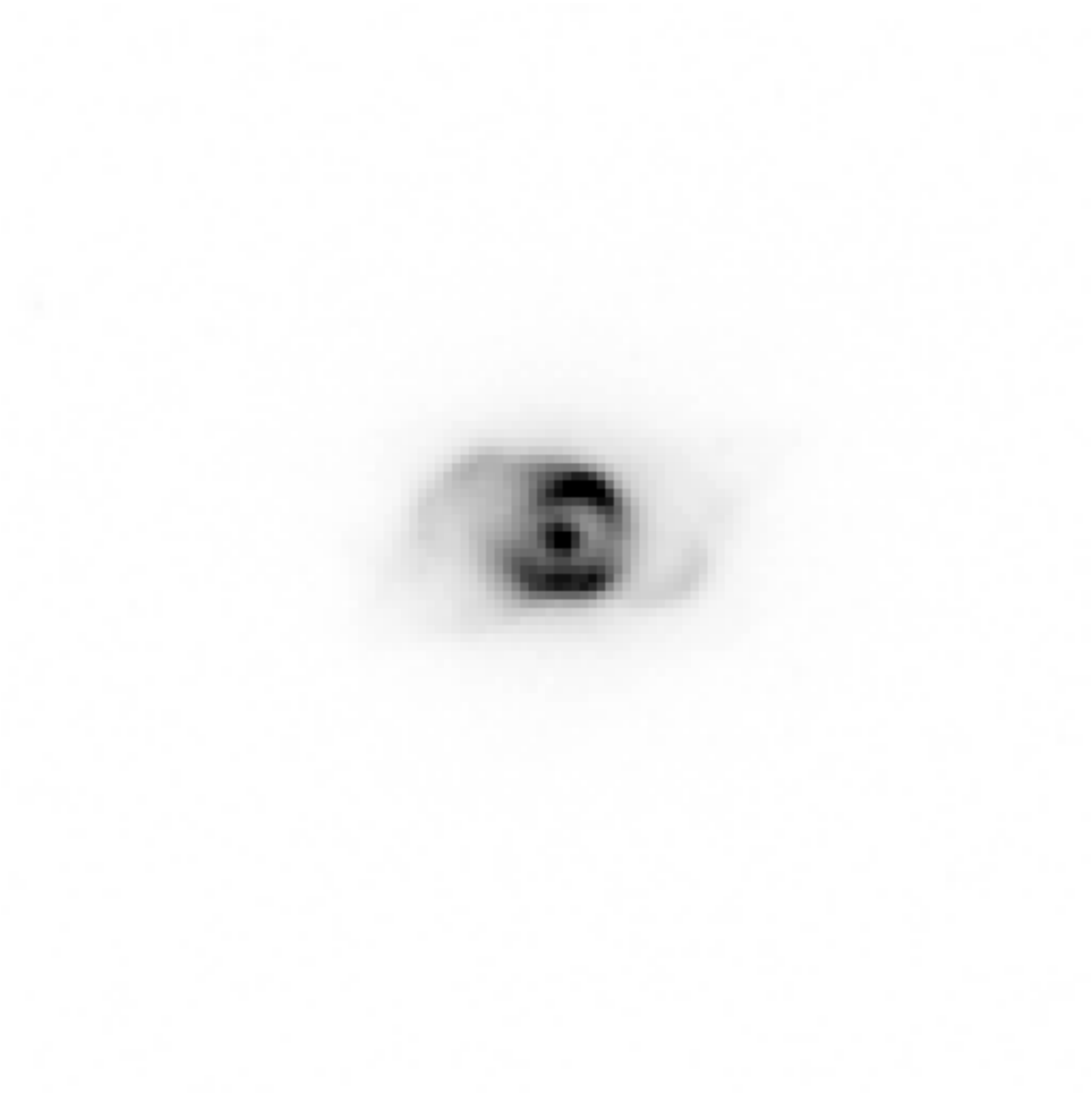} 
 \includegraphics[width=10cm]{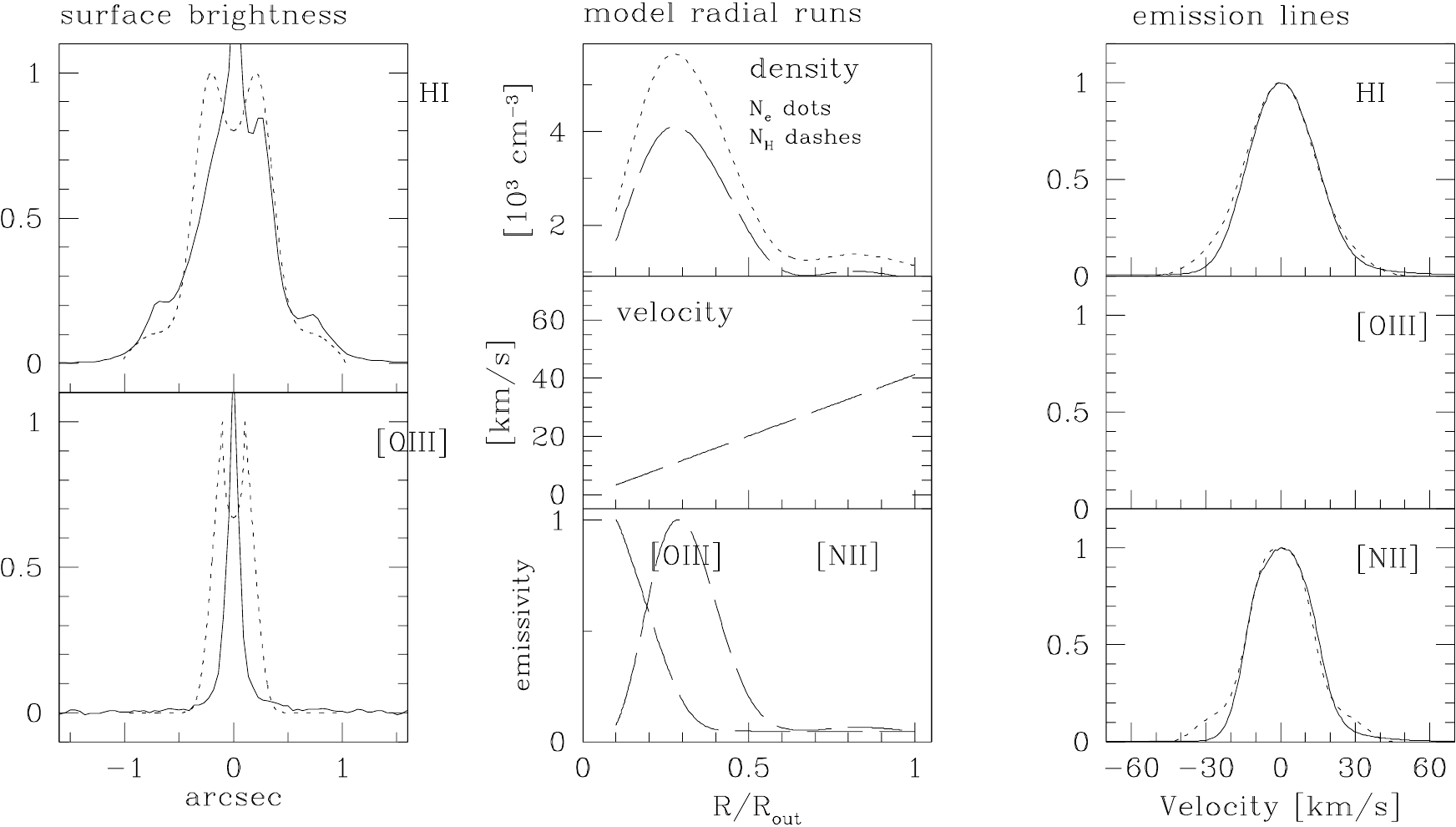}
 \caption{The nebula H\,1-43 (PN\,G\,357.1-04.7). The data are presented as in
   Fig.\,\ref{mod_fit_1}.    }
\end{figure*}

\begin{figure*}
\centering
 \includegraphics[width=5.5cm]{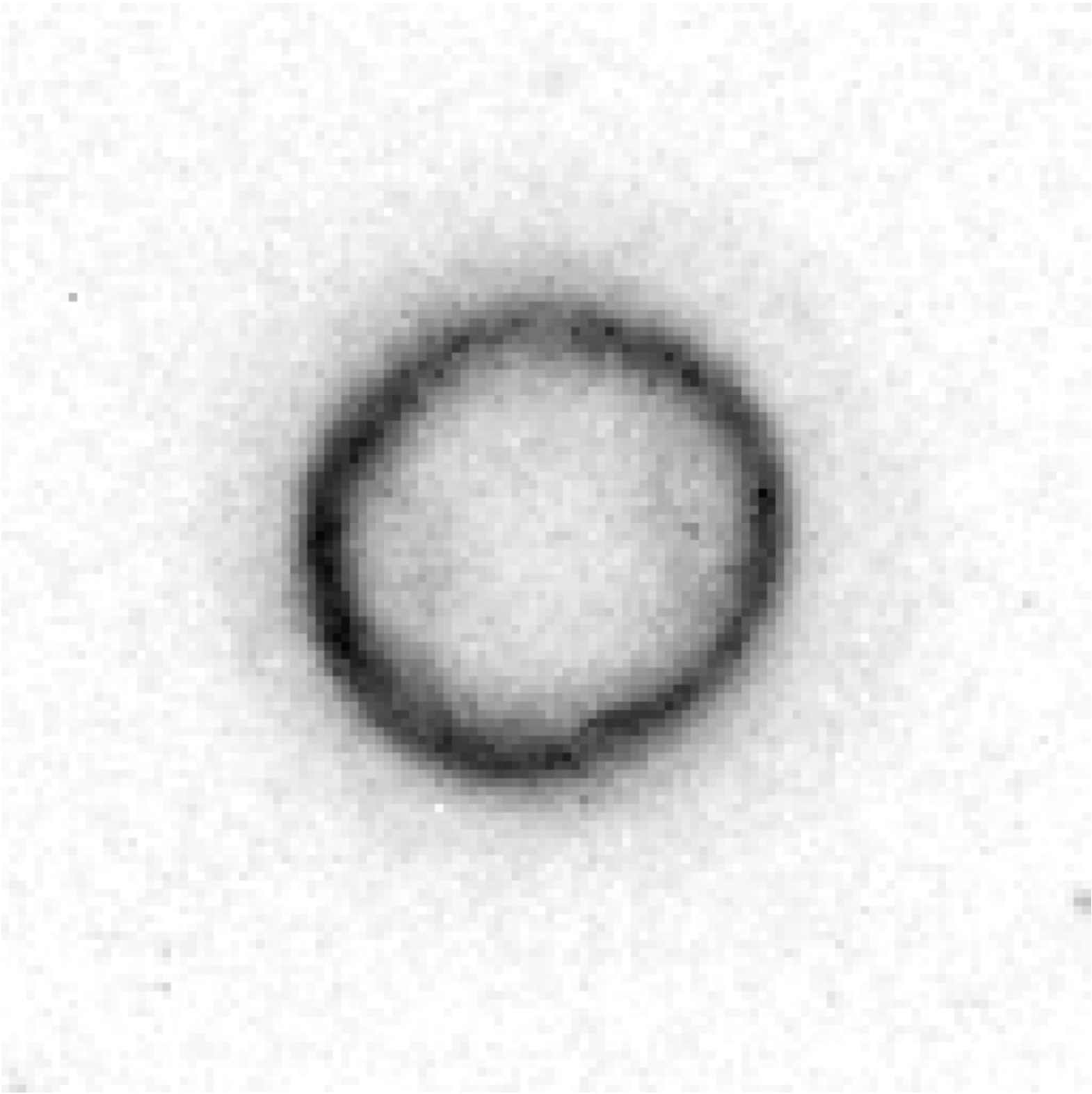} 
 \includegraphics[width=10cm]{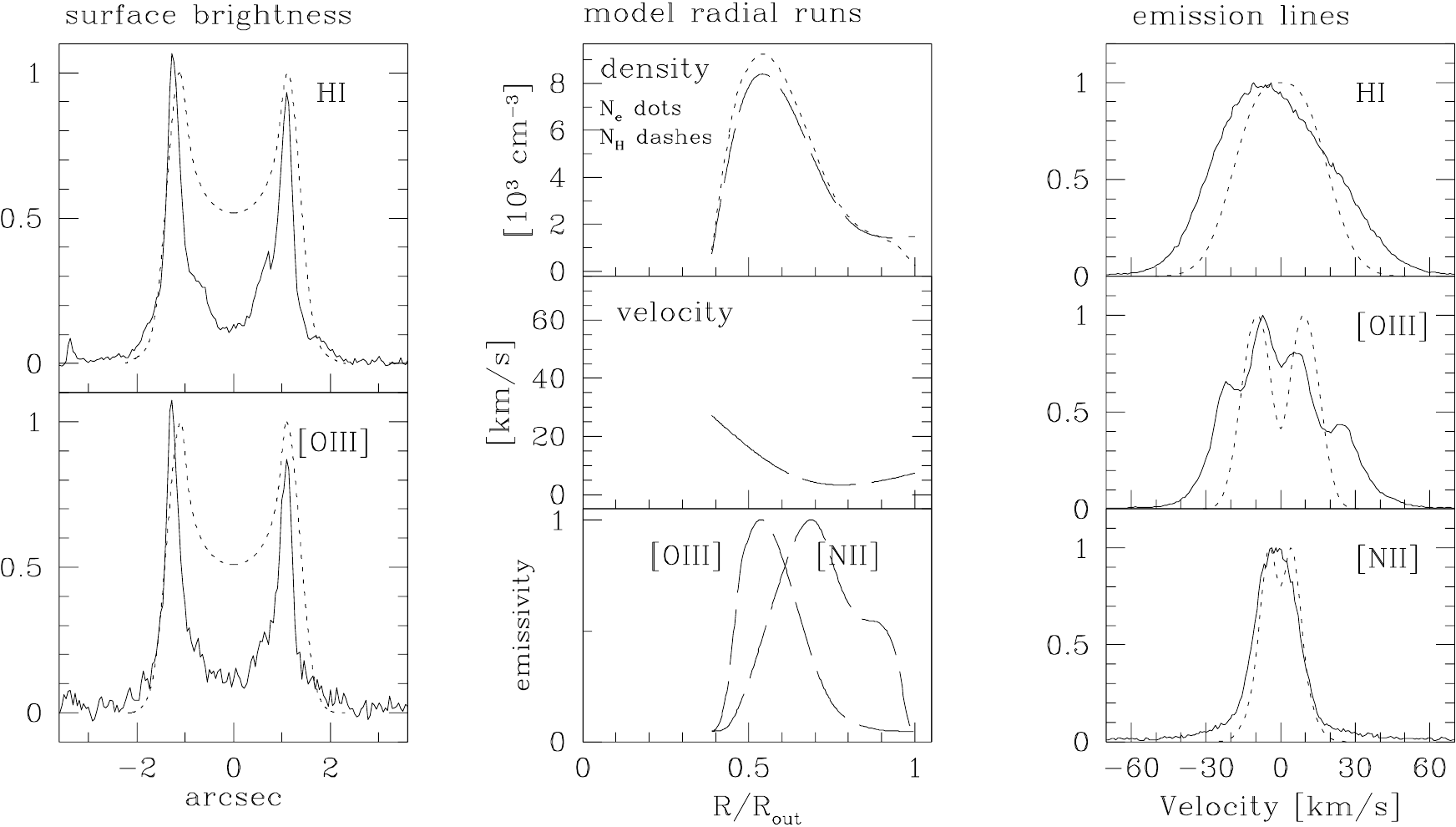}
 \caption{The nebula H\,2-13 (PN\,G\,357.2+02.0). The data are presented as in
   Fig.\,\ref{mod_fit_1}. The morphology is deceptively simple but the
   velocity profile, especially in [\ion{O}{iii}], is complex and the [\ion{O}{iii}] line is
   (unusually) wider than [\ion{N}{ii}]. Note that the profile is extracted from the
     inner arcsecond and not from the bright rim. This may be a bipolar nebula 
     seen pole-on with polar flows contributing to the line broadening.   }
\end{figure*}

\begin{figure*}
\centering
 \includegraphics[width=5.5cm]{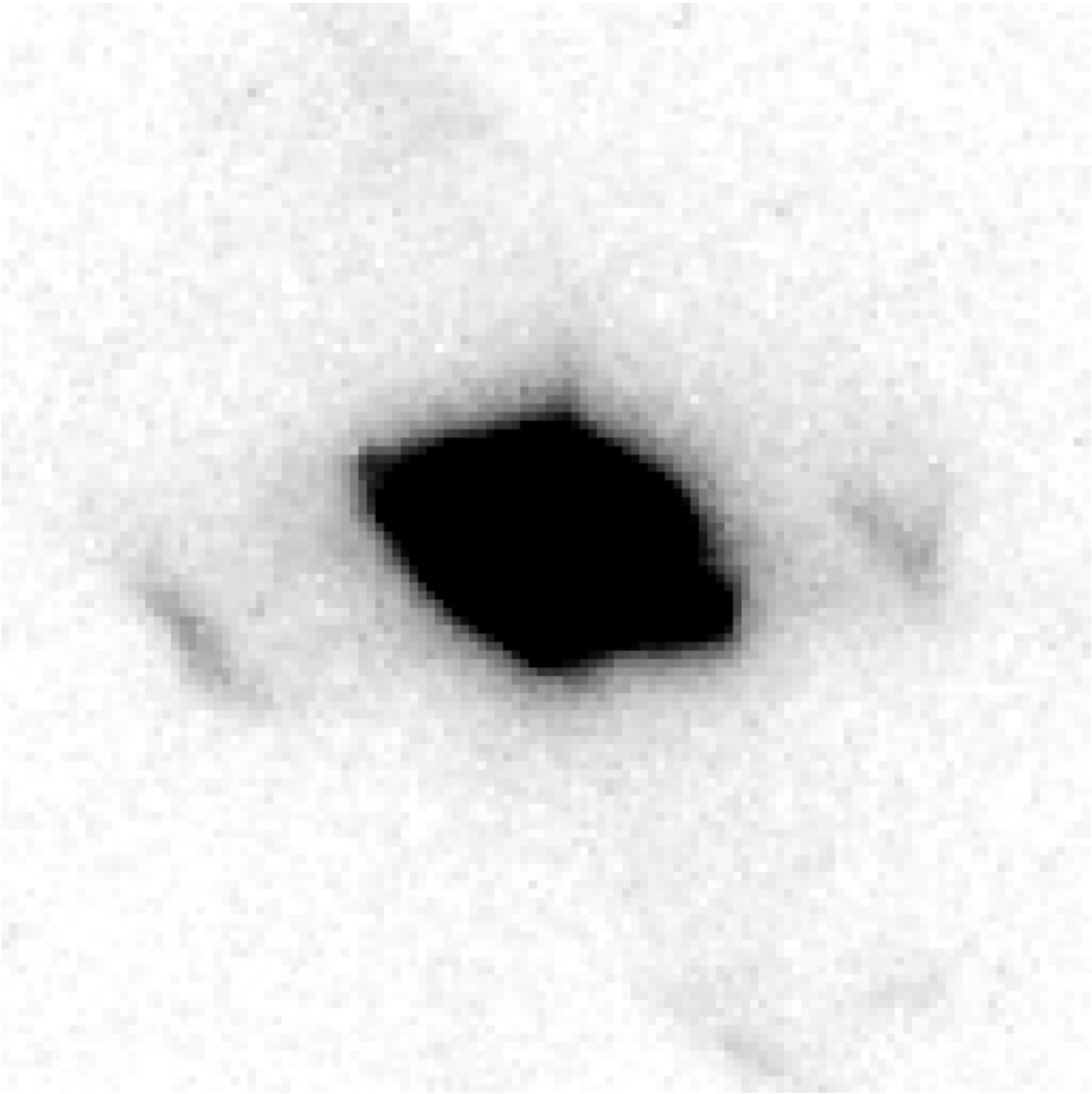} 
 \includegraphics[width=10cm]{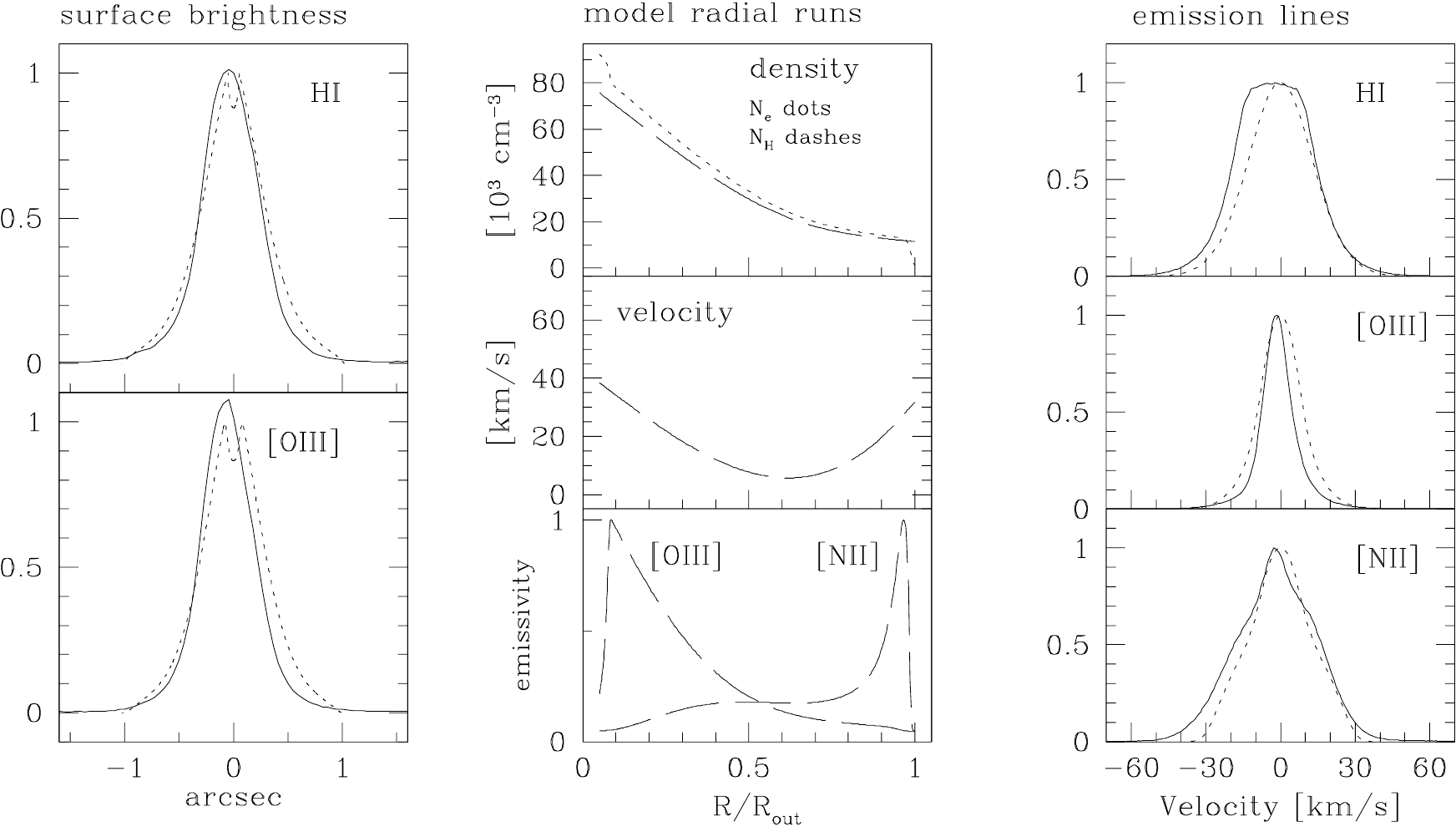}
 \caption{The nebula H\,1-46 (PN\,G\,358.5$-$04.2). The data are presented as
   in Fig.\,\ref{mod_fit_1}.  The slit is oriented along the major axis.   }
\end{figure*}

\begin{figure*}
\centering
 \includegraphics[width=5.5cm]{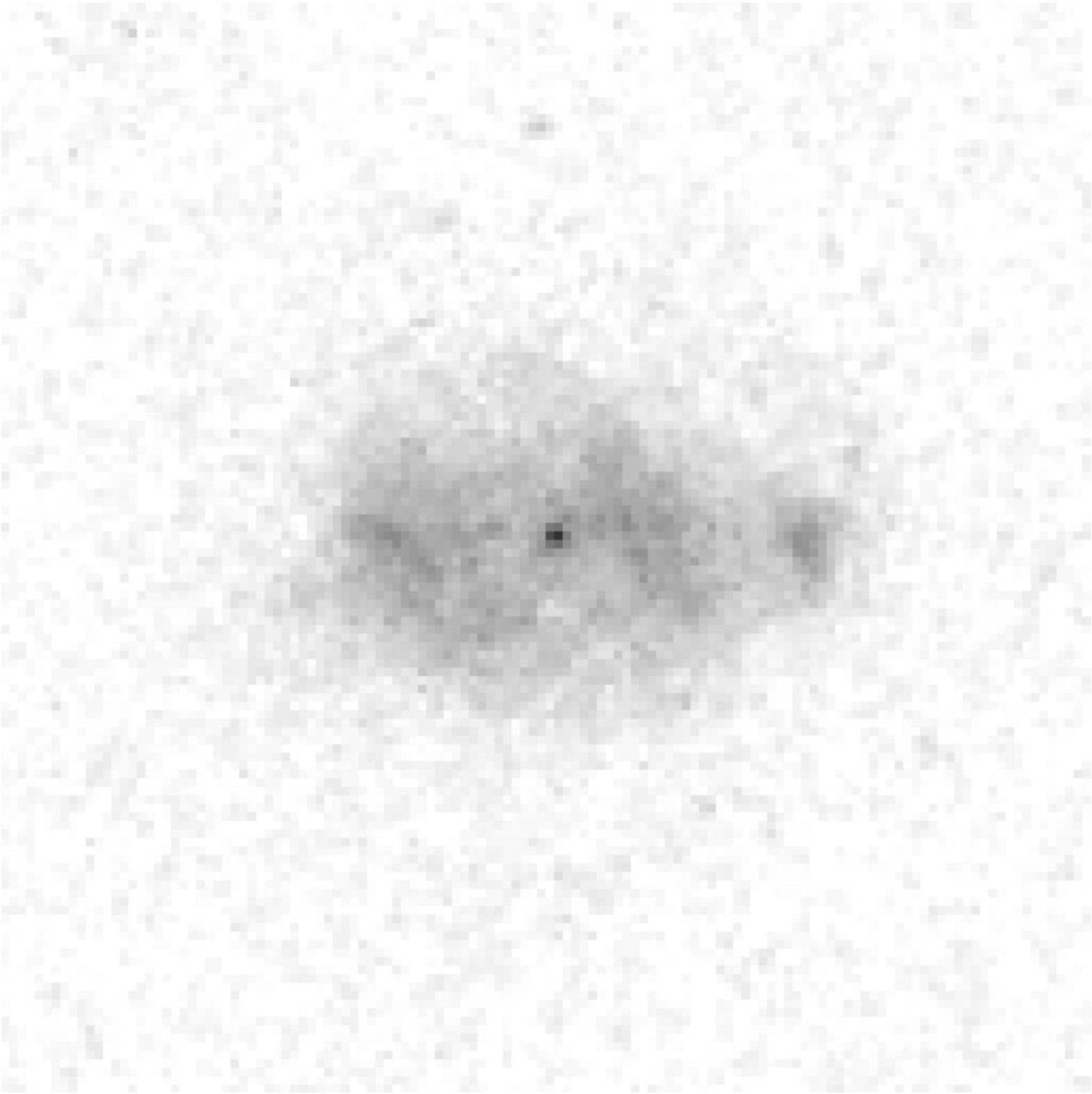} 
 \includegraphics[width=10cm]{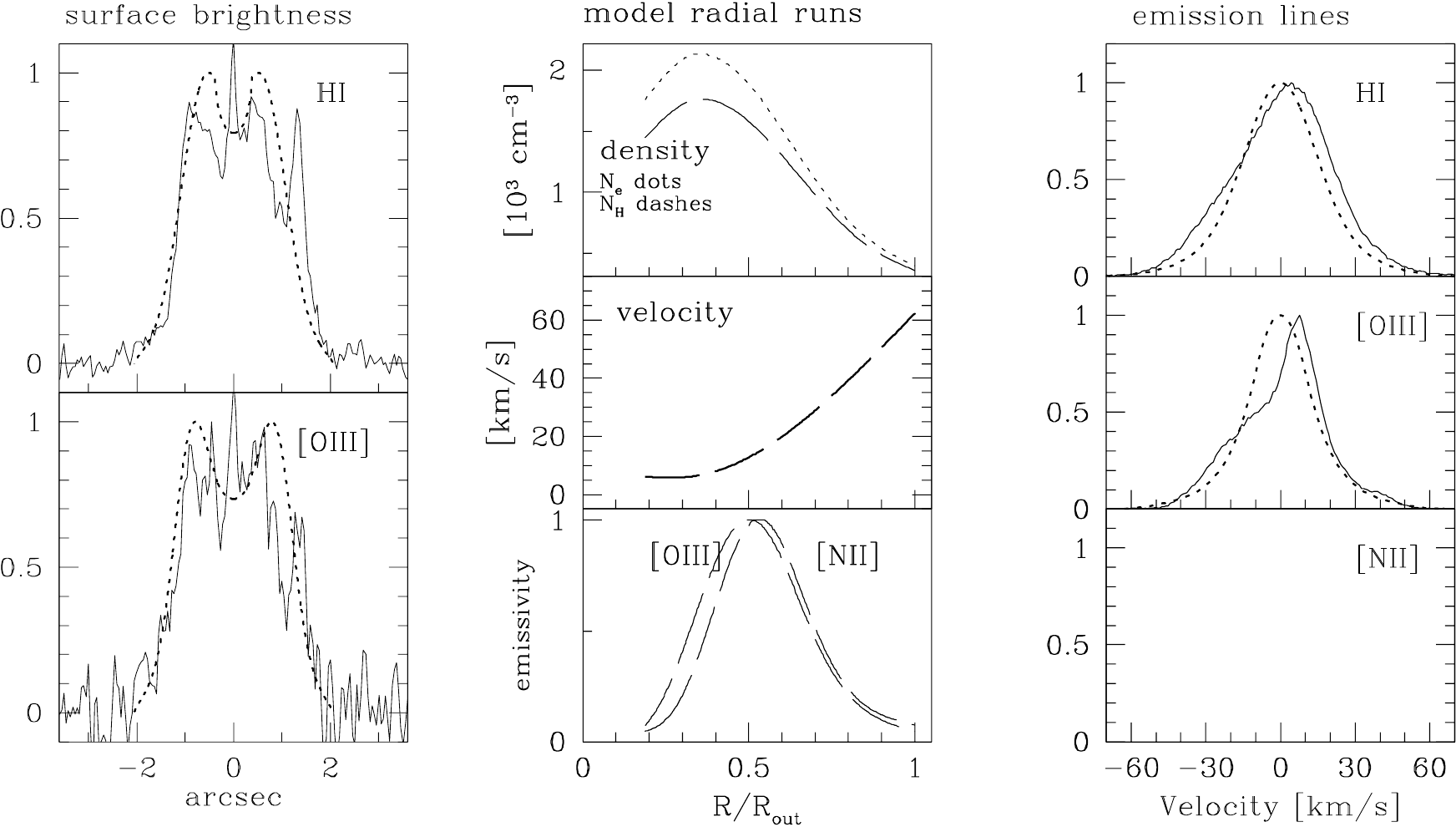}
 \caption{The nebula Al\,2-F (PN\,G\,358.5+02.9). The data are presented as in
   Fig.\,\ref{mod_fit_1}. A quadrupolar nebula.   }
\end{figure*}

\begin{figure*}
\centering
 \includegraphics[width=5.5cm]{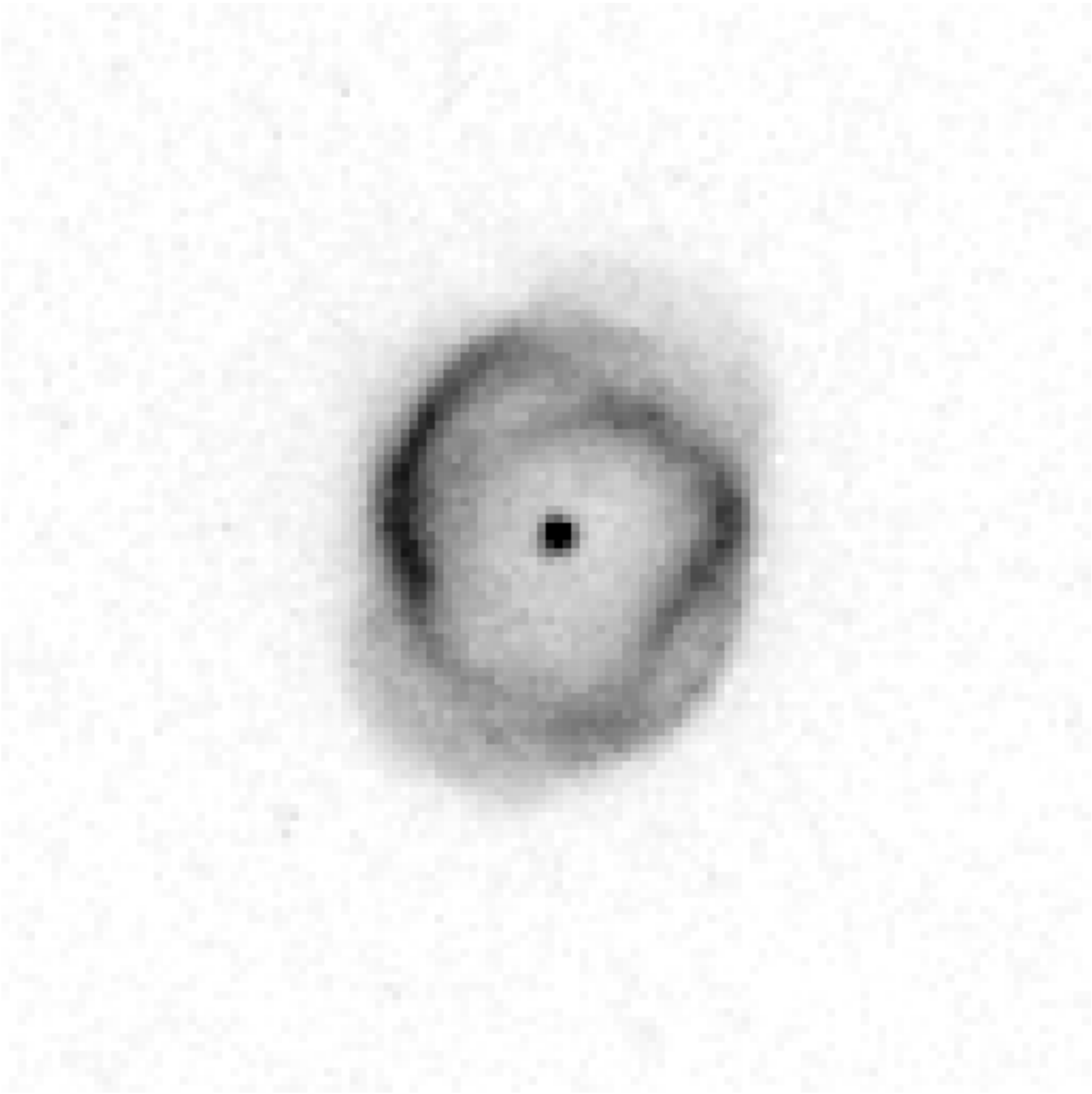} 
 \includegraphics[width=10cm]{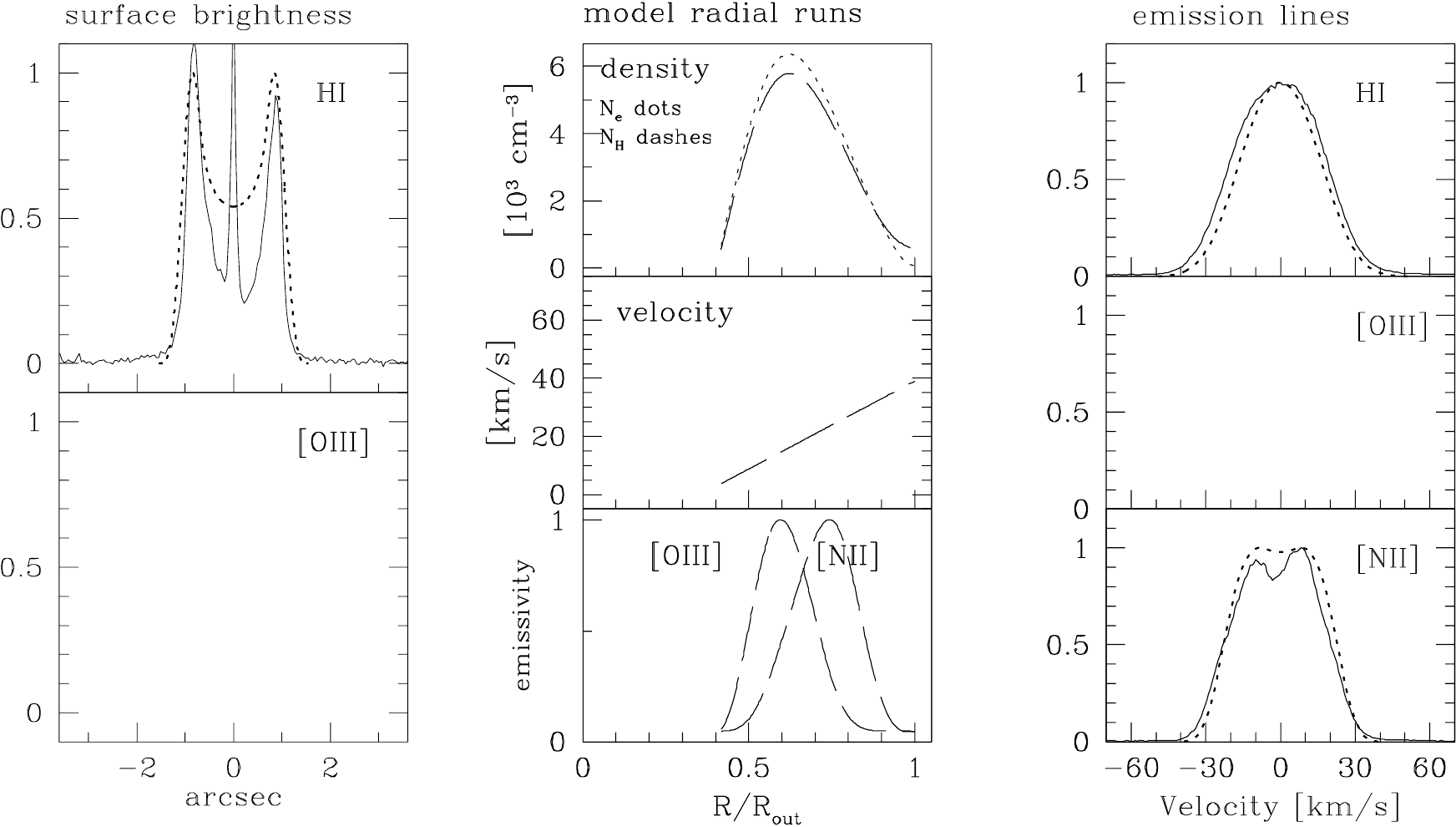}
 \caption{The nebula M\,3-40 (PN\,G\,358.7+05.2). The data are presented as in
   Fig.\,\ref{mod_fit_1}.    }
\end{figure*}

\begin{figure*}
\centering
 \includegraphics[width=5.5cm]{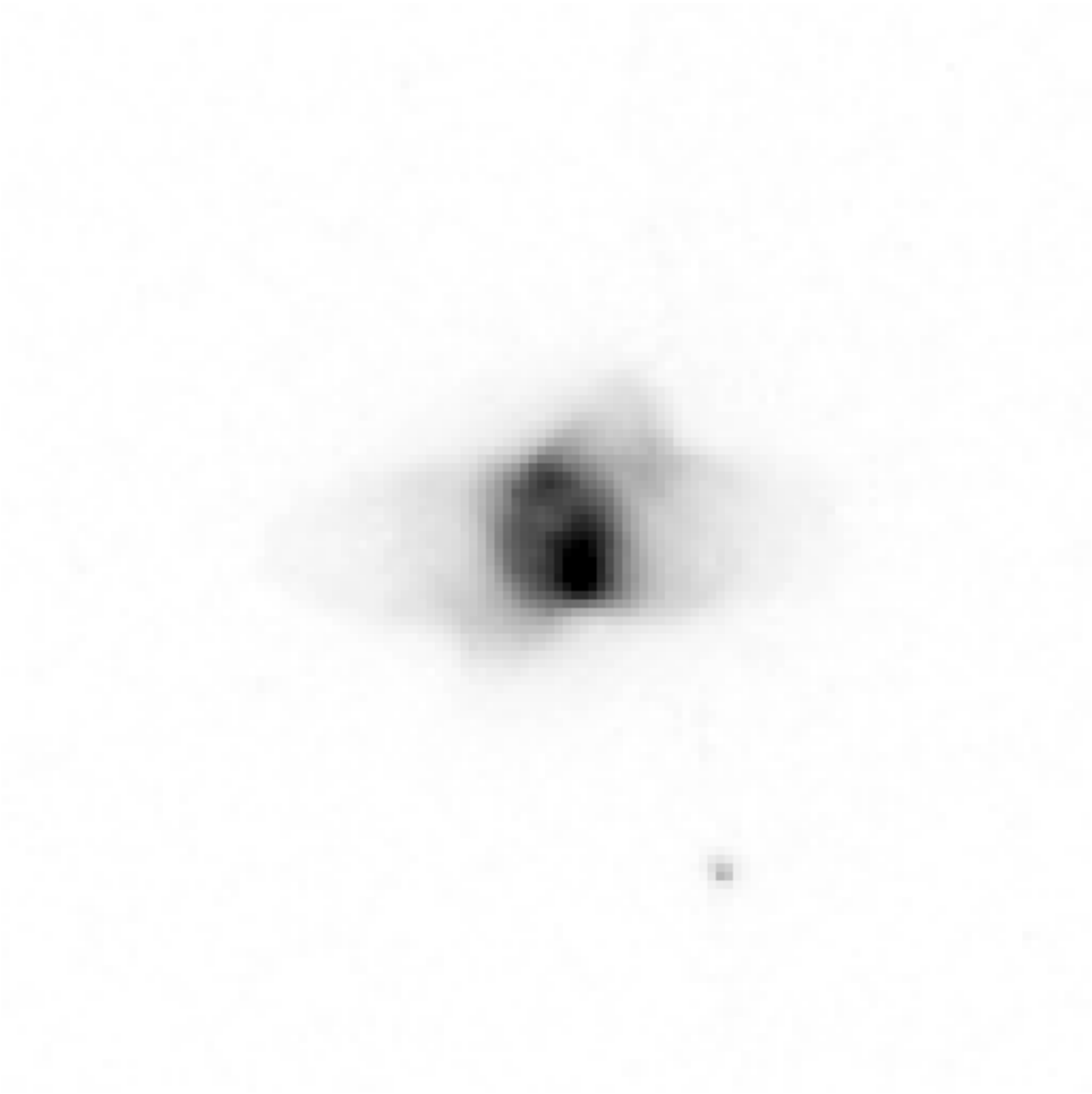} 
 \includegraphics[width=10cm]{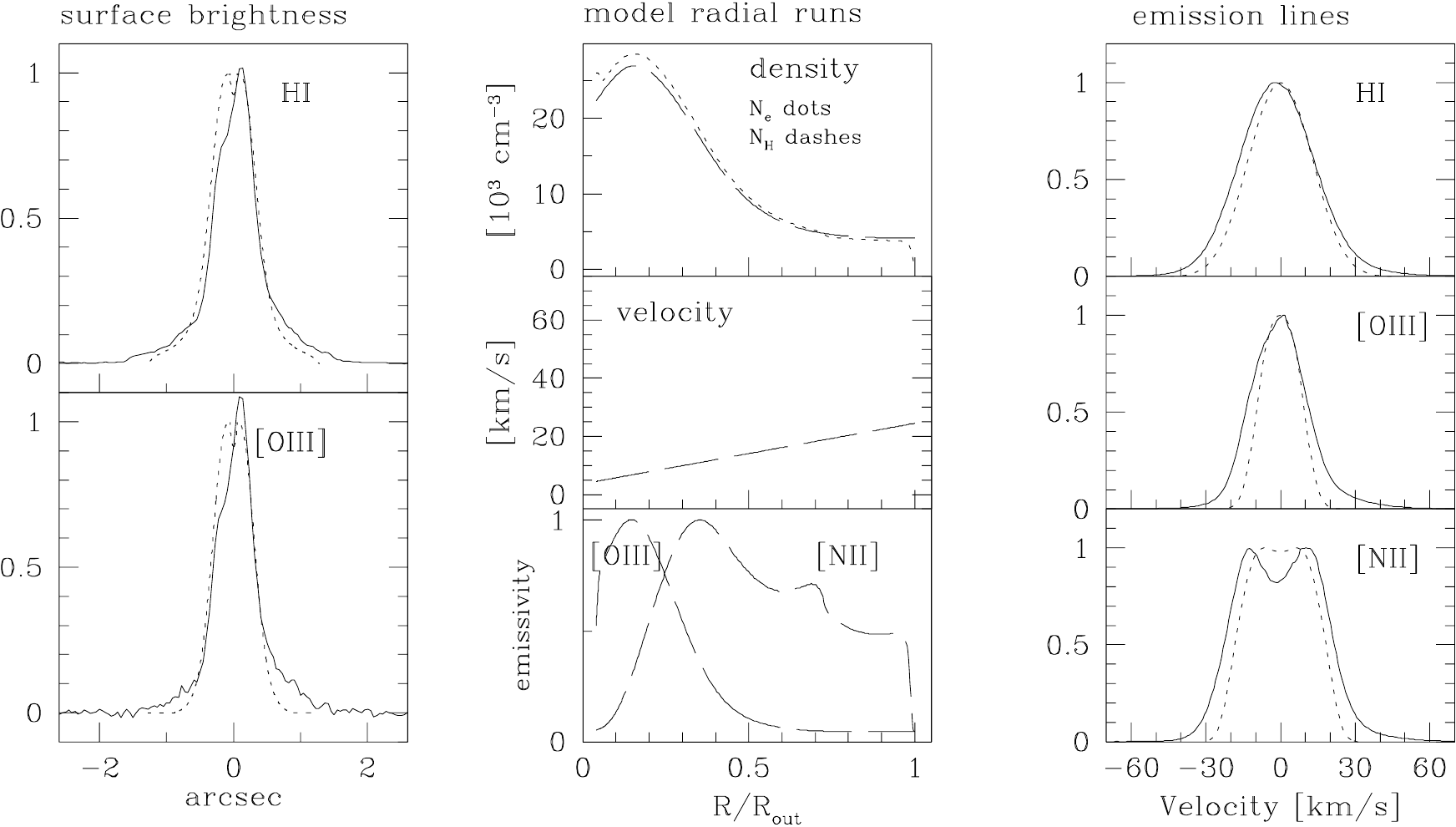}
 \caption{The nebula H\,1-19 (PN\,G\,358.9+03.4). The data are presented as in
   Fig.\,\ref{mod_fit_1}. The slit is oriented along the major axis.   }
\end{figure*}

\begin{figure*}
\centering
 \includegraphics[width=5.5cm]{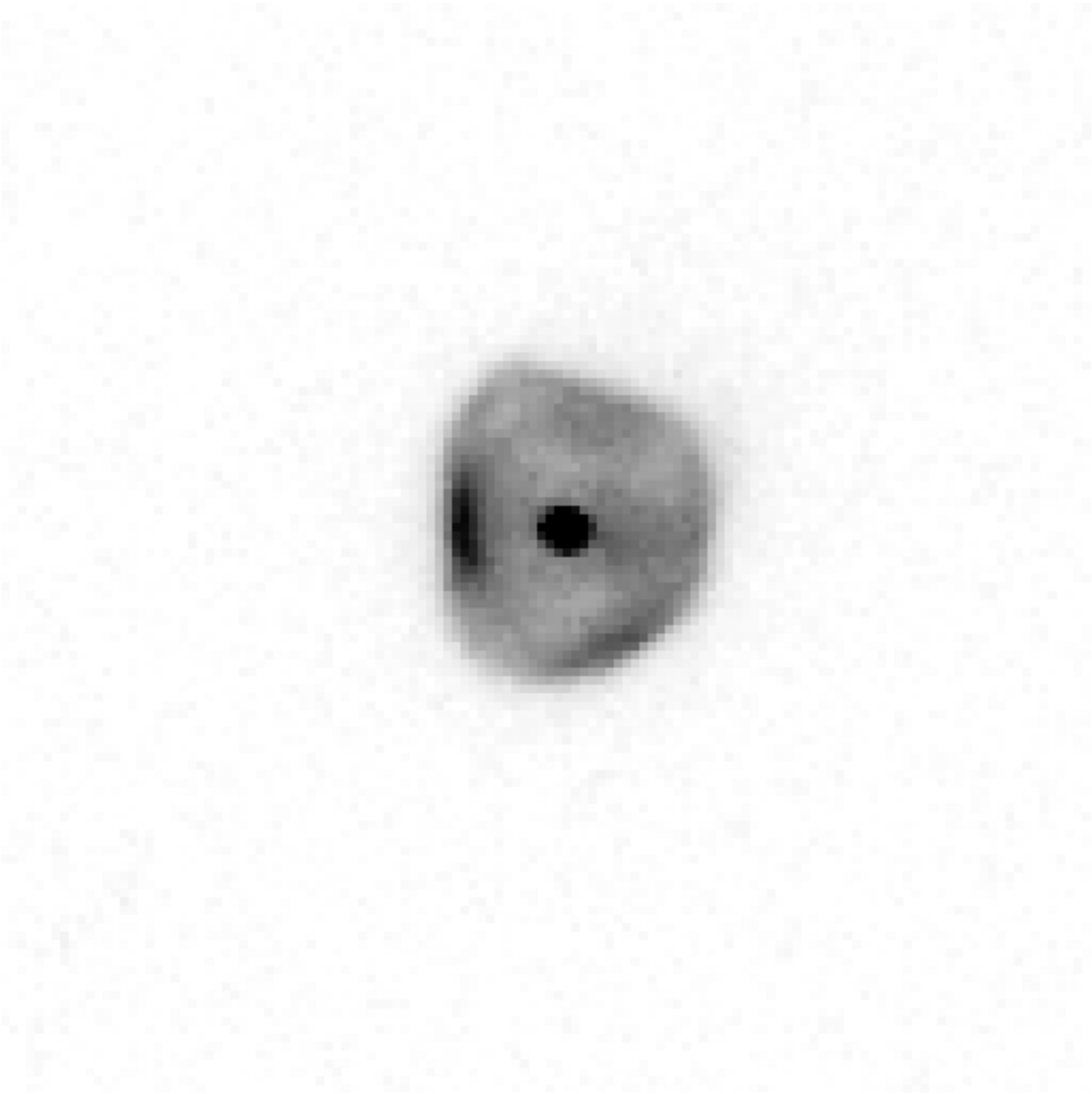} 
 \includegraphics[width=10cm]{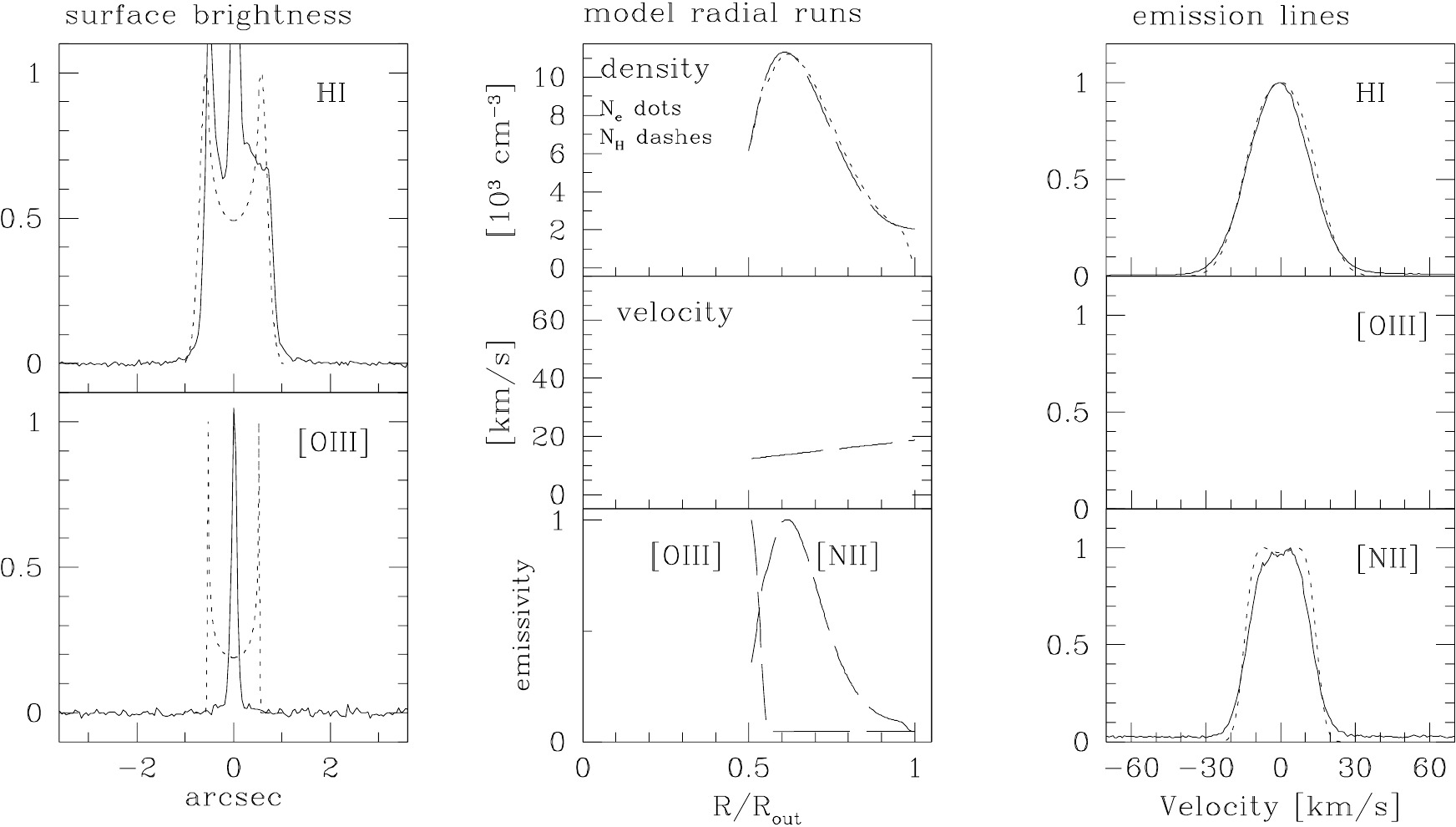}
 \caption{The nebula Th\,3-14 (PN\,G\,359.2+04.7). The data are presented as
   in Fig.\,\ref{mod_fit_1}. There is no [\ion{O}{iii}] detected: the spatial profile
   appears to show  stellar continuum only.  }
    \label{mod_fit_last}
\end{figure*}

\end{appendix}

\end{document}